%% file: main.tex
\documentclass[twocolumn]{svjour3}

\smartqed  

\input{preamble}
\input{preamble/preprint}

\journalname{Software and System Modeling}

\DeclareFieldFormat{labelnumberwidth}{#1\adddot}
\DeclareSourcemap{
  \maps[datatype=bibtex, overwrite]{
    \map{
      \perdatasource{bib/primary-studies.bib}
      \step[fieldset=KEYWORDS, fieldvalue=primary, append]
    }
  }
}

\begin{document}

\placetextbox{0.5}{0.99}{\large\colorbox{gray!3}{\textcolor{WildStrawberry}{\textbf{Author pre-print.}}}}%

\placetextbox{0.5}{0.97}{\large\colorbox{gray!3}{\textcolor{WildStrawberry}{Publication appears in the \hreff{https://www.sosym.org/}{International Journal on Software and Systems Modeling (SoSyM)}.}}}%

\placetextbox{0.5}{0.05}{\colorbox{gray!3}{\textcolor{WildStrawberry}{Author pre-print. Publication appears in} \hreff{https://www.sosym.org/}{SoSyM}.}}%

\title{Systems of Twinned Systems: A Systematic Literature Review}

\titlerunning{Systems of Twinned Systems: A Systematic Literature Review}

\author{
    Feyi Adesanya\protect\affmark[1] \and
    Kanan Castro Silva\protect\affmark[2] \and
    Valdemar V. Graciano Neto\protect\affmark[3] \and
    Istvan David\protect\affmark[1,4] 
}

\authorrunning{F. Adesanya et al.}

\institute{
    \affaddr{\affmark[1]McMaster University, Canada} \\
    \affaddr{\affmark[2]Universidade Federal do ABC, Brazil}\\
    \affaddr{\affmark[3]Universidade Federal de Goiás, Brazil}\\
    \affaddr{\affmark[4]McMaster Centre for Software Certification, Canada}\\
    \email{istvan.david@mcmaster.ca}
}

\date{Received: date / Accepted: date}

\emergencystretch 3em

\maketitle

\begin{acknowledgements}
We acknowledge the support of the Natural Sciences and Engineering Research Council of Canada (NSERC), DGECR-2024-00293.
\end{acknowledgements}

\input{sections/abstract}

\input{sections/intro}
\input{sections/background}
\input{sections/study-design}
\input{sections/classification}
\input{sections/results}
\input{sections/discussion}
\input{sections/conclusion}

\newrefcontext[labelprefix=PS]
\printbibliography[keyword={primary},title={Primary studies},resetnumbers=true]

\newrefcontext
\printbibliography[notkeyword={primary},resetnumbers=true]

\end{document}

%% file: preamble.tex
\input{preamble/packages}
\input{preamble/commands}

\input{preamble/settings}

%% file: preamble/packages.tex
\usepackage[utf8]{inputenc}
\usepackage{ifthen}
\usepackage{hyperref}
\usepackage[dvipsnames]{xcolor}

\usepackage{soul}
\usepackage[normalem]{ulem}
\usepackage{graphicx}

\usepackage{epsf,picinpar}
\usepackage{varioref}
\usepackage{fdsymbol}

\usepackage[dvipsnames]{xcolor}

\usepackage{enumerate}
\usepackage[framemethod=TikZ]{mdframed}

\usepackage{pifont}
\usepackage[pscoord]{eso-pic}
\usepackage{fancyvrb}

\usepackage[backend=biber,defernumbers=true,style=numeric,maxcitenames=1,maxbibnames=10]{biblatex}

\usepackage{booktabs}
\usepackage{tabularx}
\usepackage{multirow}
\usepackage{float}
\usepackage{array}
\usepackage{harveyballs}
\usepackage{makecell}
\usepackage{wrapfig}
\usepackage{svg}
\usepackage{orcidlink}
\usepackage{listings}
\usepackage{tablefootnote}
\usepackage[nomessages]{fp}%
\usepackage{subcaption}
\usepackage{tikz}
\usetikzlibrary{arrows}

\usepackage[color=red!50!white,textsize=tiny]{todonotes}

\usepackage{ragged2e}

%% file: preamble/commands.tex
\newcommand{\secref}[1]{Sec.~\ref{#1}}
\newcommand{\figref}[1]{Fig.~\ref{#1}}
\newcommand{\tabref}[1]{Tab.~\ref{#1}}

\newcommand*{\affaddr}[1]{#1} 
\newcommand*{\affmark}[1][*]{\textsuperscript{#1}}

\newcommand{\corner}{$\rotatebox[origin=c]{180}{$\Lsh$}$}

\newcommand{\included}[1]{\textbf{$\rightarrow${#1}}}

\newenvironment{bottomlineframe}[0]
  {\mdfsetup{
    frametitleaboveskip=-\ht\strutbox,
    frametitlealignment=\center
    }
  \begin{mdframed}[nobreak=true]%
  }
  {\end{mdframed}}

\newenvironment{conclusionframe}[1]
  {\mdfsetup{
    frametitle={\colorbox{white}{\space#1\space}},
    innertopmargin=2pt,
    frametitleaboveskip=-\ht\strutbox,
    frametitlealignment=\center
    }
  \begin{mdframed}[nobreak=true]%
  }
  {\end{mdframed}}

\newcommand\xofy[3]{%
    \FPeval{perc}{round(100.0 / #2 * #1, 1)}%
    \ifthenelse{\equal{#3}{}}{%
        {#1} of {#2} (\perc\%)%
    }%
    {%
        {#1} of {#2} {#3} (\perc\%)%
    }%
}

\newcommand\xofyp[2]{%
    \FPeval{perc}{round(100.0 / #2 * #1, 1)}%
    {#1} of {#2} -- \perc\%%
}

\newcommand\perc[2]{%
    \FPeval{perc}{round(100.0 / #2 * #1, 1)}%
    \perc%
}

\newcommand\percp[3]{%
    \FPeval{perc}{round(100.0 / #2 * #1, 1)}%
    \ifthenelse{\equal{#3}{}}{%
        \perc\% ({#1} of {#2})%
    }%
    {%
        \perc\% ({#1} of {#2} {#3})%
    }%
}

\newcommand{\rquestion}[1]{%
\noindent\textbf{\textit{#1}}\\%
}

\newcommand{\colheight}{-.25}
\newlength{\maxlen}

\newcommand{\numberofstudies}{80}

\newcommand{\maindatabar}[2][\numberofstudies]{%
    \databar[#1]{#2}{32}{barcolor}
}

\newcommand{\subdatabar}[2][\numberofstudies]{%
    \databar[#1]{#2}{36}{subbarcolor}
}

\newcommand{\databar}[4][\numberofstudies]{%
    \settowidth{\maxlen}{#3}%
    \addtolength{\maxlen}{\tabcolsep}%
    \ifthenelse{\equal{#1}{}}{%
        \FPeval{perc}{round(100.0 / \studynumberdefault * #2, 1)}%
    }%
    {%
        \FPeval{perc}{round(100.0 / #1 * #2, 1)}%
    }%
    \FPeval\result{round(\perc / #3, 1)}%
    \rlap{\color{#4}\hspace*{-.5\tabcolsep}\rule[\colheight\ht\strutbox]{\result\maxlen}{1.2\ht\strutbox}}%
    \makebox[\dimexpr\maxlen-\tabcolsep][l]{#2 (\perc\%)}%
}

%% file: preamble/settings.tex
\definecolor{highlightcolor}{rgb}{.99, 1, .0}
\sethlcolor{highlightcolor}
\definecolor{orcidlogocol}{HTML}{A6CE39}
\definecolor{challengecolor}{RGB}{66, 66, 66}
\definecolor{barcolor}{HTML}{85d4ff}
\definecolor{subbarcolor}{HTML}{d8d8d8}
\definecolor{subsubbarcolor}{HTML}{d8d8d8}
\definecolor{linkblue}{RGB}{120,106,237}
\definecolor{deepblue}{HTML}{0080C6}
\definecolor{lightblue}{HTML}{8ed4fa}

\mdfsetup{%
   backgroundcolor=gray!3,
   roundcorner=3pt%
}

%% file: preamble/preprint.tex
\usepackage{eso-pic}
\definecolor{linkblue}{HTML}{006fbd}
\newcommand{\hreffinternal}[3]{\href{#1}{\textcolor{#3}{#2}}}
\newcommand{\hreff}[2]{\hreffinternal{#1}{#2}{linkblue}}

\newcommand{\placetextbox}[3]{
  \setbox0=\hbox{#3}
  \AddToShipoutPictureFG*{
    \put(\LenToUnit{#1\paperwidth},\LenToUnit{#2\paperheight}){\vtop{{\null}\makebox[0pt][c]{#3}}}%
  }%
}%

%% file: sections/abstract.tex
\begin{abstract}
Modern systems exhibit unprecedented complexity due to their increased scale, interconnectedness, and the heterogeneity of their digital and physical components.
In response to scaling challenges, the system of systems paradigm proposes flexible aggregations of subsystems into a larger whole, while maintaining the independence of subsystems to various degrees.
In response to the cyber-physical convergence, the digital twin paradigm proposes a tight coupling between digital and physical components through computational reflection and precise control.
As these two paradigms address distinct parts of the overall challenge, combining the two promises more comprehensive methods to engineer what we call systems of twinned systems. The noticeably growing body of knowledge on systems of twinned systems calls for a review of the state of the art.
In this work, we report on our systematic literature survey of systems of twinned systems. We screened over 2\,500 potential studies, of which we included 80 and investigated them in detail.
To converge system of systems and digital twins, we derive a classification framework for systems of twinned systems that is backward compatible with the currently accepted theories of system of systems and digital twins.
\end{abstract}

\keywords{Cyber-physical systems \and Digital twins \and System of systems \and SoTS \and Systematic review}

%% file: sections/intro.tex
\section{Introduction}\label{sec:intro}
Modern engineered systems are becoming more complex as they incorporate a greater number of diverse and autonomous components. This growing complexity is widely considered as one of the defining factors of modern systems engineering practices~\cite{incose2035}.
A prominent characteristic of such complexity is the emergence of the system of systems (SoS)~\cite{kotov1999systems} paradigm.
SoS are large-scale, distributed aggregations of independently developed and managed constituent systems~\cite{maier1998architecting}.
These constituent systems maintain operational and managerial independence but may opt to collaborate in order to fulfill shared, higher-level goals~\cite{boardman2006system}. As a result the SoS paradigm embodies flexibility, scalability, and adaptability in dynamic environments~\cite{gorod2008system}.
SoS increasingly extend into the virtual domain and are often comprised of constituent systems of cyber-physical in nature~\cite{olsson2023systems}. This sets the stage for combining digital twins (DT) and digital twin systems into SoS.

DTs are fit for purpose digital representations of observable physical elements with synchronization between the element and its digital representation~\cite{ISO23247-1}, enabling simulation, monitoring, and data-driven control in one technical platform~\cite{kritzinger2018digital}. DTs have demonstrated impact across several domains, including manufacturing~\cite{leng2021digital}, smart cities~\cite{gracianoneto2023what}, and agriculture~\cite{david2023digital}. Although many current DT implementations are still domain-specific and centralized, recent research points toward more distributed, modular, and interoperable forms~\cite{david2024interoperability, aziz2022empowering}. This shift indeed promotes DTs as potential constituents in SoS.

The convergence of SoS and DTs introduces a new class of systems, in which multiple DT systems (i.e., collections of digital twins and physical twins) are integrated into a whole by SoS principles. We refer to these systems as \textbf{Systems of Twinned Systems}.

\begin{bottomlineframe}
A System of Twinned Systems (SoTS) comprises digitally twinned systems organized by system-of-systems (SoS) principles, in which digitally twinned systems may act as autonomous constituents and collaborate to achieve complex goals.
\end{bottomlineframe}

\begin{figure}[htb]
    \centering
    \includegraphics[trim=0.75cm 0.5cm 0.75cm 0.5cm, clip,width=0.75\linewidth]{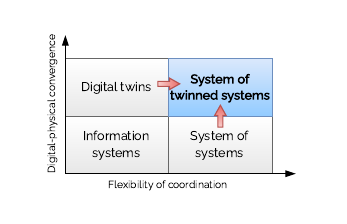}
    \caption{\vspace{-1em}Coordination vs. Convergence} 
    \label{fig:quadrants}
\end{figure}

SoTS are characterized by the duality of (i) weakly coupled constituents which, in turn, may exhibit (ii) strongly coupled digital (or cyber) and physical traits, such as DTs. Moreover, a SoTS may exhibit a hierarchical structure in which constituent systems may be SoTS themselves.
As shown in \figref{fig:quadrants}, SoTS improve over SoS by emphasizing the potential cyber-physical heterogeneity of constituents and the convergence of those cyber and physical subsystems. This, in turn, allows for stronger coupling at the constituent level, which is not present in current SoS theory. SoTS improve over DTs by allowing for the flexible coordination of otherwise highly autonomous DTs.

SoTS are different from federated and aggregated DTs in that SoTS follow SoS composition principles, which are more flexible, dynamic, and often, ad-hoc. In contrast, federated DTs are typically governed by a pre-defined set of rules ``that determine interoperability, coordination and secure communications''~\cite{vergara2023federated}.

Hereinafter, we use the term ``digitally twinned systems'' to describe the constituents and by this term, we mean any system---digital, physical, or cyber-physical---that is subject to digital twinning. This interpretation aligns with contemporary DT literature, which describes DTs as virtual representations of physical, cyber, and cyber-physical systems~\cite{barricelli2019survey}, including potentially hierarchical and multi-level system representations~\cite{grieves2017digital, tao2019digital}; and remains compatible with SoS theory, which characterizes constituent systems by their operational and managerial independence rather than by their physical, digital, or cyber-physical realization \cite{maier1998architecting}.

A pertinent example of a SoTS is a smart city where DTs of infrastructure, vehicles, and people collaborate to manage complex interactions, such as traffic optimization or energy balancing. In such a setting, e.g., vehicles can autonomously decide to be part of the traffic system or leave, impacting the overall flow of traffic. Another example is a platoon of autonomous vehicles in the traffic of a smart city. The vehicles do not have a central authority to control the platooning dynamics, e.g., enforcing safe inter-vehicle distance and headway with respect to changing road conditions. Instead, each vehicle has their own DT and these DTs work together to set the target distance and
speed, and then, each DT acts accordingly. These simple examples underscore the complexity and variety of architectures SoTS impose.

With research and development targeting SoTS on a noticeably accelerating course~\cite{olsson2023systems}, a systematic review of their engineering practices, technical characteristics, and use cases is needed.

\paragraph{Contributions}
In this article, we report on our systematic literature review of SoTS. We identify key trends and design choices in the organization of systems in such settings, SoS and DT patterns, tendencies in non-functional system properties, such as security, and outline relevant research and development directions for experts in the SoS and DT domains.

\begin{figure}[h]
    \centering
    \includegraphics[width=\linewidth]{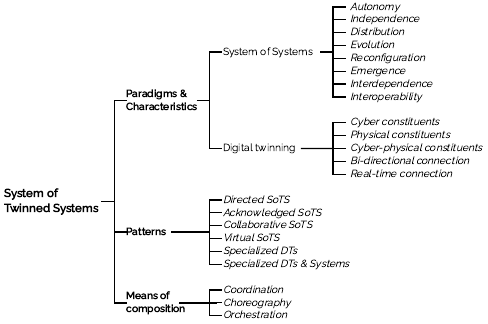}
    \caption{SoTS attributes}
    \label{fig:taxonomy}
\end{figure}

\paragraph{Taxonomy}
We set forth the taxonomy of SoTS attributes in \figref{fig:taxonomy}, derived from the findings of our work and elaborated in the remainder of this article.
\textit{Paradigms and Characteristics} define elementary traits of SoS and DTs. (See \secref{sec:background} for details.)
\textit{Patterns} define architectural variations of SoTS, derived from elementary SoS architectures under DT assumptions. \textit{Means of Composition} define mechanisms that establish a connection between subsystems. (See \secref{sec:classification} for details.)

\paragraph{Replicability} We publish a replication package containing the data and analysis scripts of our study at \url{https://zenodo.org/records/18891164}.

\paragraph{Structure} The remainder of this article is structured as follows.
In \secref{sec:background}, we review the background and the related work.
In \secref{sec:study-design}, we design a systematic literature review to study the state of the art in SoTS.
In \secref{sec:classification}, we define a classification framework for SoTS.
In \secref{sec:results}, we report the results of our review.
In \secref{sec:discussion}, we discuss the results and identify trends, tendencies, limitations and shortcomings, and key research challenges for the DT and SoS communities.
Finally, in \secref{sec:conclusion}, we draw the conclusions and identify future work.

%% file: sections/background.tex
\section{Background and related work}\label{sec:background}

In this section, we discuss the background in SoS (\secref{sec:sos}), DTs (\secref{sec:dt_background}), and the related work on combining SoS and DTs (\secref{sec:relatedwork}).

\subsection{System of systems}\label{sec:sos}
A SoS is a system composed of multiple independent systems that collaborate to achieve outcomes that no single system could accomplish alone \cite{maier1998architecting}. In ISO/IEC/IEEE~15288, a SoS is described as a configuration in which each constituent system retains its own management, goals, and resources, while coordinating within the SoS to achieve higher-level objectives \cite{7106435}.
SoS are increasingly used to manage complexity in domains where adaptability, scalability, and interoperability are essential. INCOSE identifies SoS as a key enabler for future systems, particularly in addressing global challenges that require scalable, distributed, and coordinated solutions \cite{incose2035}.

In SoS, each constituent system maintains operational and managerial independence, i.e., it can function and evolve on its own. These systems are also geographically distributed, exhibit heterogeneous capabilities, and are dynamically reconfigurable. Most importantly, SoS exhibit emergent behavior---capabilities that arise from the interaction of components, rather than being explicitly designed \cite{maier1998architecting, holland1995hidden}.

The conceptual foundations of SoS trace back to General Systems Theory by \textcite{von1950theory}, which emphasized the importance of interdependence and holism. Early theoretical contributions from \textcite{boulding1956general}, \textcite{ackoff1971towards} laid the groundwork for viewing systems as interconnected wholes. The term ``system of systems'' gained practical relevance in the 1980s and 1990s, particularly in defense and aerospace, where it was used to describe the integration of large, independent systems to enable coordinated operations and crisis response \cite{klein2013systematic, eisner1991computer, shenhar19952}.

Drawing from a wide range of prior classifications of SoS properties---including those by \textcite{keating2003system}, \textcite{boardman2006system}, \textcite{sage2001systems}, and \textcite{maier1998architecting}---\textcite{nielsen2015systems} synthesized these perspectives into a unified eight-dimensional taxonomy designed to support the analysis and engineering of complex SoS. The dimensions are the following.
\begin{description}
    \item[\bf Autonomy] is the extent to which a constituent system's behavior is governed by its own internal goals, rather than by directives from the SoS.
    \item[\bf Independence] is the ability of a constituent system to operate even when detached from the SoS.
    \item[\bf Distribution] refers to the spatial and logical separation of systems within the SoS. 
    \item[\bf Evolution] accounts for long-term change and adaptability within the system.
    \item[\bf Reconfiguration] accounts for real-time change and adaptability within the system.
    \item[\bf Emergence] is the presence of behaviors that arise through system interaction and may not be predicted by the system's design. 
    \item[\bf Interdependence] reflects mutual reliance between systems for shared objectives.
    \item[\bf Interoperability] is the ability to exchange data and services across heterogeneous systems.
\end{description}

Evidence shows that SoS are developed with increased attention to reliable and secure operation, especially in complex and changing environments. For example, \textcite{ferreira2024framework} present architectures that support fault tolerance and system recovery, while \textcite{song2023continuous} and \textcite{hyun2023timed} introduce verification methods designed for safety-critical systems. Other work, such as \textcite{wang2019motifs}, explores ways to predict reliability over time, helping to ensure reliable performance in areas such as transportation and manufacturing.

Integrating DTs and SoS offers a pathway to enhance real-time awareness and coordinated adaptation across distributed systems.

\subsection{Digital twins}\label{sec:dt_background}

A DT is a virtual representation of a physical system that maintains continuous two-way communication with its real-world counterpart \cite{kritzinger2018digital, david2024infonomics}. This characterization is aligned with ISO~23247, which describes digital twins as digital representations that remain synchronized with their target entities \cite{ISO23247-1}. This bi-directional data exchange enables real-time monitoring, simulation, and control of the physical entity. DTs are used in an array of domains, e.g., manufacturing, construction, smart cities, automotive, and avionics~\cite{liu2021review}.

The origins of the DT concept trace back to early real-time simulation, where models mirrored and predicted system behavior. During the Apollo 13 mission (1970), NASA used a virtual replica of the spacecraft to simulate scenarios and support failure recovery, an early case of real-time monitoring and decision support \cite{boschert2016digital}.
In the 1990s, \textcite{gelernter1993mirror} proposed ``mirror worlds'' in which digital representations continuously reflected their physical counterparts. In the 2000s, \textcite{grieves2005product} anchored DTs in product lifecycle management (PLM), emphasizing the bidirectional coupling of physical systems and virtual models for operation and optimization. Around the same time, \textcite{hribernik2006product} introduced the ``product avatar``, linking DTs to lifecycle integration through metadata and interoperability. NASA later advanced the concept by defining DTs as probabilistic simulation systems for predicting asset behavior and supporting health management across the lifecycle \cite{glaessgen2012digital}.

DTs have been classified both by their data flows and by their structural architectures. \textcite{kritzinger2018digital} distinguish between digital models (no automatic synchronization), digital shadows (one-way data flow from physical to digital), and full DTs (bidirectional automatic exchange). Building on this, \textcite{tao2018digital} emphasize iterative feedback loops where physical and virtual entities co-evolve through shared data flows. Complementing such taxonomies, international standards (ISO 23247 \cite{ISO23247-1}, RAMI 4.0 \cite{DIN91345}) formalize DT architectures by defining layers, entities, and interoperability requirements. Further classifications refine DT purposes, ranging from monitoring and predictive DTs to prescriptive and autonomous ones with closed-loop control \cite{verdouw2021digital}.

Since then, the scope of DTs has expanded. More recent work explores advanced forms of DTs that enable prediction, autonomous operation, and the ability to adapt and improve in response to disruptions \cite{grassi2024conceptual}. These developments position DTs not only as monitoring tools but as adaptive agents within complex cyber-physical environments.

Despite their growing adoption, DTs face several technical and organizational challenges. These include the lack of interoperability standards, concerns about data privacy and security, and difficulties in scaling DTs for large, heterogeneous systems \cite{fuller2020digital}. The absence of unified development practices further complicates cross-domain deployment.
Overcoming these barriers requires more scalable and coordinated DT architectures, pushing current efforts toward broader integration across distributed systems and domains.

\subsection{Related work}\label{sec:relatedwork}

The integration of DTs within SoS has become a topic of particular interest, and there is a growing number of secondary studies on the topic.

\textbf{Closest to our work} is the survey by \textcite{olsson2023systems}, which reviews ten studies combining DTs and SoS and identifies challenges related to interoperability and lifecycle management. Our study expands this scope through a systematic analysis of 80 primary studies, examining motivations, architectures, technical characteristics, non-functional properties, and research maturity to clarify how DT architectures support SoS characteristics in SoTS.

Several studies acknowledge that DTs can be composed into SoS and that such integrated systems may bring unique benefits. \textcite{mylonas2021digital} survey DTs in manufacturing and smart cities, and investigate their integration patterns. \textcite{semeraro2021digital} conduct a cross-domain systematic literature review of DTs and introduce the patterns of digital twinning at the unit and system level of SoS. \textcite{bottjer2023review} systematically review unit level manufacturing DTs, situating them within hierarchical system and SoS structures and identifying interoperability and standardization as prerequisites for large-scale integration. \textcite{bertoni2022designing} survey DTs in product-service systems and argue that analyzing emergent behavior and lifecycle value requires connecting multiple DTs into higher-order SoS configurations. \textcite{dalibor2022cross} present a systematic mapping study of DT research and identify systems of systems as a class of physical systems that can be digitally twinned. \textcite{klar2023digital} review existing DT maturity models and propose a framework in which interoperability across DTs represents the highest level of maturity, enabling operation as an SoS.

Despite the benefits, engineering DTs of SoS remains challenged by interoperability gaps, connectivity and privacy concerns, limited standardization, and the operational independence of constituent systems \cite{michael2022integration,dietz2020digital,borth2019digital}. These challenges make the composition of DT functions and behaviors across interconnected systems difficult in practice \cite{gill2024toward}.

%% file: sections/study-design.tex
\section{Study design}\label{sec:study-design}

We designed a study to systematically survey the literature concerned with the combination of DTs and SoS, which we refer to as systems of twinned systems. Our goal was to understand the characteristics of SoTS, their components, and constituent systems, as well as to identify the key limitations, challenges, and research opportunities in the field.

\subsection{Research questions}\label{sec:study-design-rqs}

We formulated the following research questions.

\rquestion{RQ1. \ul{Why} are DT principles and SoS principles applied jointly in the design of complex systems?}
By answering this RQ, we aim to understand the \textit{purposes}, \textit{problems}, and \textit{domains} in which SoTS are used. 
We also aim to understand whether organizing multiple DTs is a \textit{purposeful} activity, and if so, what are the \textit{motivations}, \textit{intents}, and \textit{ambitions} to do so. In particular, we are interested whether it is SoS that benefit from twinning or the other way around.

\rquestion{RQ2. \ul{How} are DT and SoS combined?}
We aim to identify \textit{architectures} along which systems, such DTs, are organized into SoS. We were interested in the \textit{nature of constituent units}: whether they are purely physical, digital, or both; as well as the \textit{type of SoS} (acknowledged, directed, etc).

\rquestion{RQ3. What are the \ul{technical characteristics of DTs} in SoTS?}
We are interested in the \textbf{details of DTs} that are combined as a SoS, such as their \textit{level of autonomy} (fully autonomous, human actuated, digital shadow, etc), \textit{services}, \textit{modeling formalisms}, etc.

\rquestion{RQ4. What are the \ul{technical characteristics of SoS} in SoTS?}
We are interested in the \textbf{details of SoS} in SoTS, such as support for typical \textit{SoS dimensions} (autonomy, belonging, etc) and the \textit{type of emergent behavior} these SoS account for (simple, weak, strong, spooky).

\rquestion{RQ5. How are \ul{non-functional properties} addressed in SoTS?}
We are particularly interested in \textbf{reliability} (i.e., continuity of correct service~\cite{avizienis2004basic}) and \textbf{security} due to their recognized importance in enabling safe and trustworthy operation in distributed and dynamic environments~\cite{fuller2020digital, incose2035, ferreira2024framework, olsson2023systems}.
We focus on how reliability and security are considered in the development and operation of SoTS by examining whether these concerns are \textit{addressed at the architectural level}, \textit{explicitly modeled}, or \textit{empirically evaluated}. 

\rquestion{RQ6. What is the level of \ul{technical and research maturity} in SoTS?}
To assess technical maturity, we rely on the Technology Readiness Level framework (TRL)~\cite{mankins2009technology}. We introduce the following clusters of levels for our purposes: \textit{Initial} (TRL 1-2); \textit{Proof-of-concept} (TRL 3-4); \textit{Demonstration prototype} (in relevant environment, TRL 5-6); \textit{Deployed prototype} (in the operating environment, TRL 7-8); \textit{Operational} (TRL 9).
To assess research maturity, we investigate how primary studies are evaluated, using the assessment framework of \textcite{petersen2015guidelines}.
As a sign of maturity, we are also interested in whether the sampled studies relied on any standards.

\rquestion{RQ7. What are the typical \ul{technological choices} to implement SoTS?}
We are interested in the technological landscape supporting the implementation of SoTS. We analyze the usage of \textbf{programming languages}, \textbf{frameworks}, and \textit{platforms}.

\subsection{Databases and search string}\label{sec:databases}

To search for potentially relevant studies, we used the key academic indexing databases: Scopus, Web of Science, ACM Digital Library, IEEE Xplore.
We considered peer-reviewed literature only. Grey literature, e.g., articles published on \textit{arXiv} and blog posts were not included. We searched in the title, abstract, and keywords of papers. Search on Scopus was limited to works from the \textit{Computer science} and \textit{Engineering} disciplines.

We constructed the search string from the key concepts of our study (\textcolor{blue}{digital twins} and \textcolor{ForestGreen}{system of systems}) and their \textcolor{RedViolet}{typical synonymous keywords} found in our preliminary investigation.

\begin{footnotesize}
\begin{Verbatim}[commandchars=\\\{\}]
(\textcolor{blue}{"digital twin*"} AND \textcolor{ForestGreen}{"system* of systems"}) OR
(\textcolor{RedViolet}{"aggregated digital twin*" OR}
    \textcolor{RedViolet}{"system of digital twins" OR}
    \textcolor{RedViolet}{"digital twin of systems" OR}
    \textcolor{RedViolet}{"system* of twinned systems"})
\end{Verbatim}
\end{footnotesize}

This is not an exhaustive list of terms, but a rather representative one and will be further compensated in the snowballing phase.

\subsection{Search and selection}

\subsubsection{Automated search}

We executed the search on September 10, 2024.
We retrieved a total of $317$ studies.
We removed duplicates using a combination of the automated and manual duplicate detection in EndNote\footnote{\url{https://endnote.com/}}. We removed 121 references and retained 196 unique references. Subsequently, we applied the exclusion criteria. The details are reported in \tabref{tab:numbers}.

\subsubsection{Selection}\label{sec:excriteria}
We used the following exclusion criteria to exclude primary studies that were not in the scope of our investigation. A primary study is excluded if it meets at least one exclusion criterion.

\begin{enumerate}[\bfseries{E}1.]
\itemsep0em
\setcounter{enumi}{-1}
    \item Not accessible (not in English or not available for download); not peer-reviewed (e.g., theses, grant proposals); not primary research (e.g., reviews, mappings).
    \item Does not discuss DT.
    \item Does not discuss SoS.
    \item Off-topic.
\end{enumerate}

E0 was trivial to evaluate and therefore, one author evaluated each study against E0 and another author validated the decisions.
In exclusion criteria E1--E3, each primary study was evaluated by two authors independently, based on the \textit{full reference} (title, authors, venue...) and the \textit{abstract}. In case of a tie, a discussion was facilitated. In \tabref{tab:numbers}, we report detailed figures of the selection and exclusion, including inter-rater agreement and reliability metrics.
We measured an inter-rater agreement (IRA) of 88.0\% and Cohen's $\kappa$ of 0.734 (substantial agreement). Most of the disagreements were due to different levels of leniency of the reviewers. We facilitated in-depth discussions to converge.

\textbf{Eventually, we arrived at 81 unique relevant references.} In the next step, these references underwent a quality assessment.

\subsection{Quality assessment}\label{sec:qa}

In line with the guidelines of \textcite{kitchenham2007guidelines}, we defined a checklist to assess the quality of primary studies. Quality criteria were derived from the research questions.
Each question was answered by ``yes'' (2 point), ``partially'' (1 points), or ``no'' (0 points), based on the full text. To retain a primary study, we required that it scored at least 1 point in each of the following quality checks:

\begin{enumerate}[\bfseries{Q}1.]
\itemsep0em
    \item SoS is clearly described.
    \item DT is clearly described.
    \item The contributions are tangible (i.e., not conceptual).
    \item Reporting quality is clear.
\end{enumerate}

The quality criteria have been evaluated by reading the full text of the primary study and judging whether the study provides sufficient and clear information that helps answer the research questions.

\input{tables/numbers}

Of the 81 tentatively included primary studies, we excluded 28 (22 due to insufficient Q1 or Q2, and 6 due to insufficient Q3; 0 studies to exclude due to insufficient Q4). This resulted in \textbf{53 primary studies} from the automated search phase, i.e., a 16.72\% overall inclusion rate.
In the next step, these 53 primary studies formed the basis of snowballing.

\subsection{Snowballing}\label{sec:snowballing}

We used \textit{forward} and \textit{backward} snowballing to enrich the corpus.
Backward snowballing was conducted in two phases. First, every reference in the previously included primary studies was assessed by title, publication venue, and date. Of the 1\,666 references, 145 seemed to be relevant for our purposes. Second, the 145 potentially relevant references underwent the same evaluation process as previously included studies, i.e., two authors applied exclusion criteria and checked the quality of works.
Forward snowballing was conducted via Google Scholar as per the recommendations of \textcite{wohlin2020guidelines}.

In the backward and forward snowballing, in total, we selected 38 potentially relevant references that underwent the same evaluation process as previous primary studies. (13 by backward snowballing and 25 by forward snowballing.)

We measured an IRA of 94.4\% and a Cohen's $\kappa$ of 0.488. We measured these numbers on the primary studies that have been reviewed by two reviewers (731 total references: 145 backward, 586 forward). The $\kappa$ was somewhat low, although by definition, it represents ``moderate'' agreement. This number was due to the ambiguity of abstracts we encountered.
Eventually, we included \textbf{27 additional primary studies}.

At the end of the first snowballing, we noted a rather low inclusion rate of 1.19\%. We interpreted this low number as sufficient evidence for saturation and we stopped with snowballing.

Eventually, we screened 2\,569 potential studies.

\begin{bottomlineframe}
In total, we included \textbf{80 primary studies}.
\end{bottomlineframe}

\subsection{Data extraction}\label{sec:data-extraction}

We extracted data from the 80 included studies into a data extraction sheet.

The data analysis included collating and summarizing the data, aiming at understanding, analyzing, and classifying the state of the art~\cite{kitchenham2007guidelines}.
We performed a combination of content analysis~\cite{franzosi2010quantitative} (mainly for categorizing and coding studies under broad thematic categories) and narrative synthesis~\cite{rodgers2009testing} (mainly for detailed explanation and interpretation of the findings coming from the content analysis).
We analyzed the extracted data to find trends and collect information about each category of the classification framework (vertical analysis).
We also explored the extracted data for possible relations across different categories of the classification framework (horizontal analysis).

Whenever possible, we started from existing categorizations or derive systematic categorizations. To characterize SoS, we chiefly relied on the taxonomy of \textcite{nielsen2015systems}. To characterize the various flavors of DTs, we invoked the works of \textcite{kritzinger2018digital} and \textcite{david2024infonomics}.

In the first phase of data extraction, we piloted the classification framework. In this phase, we discussed potential modification to the classification framework to accommodate interesting trends across the primary studies. Then, we performed the extraction. Finally, we performed the codification.

To aid independent replication, we developed Python scripts to automate these steps. The data and scripts are available in the replication package.

\subsection{Threats to validity, study quality, and limitations}\label{sec:threats}

\paragraph{Construct validity}
Our observations are artifacts of the sampled papers. Potential selection bias and missed publications may have an impact on our observations and threaten the construct validity of this study. To mitigate this threat, we employed a systematic approach in accordance with the best practices of empirical research in software engineering. Specifically, we used trusted databases, redundancy and validation in the exclusion phase~\cite{wohlin2020guidelines}, and employed snowballing to enrich our corpus~\cite{greenhalgh2005effectiveness}.

\paragraph{Internal validity}
We may have missed works due to the terminology we used. The combination of SoS and DT has had no unified definition prior to our work and thus, constructing effective search strings might not have been feasible. We mitigated this threat by an alternative, although more labor-intensive corpus construction strategy: we augmented the core keywords in the search string with synonyms, and we used snowballing.
Some threats arise from the manual evaluation of the quality criteria due to the element of subjectivity. To mitigate threats, we conducted frequent, regular, in-depth discussions about the quality criteria as we applied them to the primary studies. By that, we converged the researchers' viewpoints and eliminated a reasonable amount of the threats that may stem from inconsistent interpretations of the quality criteria.

\paragraph{Study quality}
Our work scores \percp{9}{11}{points} in the rigorous quality checklist of \textcite{petersen2015guidelines}. (Need for review: 1 point; search strategy: 3 points; evaluation of the search: 2 points (keywords from known papers; identify objective criteria for decision; add additional reviewer, resolve disagreements
between them when needed); extraction and classification: 2 points; study validity: 1 point.)\footnote{Detailed report in the replication package.} This quality score is \textit{significantly} higher than the typical values in software engineering. \textcite{petersen2015guidelines} reports a median of 33\%, with only 25\% of their sampled studies having a quality score of above 40\%. Therefore, we consider our study design of high quality.

\paragraph{Limitations} We are reasonably confident that we have mitigated the severe threats to the validity of the study; however, some limitations obviously remain and should be noted. First, SoTS is an emerging field with early-stage results, and therefore, its landscape is dynamically changing. Our work provides the first insights into this field, but given the pace of research, subsequent studies may be justified. Second, we assumed that we could gain valid and sufficient evidence of the field's tendencies through analyzing the academic body of knowledge. Given the technical nature of the field, it is possible that the grey body of knowledge contains additional useful evidence. In addition, non-technical fields may work with systems that exhibit latent SoS and DT traits, which we obviously could not detect. Thus, subsequent multi-vocal studies may be justified. Finally, while we report on technical and technological tendencies, such as typical programming languages, tools, and frameworks found in SoTS, these tendencies are pertinent to an evolving field. Thus, the technological insights are best considered as descriptive of the current state of affairs, but not indicative of universal best practices. Prospective technical contributions ought to push the technological boundaries of SoTS and therefore, subsequent tool surveys may be justified.

\subsection{Publication trends}

\input{figures/pubtrends/pubdata}

\figref{fig:pubtrends} reports the publication trends.

The number of publications (\figref{fig:pubstats}) shows an increasing trend, with a clear increase in publication output in the past four years (2024 is a partial year). After investigating the spike in publication output in 2019, we conclude that it is not a systemic phenomenon, but rather, an outlier. Overall, we observe an increasing interest in combining DT and SoS principles.
About 47\% of the sampled studies are journal articles and book chapters, suggesting relatively mature research; although the majority of sampled studies are journal or conference articles (41\% and 45\%, respectively).

The quality of reporting (\figref{fig:qscores}) is relatively high, scoring 83.4\% in our quality assessment scheme (\secref{sec:qa}). The quality of reporting on DT components is particularly high (95\%), substantially above that of SoS principles (83.8\%).
Contributions are typically tangible (75\%), with less than a quarter of the corpus being conceptual works. Finally, the reporting clarity is acceptable, scoring 80\% in our quality scheme.

\begin{bottomlineframe}
We judge the corpus to be of sufficient quality to answer the research questions with high certainty and reasonable validity.
\end{bottomlineframe}

%% file: tables/numbers.tex
\begin{table*}[t]
\centering
\setlength{\tabcolsep}{1em}
\caption{Search statistics}
\label{tab:numbers}
\footnotesize
\begin{tabular}{@{}lrrr@{\hspace{2pt}}lr@{}}
\toprule
\multicolumn{1}{c}{\textbf{Search round}} & \multicolumn{1}{c}{\textbf{All}} & \multicolumn{1}{c}{\textbf{Excluded}} & \multicolumn{2}{c}{\textbf{Included}} & \multicolumn{1}{c}{\textbf{Agreement}} \\ \midrule
\textbf{Automated search} & & & & & \\
\;\;\corner{} Duplicate removal & 317 & 121 & 196 &   &  \\
\;\;\corner{} E0 & 196 & 71 & 125 &   &  \\
\;\;\corner{} E1--E3 & 125 & 44 & \textbf{81} & & $IRA=0.88$ / $\kappa=0.734$ \\
\textbf{Quality assessment} & & & & & \\
\;\;\corner{} Q1 or Q2 insufficient & 81 & 22 & 59 &   &  \\
\;\;\corner{} Q3 insufficient & 59 & 6 & 53 &   &  \\
\;\;\corner{} Q4 insufficient & 53 & 0 & 53 &   &  \\
\textbf{Subtotal} & \textbf{317} & \textbf{264} & \included{53} & \textbf{(16.72\%)} & \\
\midrule
\textbf{Snowballing} & & & & & \\
\;\;\corner{} Backward & & & & & \\
\;\;\;\;\corner{} Selection by reference & 1\,666 & 1\,521 & 145 & & \\
\;\;\;\;\corner{} Selection by abstract & 145 & 132 & \textbf{13} & & \\
\;\;\corner{} Forward & 586 & 561 & \textbf{25} & &  \\
\;\;\corner{} Total abstracts screened & 731 & 693  & \textbf{38} & & $IRA=0.944$ / $\kappa=0.488$ \\
\textbf{Quality assessment} & & & & & \\
\;\;\corner{} Q1 or Q2 insufficient & 38 & 9 & 29 &   &  \\
\;\;\corner{} Q3 insufficient & 29 & 2 & 27 &   &  \\
\;\;\corner{} Q4 insufficient & 27 & 0 & \textbf{27} &  &  \\
\textbf{Subtotal} & \textbf{2\,252} & \textbf{2\,225} & \included{27} & \textbf{(1.19\%)} & \\
\midrule
\textbf{Total} & \textbf{2\,569} & \textbf{2\,489} & \included{80} & \textbf{(3.11\%)} & \\
\bottomrule
\end{tabular}
\end{table*}

%% file: figures/pubtrends/pubdata.tex
\begin{figure}[t]
    \centering
    \begin{subfigure}{\linewidth}
        \includegraphics[trim = 0 0.4cm 1.4cm 0.4cm,clip,width=0.9\linewidth]{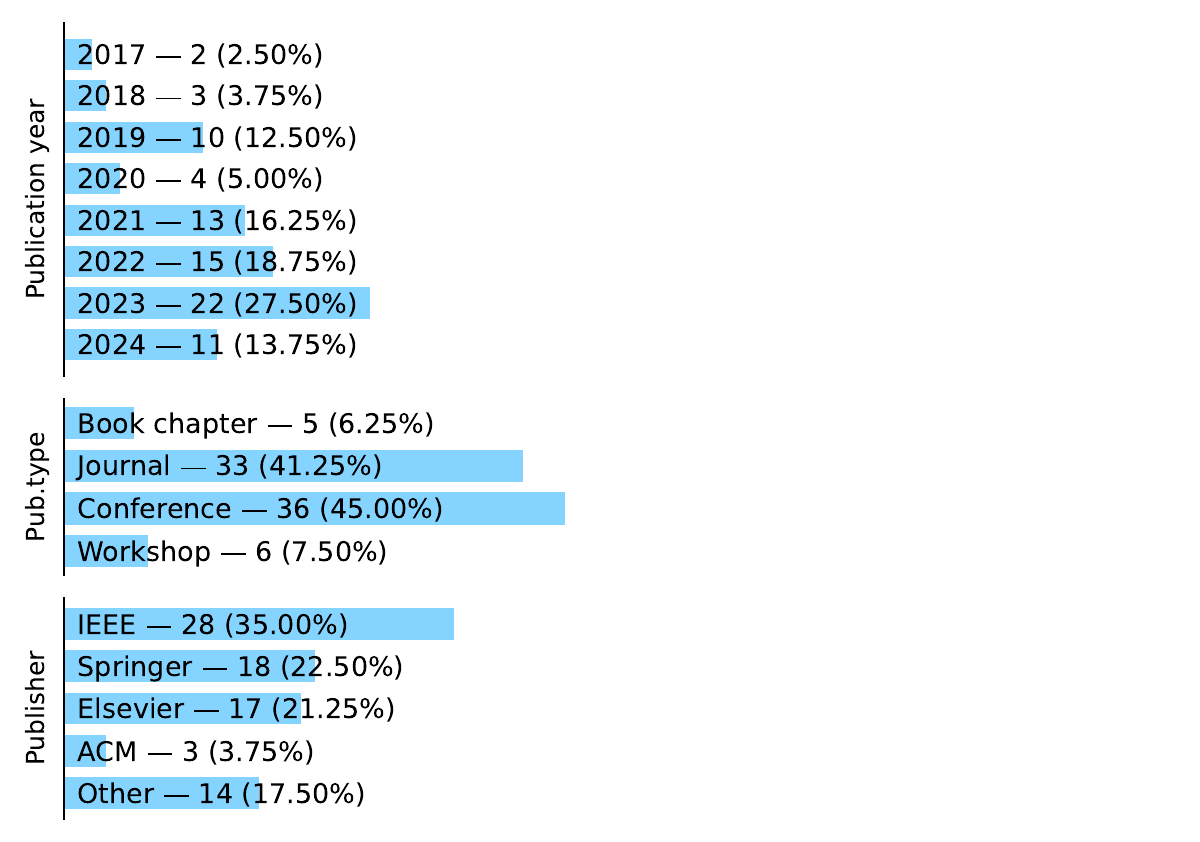}
        \caption{Scientific output (as of September 2024)}
        \label{fig:pubstats}
    \end{subfigure}\\[1em]
    \begin{subfigure}{\linewidth}
        \includegraphics[trim = 0 0.4cm 1.4cm 0.4cm,clip,width=0.9\linewidth]{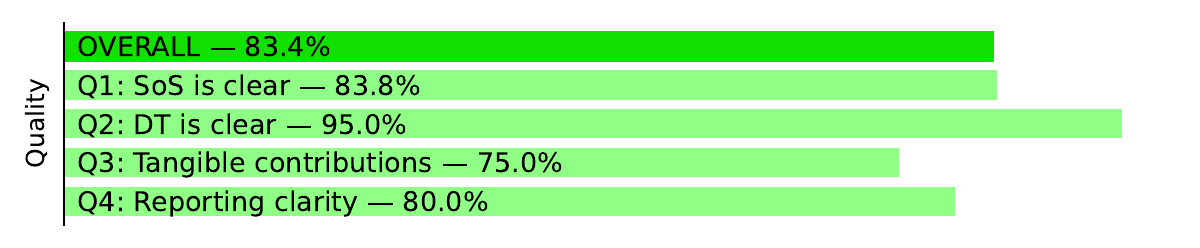}
        \caption{Quality scores}
        \label{fig:qscores}
    \end{subfigure}
    \caption{Publication trends}
    \label{fig:pubtrends}
\end{figure}

%% file: sections/classification.tex
\section{A classification framework for SoTS}\label{sec:classification}

To organize and compare the various flavors of SoTS, we devise a classification framework. We draw on the seminal works of \textcite{maier1998architecting,maier2014role} to understand how SoS are organized, and combine this theory with DT concepts~\cite{kritzinger2018digital}.

We rely on a mixed sample- and case-based generalization~\cite{wieringa2015six}. This approach is particularly useful when constructing middle-range theories that balance generality with practicality, such as in engineering sciences. In \secref{sec:study-design}, we sampled a statistically adequate corpus. Subsequently, we decomposed each study individually into architectural units as architectural abstractions allow for better judging of similarity between cases~\cite{wieringa2015six}. Finally, we identified recurring patterns.

\subsection{Core Elements}
The following section introduces the core elements that form the basis of the classification framework.

A \textit{System} is the elementary building block a SoS, generally understood as ``an assemblage of constituent systems that produces behavior or function not available from any constituent individually''~\cite{maier1998architecting}. Systems can be hierarchically \textit{composed of} other systems. In SoS, these sub-systems are referred to as the constituent systems or constituents, in short. Systems also have \textit{goal}s which drive their behavior.

At this point, we draw on the theory of DTs when we distinguish between \textit{Digital Twin}s and \textit{Physical Twin}s, i.e., the digital and physical counterparts of heterogeneous systems~\cite{kritzinger2018digital}. 

A distinguishing factor between DTs and SoS is the strength of coupling between constituent systems. SoS typically rely on \textit{weak coupling}, i.e., constituents are allowed to make individual decisions about belonging to the SoS or leaving it, and in some cases, pursue their own goals. This is in a stark contrast with the \textit{strong coupling} between a DT and its twinned counterpart, which may be physical or cyber-physical, e.g., by comprising a DT and a PT. A DT represents the prevalent state of the twinned system through computational reflection~\cite{maes1987concepts2} and controls the system through precise control.

Finally, the \textit{Controller} is a special role in SoS architectures in which constituents defer setting the goal to a higher level system. In digitally twinned systems, this controller is always the DT.

As shown in the following, coordination between systems is typically achieved either through orchestration or choreography. As defined by \textcite{peltz2003web}, orchestration inherently represents control from one party’s perspective (i.e., the controller), while choreography is a distributed approach. Therefore, orchestration is more typical in centralized SoTS architectural variants, and choreography is more typical in collaborative architectural variants.

In the following, we describe a set of architectural variants.

\subsection{Instances}\label{sec:instances}

We instantiate six architectural patterns. Four of these patterns are backward compatible with the taxonomy of \textcite{maier1998architecting,maier2014role}, who classifies SoS into directed (\secref{sec:instance-directed}), acknowledged (\secref{sec:instance-acknowledged}), collaborative (\secref{sec:instance-collaborative}), and virtual (\secref{sec:instance-virtual}) SoS.

For completeness, we discuss two additional architectural patterns: \textit{Specialized DTs} (\secref{sec:instance-spec-dt}) and \textit{Specialized DTs and Systems} (\secref{sec:instance-spec-dt-and-system}). These patterns were not observed in our empirical inquiry but are included to outline potential architectural directions for future work.

\subsubsection{Directed SoTS}\label{sec:instance-directed}

\begin{figure}[h]
    \centering
    \includegraphics[trim={0.66cm 0.25cm 1.1cm 0.25cm},clip,width=0.8\linewidth]{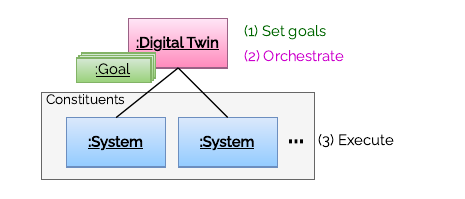}
    \caption{Directed SoTS}
    \label{fig:framework-instance-directed}
    \vspace{-1.5em}
\end{figure}

A directed SoTS (\figref{fig:framework-instance-directed}) builds on directed SoS~\cite{maes1987concepts2}, i.e., it has a central controller that sets goals and orchestrates the constituent systems as they execute their tasks in accordance with the goals. The constituents operate independently, but their normal operational mode is subordinated to the centrally managed goal. Specifically, in SoTS, the controller is a DT. This particular type of SoTS could also be labeled as a \textit{DT of DTs}, since we have one DT mirroring and monitoring an entire set of other digitally twinned constituent systems; there is a clear hierarchical relation between the controller DT and the others being monitored.

\paragraph{Example.} As an example, consider a smart city, in which a command and control (C2) center holds a virtual replica of the entire city. This DT monitors and orchestrates the entire city -- which itself may comprise other digitally twinned systems at different levels of granularity, e.g., autonomous vehicles in the traffic that is being routed by the central DT.
As another example, consider an emergency response (ER) SoS, in which a central authority, a DT monitors and orchestrates the business process that leads to accomplishing the emergency response mission. A central operations center (and a DT inside it) commands fire brigades, ambulances, and police dispatch systems. It involves vehicles, communication radios, and hospital emergency response systems, which are highly coordinated and follow direct command.

\subsubsection{Acknowledged SoTS}\label{sec:instance-acknowledged}

\begin{figure}[h]
    \centering
    \includegraphics[trim={0.66cm 0.25cm 1.1cm 0.25cm},clip,width=0.8\linewidth]{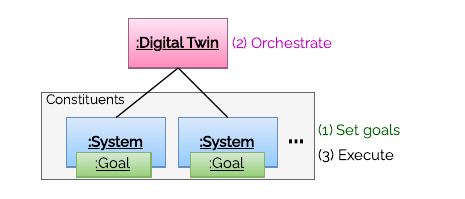}
    \caption{Acknowledged SoTS}
    \label{fig:framework-instance-acknowledged}
    \vspace{-1.5em}
\end{figure}

An acknowledged SoTS (\figref{fig:framework-instance-acknowledged}) builds on acknowledged SoS~\cite{maes1987concepts2}, i.e., it has a central controller that orchestrates the constituents, but goals are negotiated and set at the constituents' level. Thus, constituents keep their independent objectives and sustainment goals.
Similar to directed SoTS, the controller is a DT that monitors other twinned systems; however, in contrast with directed SoTS, the central DT does not have a coercive power over them, not orchestrating them. 

\paragraph{Example.} As an example, consider a smart energy grid. In such a SoS, utilities, distributed energy producers, and smart meters are managed under a central regulatory agency. This central agency can operate a DT for the entire network that orchestrates subsystems for balanced supply and demand and the resulting improved grid stability, but consumers and prosumers (consumers who also produce power) retain autonomy over their systems in terms of setting local goals, e.g., how much electricity to produce for private usage.

\subsubsection{Collaborative SoTS}\label{sec:instance-collaborative}

\begin{figure}[t]
    \centering
    \includegraphics[trim={0.66cm 0.25cm 0.75cm 0.25cm},clip,width=\linewidth]{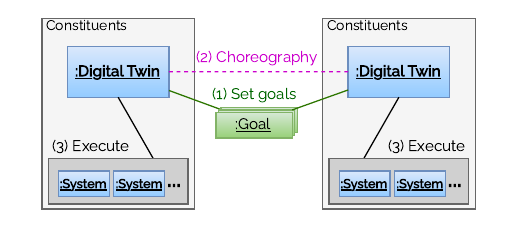}
    \caption{Collaborative SoTS}
    \label{fig:framework-instance-collaborative}
    \vspace{-1.5em}
\end{figure}

A collaborative SoTS (\figref{fig:framework-instance-collaborative}) builds on collaborative SoS~\cite{maes1987concepts2}, i.e., constituents participate in the system on a voluntary basis to collaboratively fulfill previously agreed-upon goals. The goals are centralized, but constituents \textit{choose} to participate in fulfilling those goals. 

In contrast to the previously discussed architectures, there is no central controller unit at the top level of a collaborative SoTS. (Of course constituents may be organized into a directed or acknowledged architecture, but that bears no relevance at the higher level as a constituent system is seen as a black box.) 

In the absence of a central controller, the coordination mechanism changes, too. In contrast to the previously discussed architectures, collaborative SoTS coordinate through \textit{choreography} rather than orchestration.

\paragraph{Example.} As an example, consider a platoon of autonomous vehicles in the traffic of a smart city. The vehicles do not have a central authority to control the platooning dynamics, e.g., enforcing safe inter-vehicle distance and headway with respect to changing road conditions. Instead, each vehicle has their own DT, and these DTs work together to (i) set the target distance and speed, and then (ii) each DT acts accordingly, thereby implementing the choreography.

\subsubsection{Virtual SoTS}\label{sec:instance-virtual}

\begin{figure}[h]
    \centering
    \includegraphics[trim={0.66cm 0.25cm 0.75cm 0.25cm},clip,width=\linewidth]{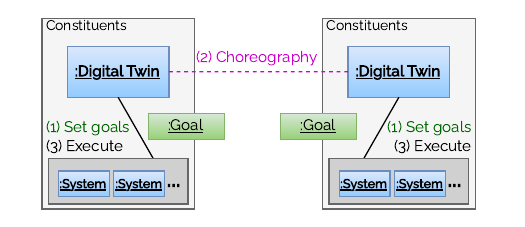}
    \caption{Virtual SoTS}
    \label{fig:framework-instance-virtual}
\end{figure}

A virtual SoTS (\figref{fig:framework-instance-virtual}) builds on virtual SoS~\cite{maes1987concepts2}, i.e., constituents participate in the system on a voluntary basis and, in contrast with collaborative architectures, they pursue their own goals rather than previously agreed-upon ones. Goals are typically negotiated on-the-fly, in accordance with the observed emergent behaviors of the SoTS. The precursors to this SoTS architectural flavor are highly autonomous DTs~\cite{david2024infonomics}, e.g., those enabled by AI components to take independent decisions~\cite{nilsson2024ai}.

\paragraph{Example.} As an example, consider an emergency response SoS in which intelligent, DT-enhanced autonomous systems (e.g., drones, autonomous ambulance, robots) detect abnormalities in the environment (e.g., a flood) and infer that they need to cooperate to form a viable and optimal strategy to achieve a joint objective (e.g., rescuing a person in a flooded area). Each constituent formulates its own goal based on prior information exchange and negotiation, and executes the joint strategy in accordance with these goals. Choreography is the key coordination mechanism as DTs of constituents act independently from each other as they pursue the joint objective.

\subsubsection{System of Specialized DTs}\label{sec:instance-spec-dt}

\begin{figure}[t]
    \centering
    \includegraphics[trim={0.66cm 0.25cm 0.5cm 0.25cm},clip,width=0.8\linewidth]{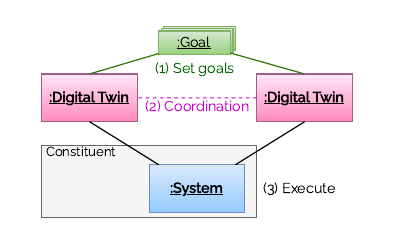}
    \caption{Specialized DTs}
    \label{fig:framework-instance-spec-dt}
\end{figure}

A system of specialized DTs (\figref{fig:framework-instance-spec-dt}) is a loosely coordinated set of DTs that twin the same constituent system. The DTs are specialized in their capabilities, which are typically complementary. Goals are typically pre-negotiated and followed by the DTs.

\paragraph{Example.} As an example, consider two disembodied agents (i.e., computer agents that have goals but have no embodied materialization) one trained in operating rescue mechanisms in disaster areas (e.g., releasing a ladder from a helicopter after detecting a human), and trained in navigating a rescue vehicle (e.g., hovering above a detected human). The agents are digitally materialized in DTs that are able to actuate different parts of the common underlying physical system, following joint goal setting. The DTs coordinate to achieve the common goal (e.g., what speed to limit a helicopter to in order to release a ladder close enough to the human), where coordination is a middle-ground between centralized orchestration and autonomous choreography.

\subsubsection{System of Specialized DTs and Specialized Systems}\label{sec:instance-spec-dt-and-system}

\begin{figure}[h]
    \centering
    \includegraphics[trim={0.66cm 0.25cm 0.75cm 0.25cm},clip,width=\linewidth]{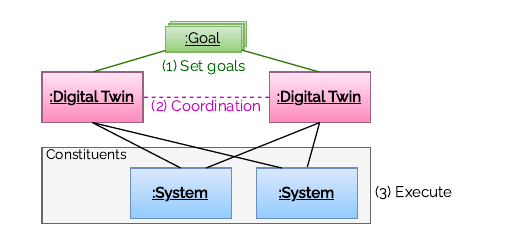}
    \caption{Specialized DTs and Specialized Systems}
    \label{fig:framework-instance-spec-dt-and-system}
\end{figure}

A system of specialized DTs and specialized systems (\figref{fig:framework-instance-spec-dt-and-system}) is a loosely coordinated set of DTs that twin multiple constituent systems and the sets of twinned systems might overlap. Similar to the previous case, the DTs are specialized in their capabilities; but in addition, the constituent systems might be specialized as well. Similar to the previous case, goals are typically pre-negotiated and followed by the DTs.

\paragraph{Example.} As an example, consider a cyber-physical system (CPS) with mechanical and electrical physical components, which are twinned in an electro-mechanical \textit{safety} DT and an electro-mechanical \textit{performance} DT. The two DTs set a joint goal to operate the CPS in a performant but safe way, and control the physical systems by actuating them in coordination.

%% file: sections/results.tex
\input{tables/rq1/motivations}
\input{tables/rq1/intentsTable}

\section{Results}\label{sec:results}
In this section, we report the key findings of our study on the state-of-the-art of SoTS.

\subsection{Why are DT and SoS principles combined? (RQ1)}\label{sec:results-rq1}

We address why SoS and DT principles are combined by analyzing the motivations (\secref{sec:motivations}), integration intents, and application domains (\secref{sec:intents}) of SoTS.

\subsubsection{Motivations}\label{sec:motivations}

As shown in \tabref{tab:motivations-table}, most studies develop SoTS to support optimization, integration, validation, or maintainability.
Optimization is the most common motivation (\xofyp{30}{80}). SoTS enable detailed monitoring of components while improving system-level awareness to support decision-making and control. In one example, SoTS coordinate UAV landings on USVs to minimize operation time \cite{li2022cognitive}.
Integration is a motivation in (\xofyp{25}{80}) studies. SoTS connect heterogeneous systems by coupling multiple DTs and enabling communication across distributed constituents. In power systems, for example, SoTS support coordinated operation across grid elements \cite{monsalve2021novel}.
Validation is addressed in (\xofyp{15}{80}) studies. SoTS enable risk-free testing by simulating system behaviors that are costly or unsafe to observe physically. This includes modeling interactions between autonomous subsystems to validate scenarios like advanced driver assistance in cars \cite{dahmen2022modeling}.
Maintainability appears in (\xofyp{10}{80}) studies.

\subsubsection{Intents and application domains}\label{sec:intents}

We distinguish between two intents in SoTS: (i) twinning a SoS, where a DT is used to represent and reason about the SoS as a whole; and (ii) combining DTs into a SoS, where multiple DTs are integrated. As shown in \tabref{tab:intents-table}, the latter is more common (\xofyp{49}{80}).

\begin{figure}[t]
    \centering
    \includegraphics[width=\linewidth]{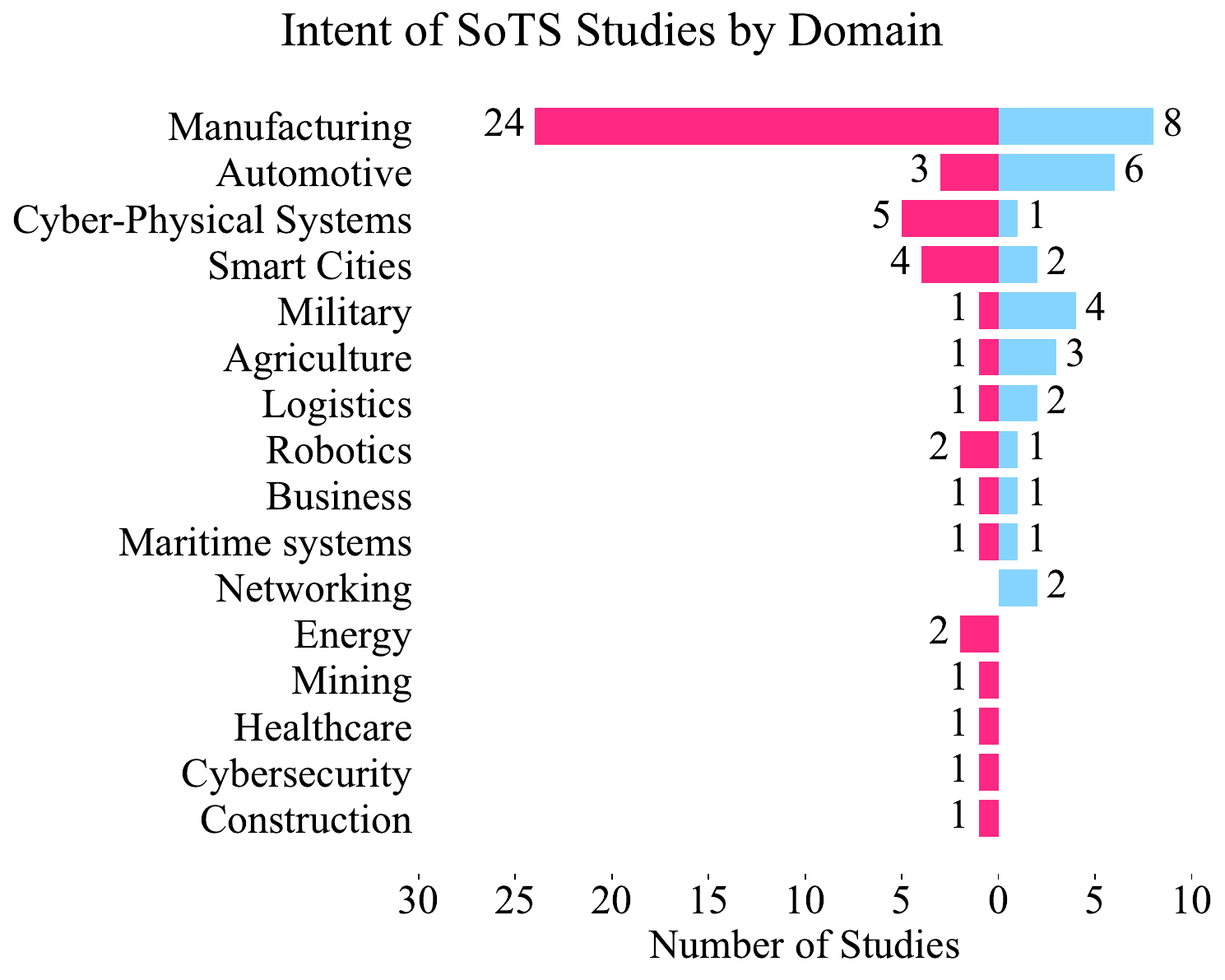}
    \caption{Intent of SoTS vs Domain\vspace{-1em}}
    \caption*{{\color[HTML]{FF2882}\ding{110}} Combining DTs into a SoS~~~{\color[HTML]{85d4ff}\ding{110}} Twinning a SoS }
    \label{fig:sos-dt-intent}
\end{figure}

\figref{fig:sos-dt-intent} breaks down these numbers by application domain. 
Manufacturing dominates both approaches (\xofyp{24}{80} and \xofyp{8}{80}), with most studies using SoTS to coordinate machines and production lines \cite{villalonga2021decision-making, novak2022digitalized}.
Smart cities (\xofyp{4}{80}) more often adopt DT combination approach to coordinate distributed services and infrastructure across urban subsystems.
Automotive (\xofyp{6}{80}) and military systems (\xofyp{4}{80}) more often rely on twinning to support global system awareness. 

Some domains appear exclusively under one approach. For example, networking appears only under twinning an SoS, where studies focus on holistic oversight of large-scale, dynamic communication infrastructures \cite{dobie2024network, priyanta2024is}. Energy, mining, healthcare, cybersecurity, and construction appear only under DT combination. 
Other domains, e.g., smart cities and logistics show both approaches.

As shown in \tabref{tab:domains-table}, the most frequently encountered domain is manufacturing (\xofyp{32}{80}), where SoTS are used to coordinate production lines and factory systems~\cite{gollner2022collaborative}. The automotive domain (\xofyp{9}{80}) applies SoTS for simulation-based testing, diagnostics, and control of vehicle subsystems~\cite{potteiger2023live}. In smart cities applications (\xofyp{6}{80}), SoTS support the modeling and integration of urban infrastructure~\cite{li2024comprehensive, human2023design}. The cyber-physical systems domain (\xofyp{6}{80}) focuses on managing real-time interaction between distributed physical processes and digital components~\cite{alam2017c2ps, stary2022privacy}.

Military, agriculture, logistics, and robotics applications appear in fewer than 6 studies each. The remaining 15\% span maritime, healthcare, construction, energy, and networking domains.

\input{tables/rq1/domainsTable}

\begin{conclusionframe}{RQ1: Why are DT and SoS combined}
SoTS are developed to support optimization, integration, validation, and maintainability in complex systems. Manufacturing is the most common application domain, followed by automotive and smart cities.
\end{conclusionframe}

\subsection{How are DT and SoS combined? (RQ2)}\label{results-rq2}

To understand how DT and SoS are combined, we analyze architectures (\secref{sec:architecture}) and types of constituent units (\secref{sec:constituents}) represented in SoTS.

\subsubsection{Architecture Configurations}\label{sec:architecture}

\input{tables/rq2/sotsTypeTable}

We applied our SoTS classification framework (\secref{sec:classification}) to categorize the studies into distinct architectural types. These types reflect the degree of autonomy, goal alignment, and coordination mechanisms between constituent systems, with DTs acting as either orchestrators or peers. The distribution of studies across types is summarized in \tabref{tab:sots-type-table}.

The majority of studies followed an Acknowledged SoTS architecture (\xofyp{31}{80}). In these systems, a central DT facilitates coordination, but each constituent retains managerial independence and negotiates its own goals. For instance, \textcite{li2022cognitive} implements a cognitive twin that synthesizes simulations and provides recommendations to UAVs and USVs, which maintain control over their own missions. Similarly, in \textcite{monsalve2021novel}, a Digital Twin Master (DTM) oversees synchronization and data flow across grid simulations, while each local Digital Twin Client (DTC) retains its own model and operational logic.

A comparable number of studies implement a Directed SoTS  (\xofyp{26}{80}). These systems are governed by a central DT that imposes goals and orchestrates constituent behavior. In \textcite{reiche2021digital}, the Digital Twin of a System (DTS) aggregates and controls individual machine twins, using a dedicated interface (DTS2DT) to monitor operations, issue commands, and maintain an integrated simulation of the whole unit. Similarly, \textcite{li2024comprehensive} introduces an infrastructure DT that coordinates multiple civil subsystems under a unified scenario-based control structure.

Collaborative SoTS architectures were found in \xofyp{19}{80} studies. These systems are formed through voluntary cooperation among DTs, with no centralized controller enforcing goals. \textcite{vogel-heuser2021approach} presents a decentralized manufacturing system composed of DTs instantiated as autonomous agents. Each agent voluntarily engages in shared production tasks through local negotiation without relying on centralized orchestration. Additionally, \textcite{chen2018digital} describes a fleet of connected vehicles, each sharing its own behavioral DT to support collective driving decisions without central command. Coordination emerges dynamically through peer-to-peer risk assessments.

Some studies qualify as Virtual SoTS (\xofyp{4}{80}), where constituents join voluntarily, pursue independent goals, and coordinate dynamically without centralized control. \textcite{pickering2023towards} presents the MAS-H platform, where independent stakeholders operate autonomously while dynamically coordinating through an open DT and modular infrastructure. Goals such as labor efficiency or sustainability emerge from voluntary collaboration rather than centralized directives. Similarly, \textcite{esterle2021digital} explores a system of autonomous cyber-physical entities that self-integrate during encounters. Coordination arises through dynamic model exchange and adaptation using DTs, without pre-defined tasks.

\begin{figure}[t]
    \centering
    \includegraphics[width=\linewidth]{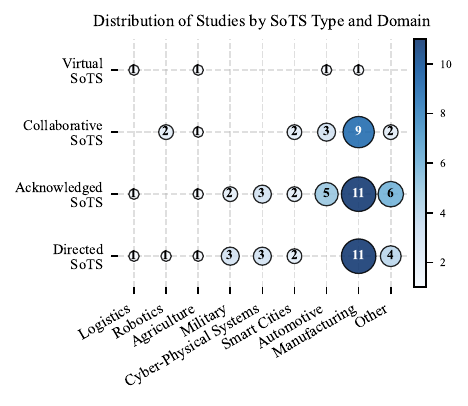}
    \caption{Distribution of SoTS types across domains}
    \label{fig:domain-sots-bubble}
\end{figure}

\figref{fig:domain-sots-bubble} shows the distribution of SoTS types across domains. Manufacturing accounts for the largest number of studies and spans mainly collaborative, acknowledged, and directed SoTS (\xofy{32}{80}), with acknowledged and directed configurations being most frequent. Virtual SoTS are rare (\xofy{4}{80}) and appear only in a small number of domains, including automotive, manufacturing, agriculture, and logistics. Military studies are limited to acknowledged and directed SoTS, while robotics studies include collaborative and directed architectures.

\subsubsection{Constituent units}\label{sec:constituents}

\input{tables/rq2/constituentUnitsTable}

\tabref{tab:constituent-units-table} summarizes the types of constituent units in SoTS. Most studies (\xofyp{62}{80}) focus on physical systems, e.g., machines, vehicles, or industrial assets. These DTs support monitoring, control, and optimization at the asset or network level \cite{reiche2021digital, kruger2022towards}.
Cyber-Physical Systems (CPS) appear in (\xofyp{9}{80}) studies, where emphasis is placed on cross-domain interoperability and reusable architectures \cite{marah2023architecture, mahoro2023articulating}.
Cyber-Physical-Human Systems (CPHS) are considered in (\xofyp{7}{80}) studies, incorporating human interaction or oversight. Examples include human-robot collaboration and adaptive mission planning \cite{savur2019hrc-sos, folds2019digital}.
Only (\xofyp{2}{80}) studies address enterprise systems, modeling organizational entities, e.g., departments or administrative units as DTs \cite{kulkarni2019towards, maheshwari2022digital}.

\begin{conclusionframe}{RQ2: How DTs and SoS are combined}
Most SoTS adopt centralized architectures, with DTs coordinating physical systems via Acknowledged or Directed patterns. Decentralized forms like Collaborative and Virtual SoTS are less common. Constituents are primarily physical assets, with limited use of cyber-physical systems, cyber-physical-human systems, or enterprise-level twins.
\end{conclusionframe}

\subsection{What are the characteristics of DTs that are combined with SoS? (RQ3)}\label{results-rq3}

To find the characteristics of DTs used in SoTS we analyze their levels of autonomy (\secref{sec:level-of-autonomy}), the services they provide (\secref{sec:dt-services}), and the modeling and simulation techniques applied (\secref{sec:modeling-methods}).

\begin{figure*}[t]
    \centering
    \includegraphics[width=\linewidth]{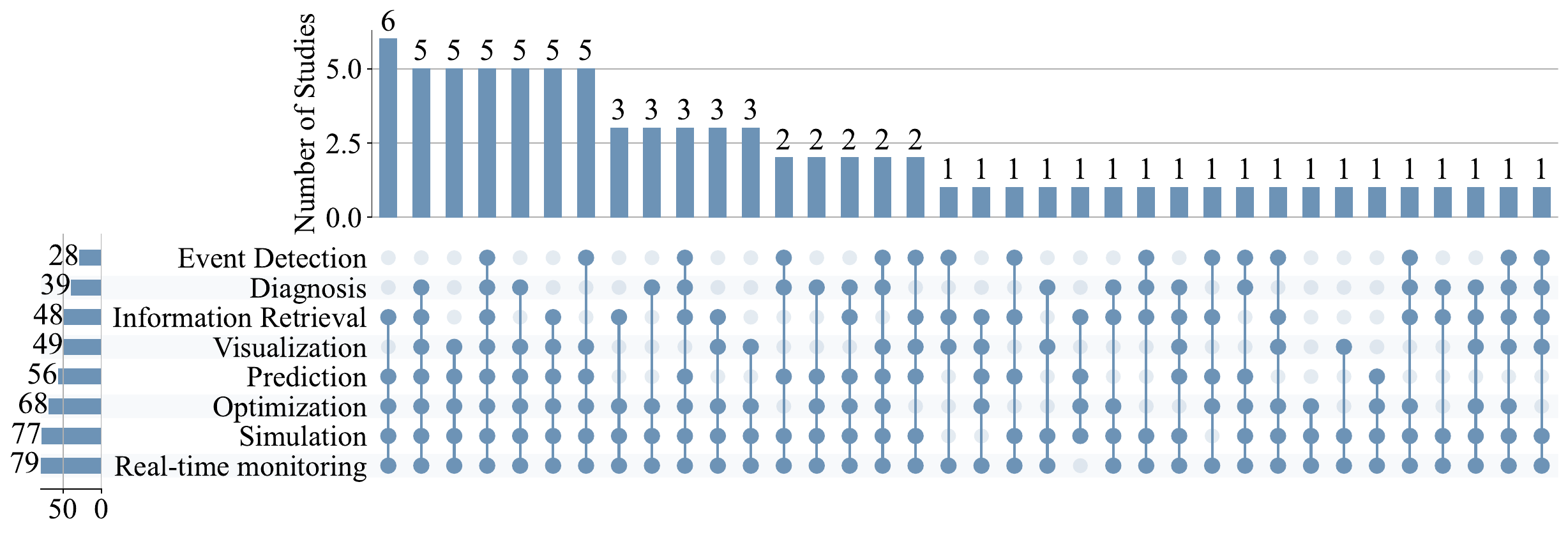}
    \caption{Combinations of DT services supported across reviewed SoTS studies}
    \label{fig:dt-services}
\end{figure*}

\subsubsection{Level of autonomy}\label{sec:level-of-autonomy}

\input{tables/rq3/autonomyTable}

\tabref{tab:autonomy-table} summarizes the autonomy levels in SoTS DTs in accordance with the classifications of \textcite{kritzinger2018digital} and \textcite{david2024infonomics}. 
Most studies (\xofyp{66}{80}) implement fully autonomous DTs for independent monitoring, control, or decision-making.
Digital shadows, passive representations without autonomy, appear in (\xofyp{6}{80}) studies. \textcite{hofmeister2024semantic} use them as data layers for agents assessing environmental risks.
Human-supervised DTs appear in (\xofyp{4}{80}) studies and human-actuated DTs in (\xofyp{3}{80}), typically in safety-critical contexts. For example, \textcite{folds2019digital} use a supervised DT for mission adaptation in a cyber-physical-human system.
Only one study uses a digital model (\xofyp{1}{80}), representing static models, for enterprise-level planning rather than real-time operation \cite{kulkarni2019towards}.

\subsubsection{DT services}\label{sec:dt-services}

\input{tables/rq3/dtServicesTable}

\tabref{tab:dt-services-table} summarizes the services provided by DTs in SoTS configurations. As shown in \figref{fig:dt-services}, most studies combine multiple services rather than using them in isolation.

The most widely used services are real-time monitoring (\xofyp{79}{80}), simulation (\xofyp{77}{80}), and optimization (\xofyp{68}{80}). Prediction (\xofyp{56}{80}), visualization (\xofyp{49}{80}), and information retrieval (\xofyp{48}{80}) are also frequently integrated.

The most common service combination, observed in \xofyp{6}{80} studies, includes real-time monitoring, simulation, optimization, prediction, and information retrieval, supporting both continuous system supervision and proactive planning. Other studies incorporate varied combinations, typically coupling the core services (monitoring, simulation, optimization, and prediction) with additional functionalities, e.g., visualization, information retrieval, diagnosis, and event detection.

\subsubsection{Modeling and simulation formalisms and techniques}\label{sec:modeling-methods}

\input{tables/rq3/hierarchicalModelingMethodsTable}

\tabref{tab:modeling-methods-structured-table} summarizes the modeling and simulation formalisms used in SoTS studies. Architectural and structural methods are most common (\xofyp{31}{80}), with UML (\xofyp{12}{80}) and SysML (\xofyp{11}{80}) for system specification. Spatial and visual models appear in (\xofyp{24}{80}) studies, including CAD (\xofyp{12}{80}) and 3D modeling (\xofyp{10}{80}) for physical layout and geometry. Mathematical and statistical models (\xofyp{23}{80}) support dynamics and uncertainty, often using Bayesian networks (BN) or general equations.
Ontological methods (\xofyp{19}{80}) address semantic integration via Web Ontology Language (OWL) and AutomationML. Formal methods (\xofyp{14}{80}) use Finite State Machines (FSM) and Fault Tree Analysis (FTA) for verification. AI/ML (\xofyp{13}{80}) enable adaptive learning. Continuous simulation methods (\xofyp{12}{80}) and agent-based simulations (\xofyp{10}{80}) model physical dynamics and interactions. Discrete-event simulation methods (\xofyp{8}{80}) are used for workflow and performance analysis.

At this point, we recall the previously discussed inherent limitation of the study (\secref{sec:threats}) that the technical and technological choices we observe are not necessarily universal best practices to implement SoTS, they are mere tendencies in an emerging field. With the maturation of the field, better technical and technological choices may emerge.

\begin{conclusionframe}{RQ3: Characteristics of DTs in SoTS}
Most SoTS use fully autonomous DTs that provide monitoring, simulation, prediction, and optimization services. Modeling approaches vary, with architectural, visual, and mathematical formalisms being the most frequently used.
\end{conclusionframe}

\subsection{What are the characteristics of SoS that are combined with DTs? (RQ4)}\label{results-rq4}

To identify the characteristics of SoS used in SoTS, we analyze their supported SoS dimensions (\secref{sec:dimensions}) and the forms of emergent behavior they exhibit (\secref{sec:emergence-type}).

\subsubsection{Dimensions of SoS}\label{sec:dimensions}

\begin{figure*}[htb]
    \centering
     \includegraphics[width=0.9\linewidth]{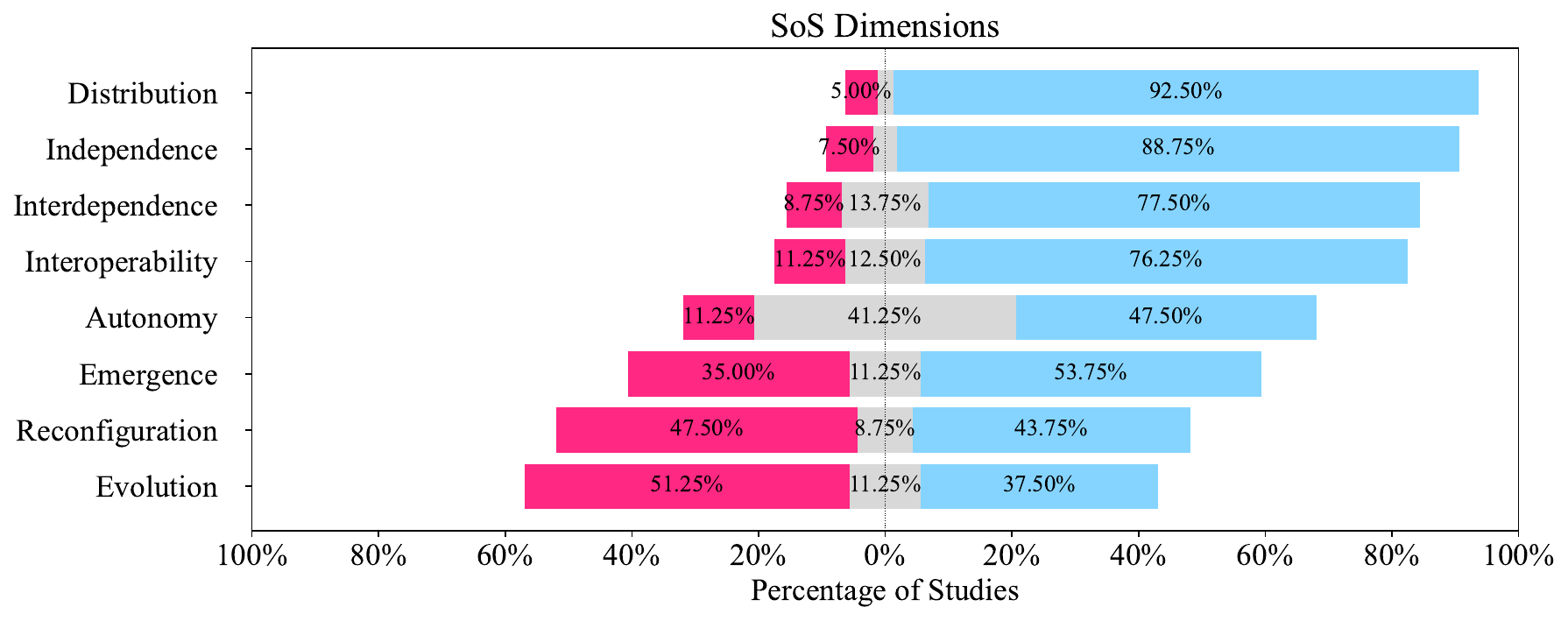}
    \caption{SoS dimensions: 
     {\color[HTML]{ff2882}\ding{108}} No,  
     {\color[HTML]{d8d8d8}\ding{108}} Partial, 
     {\color[HTML]{85d4ff}\ding{108}} Yes
    } 
    \label{fig:sos-dim}
\end{figure*}

\figref{fig:sos-dim} shows the SoS dimensions addressed in the studies. We follow the coding of SoS dimensions of \textcite{nielsen2015systems} and mark each SoS property either as explicitly present, partially evident, or absent. The system in the primary study is still considered a SoS as long as the primary study positions it as such.

The most frequently reported and investigated dimensions are distribution (\xofyp{74}{80}) and independence (\xofy{71}{80}). Interdependence (\xofyp{62}{80}) and interoperability (\xofyp{61}{80}) appear frequently as well, highlighting the perceived importance of coordination and information exchange in SoTS. Autonomy (\xofyp{38}{80} present and \xofyp{33}{80} partially present) and emergence (\xofyp{43}{80}) are less frequently addressed, with a substantial number of studies only partially addressing these properties. Reconfiguration (\xofyp{35}{80}) and evolution (\xofyp{30}{80}) are the least acknowledged SoS dimensions.

\subsubsection{Emergence type}\label{sec:emergence-type}

\tabref{tab:emergence-type-table} shows the types of emergent behavior reported in the studies. Weak emergence is most common (\xofyp{30}{80}). It involves behaviors that appear in system-level simulations but not in isolated components. \textcite{malayjerdi2022combined} demonstrate this through vehicle safety testing in software-in-the-loop setups. 
Simple emergence appears in (\xofyp{16}{80}) studies. It involves predictable interactions, e.g., in \textcite{zhang2022multi-scale}’s DT framework for shop floor coordination. 
Strong emergence is rare (\xofyp{6}{80}). It captures behaviors not predictable from subsystems. Examples include SoS simulations in mining \cite{bertoni2022digital} and automotive systems \cite{dahmen2022modeling}. (\xofyp{28}{80}) studies do not address emergent behaviors at all.

\begin{conclusionframe}{RQ4: Characteristics of SoS in SoTS}
SoTS support architectural SoS dimensions (distribution, independence, interdependence, and interoperability) but rarely address dynamical aspects (emergence, reconfiguration, and evolution). Emergent behavior is addressed in two thirds of the studies, most often as weak emergence, and many studies do not consider emergence at all.
\end{conclusionframe}

\input{tables/rq4/emergenceTable}

\subsection{How are non-functional properties addressed in systems that combine SoS and DT? (RQ5)}\label{results-rq5}

To understand how non-functional properties---specifically, reliability (i.e., continuity of correct service~\cite{avizienis2004basic}) and security---are handled in SoTS, we analyze whether such considerations are addressed in the system requirements, architecture, or implementation, and whether they are evaluated or validated.

\input{tables/rq5/reliabilityTable}
\input{tables/rq5/securityTable}

As shown in \tabref{tab:reliability-table} and \tabref{tab:security-table}, reliability as a concern appears in (\xofyp{41}{80}) studies, mostly addressed at the architectural level. Some typical architecture-level mechanisms for reliability of SoTS include fallback to local or lightweight DTs during communication loss~\cite{alam2017c2ps, larsen2024towards}, asynchronous communication for handling intermittent updates~\cite{acharya2023twins, liu2020web-based}, and runtime fault recovery~\cite{esterle2021digital, villalonga2021decision-making}. However, only (\xofyp{2}{80}) studies formally model reliability, and only (\xofyp{3}{80}) validate it. Those that do, typically validate reliability through simulation or fault injection \cite{park2020digital, saraeian2022digital}.

Security is covered architecturally in (\xofyp{19}{80}) studies, often through secure communication, access control, or authentication \cite{aziz2022empowering, acharya2023twins, dobie2024network}. Just (\xofyp{2}{80}) studies model security explicitly, and (\xofyp{3}{80}) perform validation through threat simulation or attack injection \cite{malayjerdi2022combined, stary2022privacy}.

These two concerns remain central, but they represent only part of the broader quality landscape. ISO/IEC 25010 outlines other key properties, e.g., maintainability, interoperability, and usability.

\begin{conclusionframe}{RQ5: NFPs focused on in SoTS}
Reliability is frequently addressed through architectural strategies, but rarely formalized or evaluated. Security is less commonly treated, and most studies lack explicit modeling or validation.
\end{conclusionframe}

\subsection{What is the level of technical and research maturity in SoTS? (RQ6)}\label{results-rq6}

To assess the maturity of SoTS research, we analyzed the TRLs and contribution types of studies (\secref{sec:trl-contribution}), assessment strategies (\secref{sec:evaluation}), and the role of standardization (\secref{sec:standards}). Note that due to the rigorous study design (i.e., the exclusion of shallow contributions), the following results may or may not be representative of the state-of-the-art.

\subsubsection{Technology readiness levels and contribution types}\label{sec:trl-contribution}

\input{tables/rq6/trlTable}
\input{tables/rq6/contributionTypeTable}

\begin{figure}[t]
    \centering
    \includegraphics[width=\linewidth]{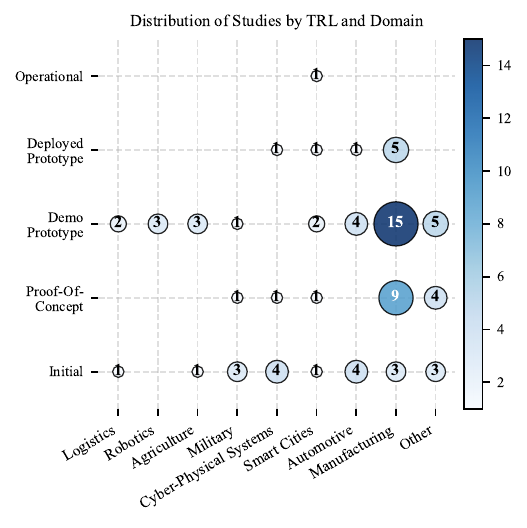}
    \caption{Distribution of TRL across domains}
    \label{fig:domain-trl-bubble}
\end{figure}

\begin{figure*}[t]
    \centering
    \includegraphics[width=0.9\linewidth]{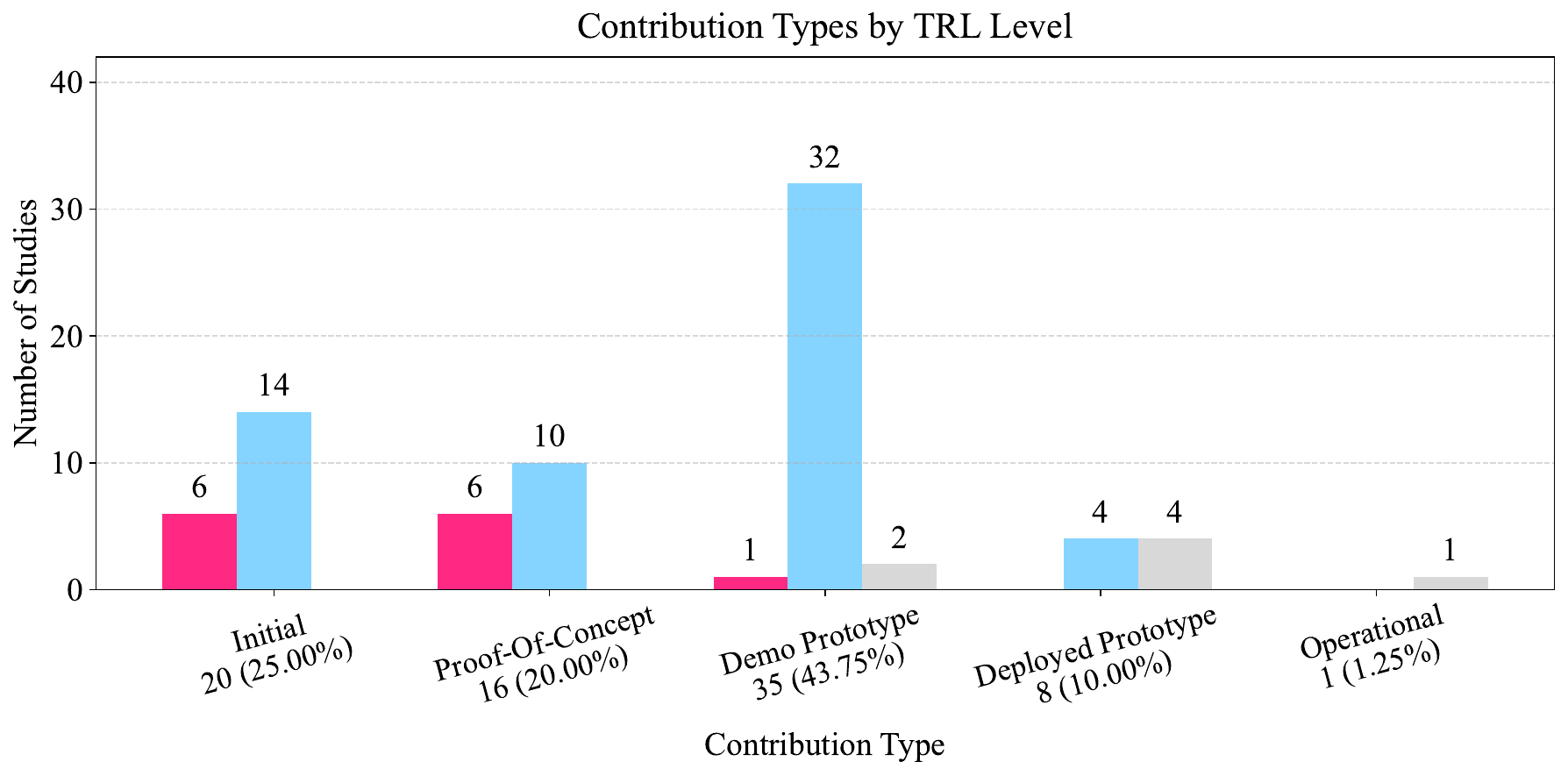}
    \caption{Distribution of contribution types across TRLs}
    \caption*{{
    \color[HTML]{FF2882}\ding{110}} Conceptual
    ~~~{\color[HTML]{85d4ff}\ding{110}} Technical
    ~~~{\color[HTML]{d8d8d8}\ding{110}} Case Study
    }
    \label{fig:trl-v-cont}
\end{figure*}

\tabref{tab:trl-table} shows that most studies operate at lower-to-mid maturity, with demonstrative prototypes being the most common stage (\xofyp{35}{80}), followed by initial (\xofyp{20}{80}) and proof-of-concept efforts (\xofyp{16}{80}). Only a few studies report deployed prototypes (\xofyp{8}{80}) or fully operational systems (\xofyp{1}{80}).

A cross-domain view of technology readiness levels (\figref{fig:domain-trl-bubble}) shows that low-to-mid TRL contributions dominate across all domains. Manufacturing accounts for the largest share of demonstrative prototype (\xofy{15}{35}) and proof-of-concept (\xofy{9}{16}) studies, while domains such as military, agriculture, robotics, and logistics are limited to early-stage TRLs, with no operational systems reported outside manufacturing.

\input{tables/rq6/hierarchicalEvaluationTable}

In terms of contribution types (\tabref{tab:contribution-type-table}), the vast majority are technical contributions (\xofyp{60}{80}), often proposing new architectures or implementations. Conceptual works (\xofyp{13}{80}) make up a smaller portion of the sample, and case studies are underrepresented (\xofyp{7}{80}).

As illustrated in \figref{fig:trl-v-cont}, technical contributions are the most typical across all TRLs but especially in demo prototypes and initial stages. Conceptual works appear mostly at lower TRLs. Case studies, i.e., in-depth investigations of a real-world, contemporary phenomena within their actual contexts~\cite{crowe2011case}, are rarely found and only emerge beyond the initial and proof-of-concept stages. This reflects a strong emphasis on engineering feasibility but limited real-world validation.

\subsubsection{Evaluation}\label{sec:evaluation}

\tabref{tab:evaluation-structured-table} shows that validation research (\xofyp{72}{80}) dominates the sample, mainly through prototyping (\xofyp{36}{80}), simulation (\xofyp{16}{80}), and conceptual design validation (\xofyp{13}{80}). Along with laboratory experiments (\xofyp{4}{80}) and mathematical analysis (\xofyp{3}{80}). For example, \textcite{hatledal2020co-simulation} and \textcite{chen2018digital} use simulation to validate co-simulated and behavior-predictive DTs, respectively. \textcite{larsen2024towards} prototype a DTaaS platform for robot composition, while \textcite{redelinghuys2020six-layer} validate architecture designs through structured frameworks and applied case studies. Mathematical analysis is used in \textcite{mahoro2023articulating} to formalize graph-based synchronization across DT layers. \textcite{savur2019hrc-sos} conduct laboratory experiments to evaluate a human-robot collaboration system through physical trials.

In contrast, evaluation research appears in only (\xofyp{8}{80}) studies. \textcite{ashtaritalkhestani2019architecture} conduct an industrial case study to assess DT-based automation, and \textcite{bertoni2022digital} apply action research to support planning in mining operations using an operational DT.

\subsubsection{Standards}\label{sec:standards}

\input{tables/rq6/standards}
\input{tables/rq6/dtOrSoSRelated}
\input{tables/rq7/hierarchicalProgrammingLanguagesTable}

\tabref{tab:standards-table} summarizes standards referenced across the studies. Open Platform Communications Unified Architecture (OPC UA) is the most used (\xofyp{13}{80}), supporting secure communication and hierarchical data exchange \cite{dobie2024network, joseph2021aggregated}. IEC 63278 (Asset Administration Shell) appears in (\xofyp{8}{80}) studies for asset representation and interoperability \cite{gill2022method, gollner2022collaborative}. Reference Architectural Model Industrie 4.0 (RAMI 4.0) is cited in (\xofyp{4}{80}) studies to guide structured DT integration \cite{binder2021utilizing}. Other domain-specific standards include VANET, IPv6 \cite{alam2017c2ps}, ISO/IEC/IEEE 15288 \cite{altamiranda2019system}, ISA-95 \cite{dobie2024network}, IEC 61850 \cite{jiang2022novel}, and IEEE 1451 \cite{li2022cognitive}. Security-related standards include GDPR \cite{stary2022privacy} and OAuth 2.0 \cite{human2023design, jiang2022novel}.

\input{tables/rq7/hierarchicalFrameworksTable}

\tabref{tab:dt-or-sos-related-table} shows that most standards are applied in DT specific contexts (\xofyp{18}{80}), fewer relate to SoS (\xofyp{10}{80}), and only (\xofyp{6}{80}) support both. DT-oriented examples include the use of OPC UA and RAMI 4.0 for modeling and communication \cite{binder2021utilizing}. SoS-focused rely on NATO and SISO standards to support coordination and mission-level system integration\cite{barden2022academic}. \textcite{vermesan2021internet} present a combined view, applying both DT and SoS-relevant standards in the Internet of Vehicles (IoV) context.
In total, \xofy{36}{80}{} unique studies rely on a standard, i.e., the majority of the sampled studies does not adhere to standards.

\begin{conclusionframe}{RQ6: Maturity of SoTS research}
SoTS research in our sample, even after rigorous quality criteria, is situated largely at low-to-mid TRLs, with demo prototypes and proof-of-concept efforts being the most common.
Validation is primarily conducted through prototyping and simulation, with limited empirical evaluation. Standards are inconsistently applied and tend to focus on DT-specific components, with few addressing SoS integration or supporting both layers.
\end{conclusionframe}

\subsection{What technology is used to implement systems that combine SoS and DT? (RQ7)}\label{results-rq7}

To understand what technologies support the implementation of SoTS, we examine the programming languages and data formats used (\secref{sec:programming-languages}), as well as the development frameworks and platforms adopted across studies (\secref{sec:frameworks}).

\subsubsection{Programming languages and formats}\label{sec:programming-languages}

\tabref{tab:programming-languages-structured-table} shows that most studies rely on general-purpose programming languages (\xofyp{36}{80}), particularly Python (\xofyp{22}{80}) and Java (\xofyp{14}{80}), reflecting their flexibility in data processing and simulation. Languages like JavaScript, C++, and C\# appear less frequently.
Data representation formats are used in (\xofyp{12}{80}) studies, with XML (\xofyp{9}{80}) and JSON (\xofyp{5}{80}) supporting structured data exchange.
Markup and styling languages, e.g., HTML and CSS, appear in (\xofyp{4}{80}) cases each, usually for visualization or web-based system interfaces.

\subsubsection{Frameworks and platforms}\label{sec:frameworks}

\tabref{tab:frameworks-structured-table} shows that most studies use modeling and simulation tools (\xofyp{35}{80}), notably MATLAB (\xofyp{10}{80}), Gazebo, Modelica, and Simulink (each \xofyp{4}{80}), supporting system dynamics and co-simulation.
Data management tools appear in (\xofyp{19}{80}) studies, with MongoDB (\xofyp{6}{80}) leading. Other tools like PostgreSQL, Redis, and Protégé support storage, synchronization, and ontology modeling.
Visualization tools are also common (\xofyp{19}{80}), with Unity (\xofyp{5}{80}) and platforms like WebGL and Kinect enabling interactive 3D or AR interfaces.
DT and IoT platforms are used in (\xofyp{15}{80}), including Eclipse Ditto and ROS (each \xofyp{4}{80}) supporting twin orchestration, and device interoperability.
Systems engineering tools (\xofyp{11}{80}), like Cameo Systems Modeler, Metasonic Suite, and Enterprise Architect, support architectural modeling.
Other categories include web/app frameworks (\xofyp{10}{80}), cloud and DevOps tools (\xofyp{8}{80}) like Docker and Azure, and analytics platforms (\xofyp{7}{80}), e.g., Grafana and Jupyter Lab for monitoring and machine learning (ML).

\begin{conclusionframe}{RQ7: Technologies used in SoTS}
Systems combining SoS and DT use diverse technologies, with Python and Java as primary languages and XML/JSON for data formatting. The frameworks used focus on supporting simulation, data management, and systems engineering.
\end{conclusionframe}

%% file: tables/rq1/motivations.tex
\begin{table*}[]
            \centering
            \caption{Motivations for combining DT and SoS}
            \label{tab:motivations-table}
            \begin{tabular}{@{}p{4cm}l p{11.5cm}@{}}
            \toprule
            \multicolumn{1}{c}{\textbf{Motivation}} & 
            \multicolumn{1}{c}{\textbf{Frequency}} & 
            \multicolumn{1}{c}{\textbf{Studies}} \\ 
            \midrule
            Optimization & \maindatabar{30} & \cite{bao2024digital}, \cite{becue2018cyberfactory}, \cite{chavezbaliguat2023digital}, \cite{chen2018digital}, \cite{duan2023digital}, \cite{folds2019digital}, \cite{gill2022method}, \cite{howard2021greenhouse}, \cite{jirsa2024use}, \cite{joseph2021aggregated}, \cite{kruger2022towards}, \cite{lee2022simulation}, \cite{li2022cognitive}, \cite{li2024comprehensive}, \cite{lippi2023enabling}, \cite{liu2020web-based}, \cite{maheshwari2022digital}, \cite{malayjerdi2022combined}, \cite{novak2022digitalized}, \cite{park2020digital}, \cite{pickering2023towards}, \cite{pillai2023digital}, \cite{potteiger2023live}, \cite{priyanta2024is}, \cite{somma2023digital}, \cite{vermesan2021internet}, \cite{villalonga2021decision-making}, \cite{wullink2024foundational}, \cite{zhang2022multi-scale}, \cite{zhang2021bi-level} \\
Integration & \maindatabar{25} & \cite{acharya2023twins}, \cite{alam2017c2ps}, \cite{ashtaritalkhestani2019architecture}, \cite{aziz2022empowering}, \cite{bellavista2023requirements}, \cite{demir2023vertically-integrated}, \cite{dobie2024network}, \cite{doubell2023digital}, \cite{esterle2021digital}, \cite{gil2023modeling}, \cite{gil2024integrating}, \cite{gollner2022collaborative}, \cite{hatledal2020co-simulation}, \cite{heininger2021capturing}, \cite{heithoff2023challenges}, \cite{hofmeister2024cross-domain}, \cite{hofmeister2024semantic}, \cite{human2023design}, \cite{larsen2024towards}, \cite{mahoro2023articulating}, \cite{monsalve2021novel}, \cite{reiche2021digital}, \cite{savur2019hrc-sos}, \cite{stary2022privacy}, \cite{vogel-heuser2021approach} \\
Validation & \maindatabar{15} & \cite{bertoni2022digital}, \cite{clark2021chapter}, \cite{coupaye2023graph-based}, \cite{dahmen2022modeling}, \cite{dickopf2019holistic}, \cite{ehemann2023digital}, \cite{kulkarni2019towards}, \cite{marah2023architecture}, \cite{mavromatis2024umbrella}, \cite{oquendo2019dealing}, \cite{redelinghuys2020six-layer}, \cite{samak2023autodrive}, \cite{schluse2017experimentable}, \cite{wagner2023using}, \cite{wang2024construction} \\
Maintainability & \maindatabar{10} & \cite{altamiranda2019system}, \cite{barden2022academic}, \cite{binder2021utilizing}, \cite{hatakeyama2018systems}, \cite{jiang2022novel}, \cite{kutzke2021subsystem}, \cite{lopez2023modeling}, \cite{parri2021framework}, \cite{parri2019jarvis}, \cite{saraeian2022digital} \\
\bottomrule
            \end{tabular}
            \end{table*}

%% file: tables/rq1/intentsTable.tex
\begin{table*}[]
            \centering
            \caption{Intents of combining DT and SoS}
            \label{tab:intents-table}
            \begin{tabular}{@{}p{4cm}l p{11.5cm}@{}}
            \toprule
            \multicolumn{1}{c}{\textbf{Intent}} & 
            \multicolumn{1}{c}{\textbf{Frequency}} & 
            \multicolumn{1}{c}{\textbf{Studies}} \\ 
            \midrule
            Combining DTs into a SoS & \maindatabar{49} & \cite{acharya2023twins}, \cite{alam2017c2ps}, \cite{altamiranda2019system}, \cite{ashtaritalkhestani2019architecture}, \cite{aziz2022empowering}, \cite{barden2022academic}, \cite{becue2018cyberfactory}, \cite{bellavista2023requirements}, \cite{bertoni2022digital}, \cite{chen2018digital}, \cite{coupaye2023graph-based}, \cite{dahmen2022modeling}, \cite{demir2023vertically-integrated}, \cite{dickopf2019holistic}, \cite{doubell2023digital}, \cite{ehemann2023digital}, \cite{esterle2021digital}, \cite{gil2023modeling}, \cite{gil2024integrating}, \cite{gill2022method}, \cite{gollner2022collaborative}, \cite{hatakeyama2018systems}, \cite{heininger2021capturing}, \cite{heithoff2023challenges}, \cite{hofmeister2024cross-domain}, \cite{hofmeister2024semantic}, \cite{howard2021greenhouse}, \cite{human2023design}, \cite{jiang2022novel}, \cite{jirsa2024use}, \cite{joseph2021aggregated}, \cite{kruger2022towards}, \cite{kulkarni2019towards}, \cite{larsen2024towards}, \cite{li2022cognitive}, \cite{lippi2023enabling}, \cite{liu2020web-based}, \cite{mahoro2023articulating}, \cite{marah2023architecture}, \cite{mavromatis2024umbrella}, \cite{monsalve2021novel}, \cite{parri2019jarvis}, \cite{redelinghuys2020six-layer}, \cite{reiche2021digital}, \cite{schluse2017experimentable}, \cite{stary2022privacy}, \cite{villalonga2021decision-making}, \cite{vogel-heuser2021approach}, \cite{wullink2024foundational} \\
Twinning a SoS & \maindatabar{31} & \cite{bao2024digital}, \cite{binder2021utilizing}, \cite{chavezbaliguat2023digital}, \cite{clark2021chapter}, \cite{dobie2024network}, \cite{duan2023digital}, \cite{folds2019digital}, \cite{hatledal2020co-simulation}, \cite{kutzke2021subsystem}, \cite{lee2022simulation}, \cite{li2024comprehensive}, \cite{lopez2023modeling}, \cite{maheshwari2022digital}, \cite{malayjerdi2022combined}, \cite{novak2022digitalized}, \cite{oquendo2019dealing}, \cite{park2020digital}, \cite{parri2021framework}, \cite{pickering2023towards}, \cite{pillai2023digital}, \cite{potteiger2023live}, \cite{priyanta2024is}, \cite{samak2023autodrive}, \cite{saraeian2022digital}, \cite{savur2019hrc-sos}, \cite{somma2023digital}, \cite{vermesan2021internet}, \cite{wagner2023using}, \cite{wang2024construction}, \cite{zhang2022multi-scale}, \cite{zhang2021bi-level} \\
\bottomrule
            \end{tabular}
            \end{table*}

%% file: tables/rq1/domainsTable.tex
\begin{table*}[]
            \centering
            \caption{Application domains}
            \label{tab:domains-table}
            \begin{tabular}{@{}p{4cm}l p{11.5cm}@{}}
            \toprule
            \multicolumn{1}{c}{\textbf{Domain}} & 
            \multicolumn{1}{c}{\textbf{Frequency}} & 
            \multicolumn{1}{c}{\textbf{Studies}} \\ 
            \midrule
            Manufacturing & \maindatabar{32} & \cite{acharya2023twins}, \cite{ashtaritalkhestani2019architecture}, \cite{aziz2022empowering}, \cite{bao2024digital}, \cite{barden2022academic}, \cite{bellavista2023requirements}, \cite{binder2021utilizing}, \cite{demir2023vertically-integrated}, \cite{duan2023digital}, \cite{ehemann2023digital}, \cite{esterle2021digital}, \cite{gill2022method}, \cite{gollner2022collaborative}, \cite{jirsa2024use}, \cite{joseph2021aggregated}, \cite{kruger2022towards}, \cite{kutzke2021subsystem}, \cite{larsen2024towards}, \cite{lippi2023enabling}, \cite{liu2020web-based}, \cite{marah2023architecture}, \cite{monsalve2021novel}, \cite{novak2022digitalized}, \cite{park2020digital}, \cite{parri2019jarvis}, \cite{redelinghuys2020six-layer}, \cite{reiche2021digital}, \cite{schluse2017experimentable}, \cite{villalonga2021decision-making}, \cite{vogel-heuser2021approach}, \cite{zhang2022multi-scale}, \cite{zhang2021bi-level} \\
Automotive & \maindatabar{9} & \cite{chen2018digital}, \cite{dahmen2022modeling}, \cite{heithoff2023challenges}, \cite{malayjerdi2022combined}, \cite{oquendo2019dealing}, \cite{pillai2023digital}, \cite{potteiger2023live}, \cite{samak2023autodrive}, \cite{vermesan2021internet} \\
Cyber-Physical Systems & \maindatabar{6} & \cite{alam2017c2ps}, \cite{coupaye2023graph-based}, \cite{li2022cognitive}, \cite{mahoro2023articulating}, \cite{parri2021framework}, \cite{stary2022privacy} \\
Smart Cities & \maindatabar{6} & \cite{hofmeister2024cross-domain}, \cite{hofmeister2024semantic}, \cite{human2023design}, \cite{li2024comprehensive}, \cite{mavromatis2024umbrella}, \cite{somma2023digital} \\
Military & \maindatabar{5} & \cite{folds2019digital}, \cite{hatakeyama2018systems}, \cite{lee2022simulation}, \cite{lopez2023modeling}, \cite{wang2024construction} \\
Agriculture & \maindatabar{4} & \cite{chavezbaliguat2023digital}, \cite{howard2021greenhouse}, \cite{pickering2023towards}, \cite{saraeian2022digital} \\
Logistics & \maindatabar{3} & \cite{clark2021chapter}, \cite{doubell2023digital}, \cite{wagner2023using} \\
Robotics & \maindatabar{3} & \cite{gil2023modeling}, \cite{gil2024integrating}, \cite{savur2019hrc-sos} \\
Other & \maindatabar{12} & \cite{altamiranda2019system}, \cite{becue2018cyberfactory}, \cite{bertoni2022digital}, \cite{dickopf2019holistic}, \cite{dobie2024network}, \cite{hatledal2020co-simulation}, \cite{heininger2021capturing}, \cite{jiang2022novel}, \cite{kulkarni2019towards}, \cite{maheshwari2022digital}, \cite{priyanta2024is}, \cite{wullink2024foundational} \\
\bottomrule
            \end{tabular}
            \end{table*}

%% file: tables/rq2/sotsTypeTable.tex
\begin{table*}[]
            \centering
            \caption{SoTS Type}
            \label{tab:sots-type-table}
            \begin{tabular}{@{}p{4cm}l p{11.5cm}@{}}
            \toprule
            \multicolumn{1}{c}{\textbf{SoTS}} & 
            \multicolumn{1}{c}{\textbf{Frequency}} & 
            \multicolumn{1}{c}{\textbf{Studies}} \\ 
            \midrule
            Acknowledged SoTS & \maindatabar{31} & \cite{altamiranda2019system}, \cite{becue2018cyberfactory}, \cite{bellavista2023requirements}, \cite{bertoni2022digital}, \cite{binder2021utilizing}, \cite{coupaye2023graph-based}, \cite{demir2023vertically-integrated}, \cite{ehemann2023digital}, \cite{folds2019digital}, \cite{gill2022method}, \cite{heininger2021capturing}, \cite{heithoff2023challenges}, \cite{human2023design}, \cite{kruger2022towards}, \cite{li2022cognitive}, \cite{lippi2023enabling}, \cite{lopez2023modeling}, \cite{maheshwari2022digital}, \cite{mahoro2023articulating}, \cite{malayjerdi2022combined}, \cite{monsalve2021novel}, \cite{parri2019jarvis}, \cite{pillai2023digital}, \cite{potteiger2023live}, \cite{priyanta2024is}, \cite{redelinghuys2020six-layer}, \cite{samak2023autodrive}, \cite{saraeian2022digital}, \cite{somma2023digital}, \cite{wagner2023using}, \cite{zhang2021bi-level} \\
Directed SoTS & \maindatabar{26} & \cite{alam2017c2ps}, \cite{ashtaritalkhestani2019architecture}, \cite{aziz2022empowering}, \cite{chavezbaliguat2023digital}, \cite{dickopf2019holistic}, \cite{dobie2024network}, \cite{doubell2023digital}, \cite{duan2023digital}, \cite{gil2024integrating}, \cite{hatakeyama2018systems}, \cite{hofmeister2024semantic}, \cite{jiang2022novel}, \cite{kulkarni2019towards}, \cite{kutzke2021subsystem}, \cite{larsen2024towards}, \cite{lee2022simulation}, \cite{li2024comprehensive}, \cite{novak2022digitalized}, \cite{park2020digital}, \cite{parri2021framework}, \cite{reiche2021digital}, \cite{schluse2017experimentable}, \cite{stary2022privacy}, \cite{villalonga2021decision-making}, \cite{wang2024construction}, \cite{zhang2022multi-scale} \\
Collaborative SoTS & \maindatabar{19} & \cite{acharya2023twins}, \cite{bao2024digital}, \cite{barden2022academic}, \cite{chen2018digital}, \cite{dahmen2022modeling}, \cite{gil2023modeling}, \cite{gollner2022collaborative}, \cite{hatledal2020co-simulation}, \cite{hofmeister2024cross-domain}, \cite{howard2021greenhouse}, \cite{jirsa2024use}, \cite{joseph2021aggregated}, \cite{liu2020web-based}, \cite{marah2023architecture}, \cite{mavromatis2024umbrella}, \cite{savur2019hrc-sos}, \cite{vermesan2021internet}, \cite{vogel-heuser2021approach}, \cite{wullink2024foundational} \\
Virtual SoTS & \maindatabar{4} & \cite{clark2021chapter}, \cite{esterle2021digital}, \cite{oquendo2019dealing}, \cite{pickering2023towards} \\
\bottomrule
            \end{tabular}
            \end{table*}

%% file: tables/rq2/constituentUnitsTable.tex
\begin{table*}[]
            \centering
            \caption{Constituent units}
            \label{tab:constituent-units-table}
            \begin{tabular}{@{}p{4cm}l p{11.5cm}@{}}
            \toprule
            \multicolumn{1}{c}{\textbf{Constituent Unit}} & 
            \multicolumn{1}{c}{\textbf{Frequency}} & 
            \multicolumn{1}{c}{\textbf{Studies}} \\ 
            \midrule
            Physical Systems & \maindatabar{62} & \cite{acharya2023twins}, \cite{altamiranda2019system}, \cite{ashtaritalkhestani2019architecture}, \cite{aziz2022empowering}, \cite{bao2024digital}, \cite{barden2022academic}, \cite{becue2018cyberfactory}, \cite{bellavista2023requirements}, \cite{bertoni2022digital}, \cite{binder2021utilizing}, \cite{chavezbaliguat2023digital}, \cite{chen2018digital}, \cite{coupaye2023graph-based}, \cite{dahmen2022modeling}, \cite{demir2023vertically-integrated}, \cite{dobie2024network}, \cite{doubell2023digital}, \cite{duan2023digital}, \cite{ehemann2023digital}, \cite{esterle2021digital}, \cite{gil2023modeling}, \cite{gill2022method}, \cite{gollner2022collaborative}, \cite{hatledal2020co-simulation}, \cite{heininger2021capturing}, \cite{heithoff2023challenges}, \cite{hofmeister2024cross-domain}, \cite{hofmeister2024semantic}, \cite{howard2021greenhouse}, \cite{human2023design}, \cite{jiang2022novel}, \cite{jirsa2024use}, \cite{joseph2021aggregated}, \cite{kruger2022towards}, \cite{kutzke2021subsystem}, \cite{larsen2024towards}, \cite{lee2022simulation}, \cite{li2022cognitive}, \cite{li2024comprehensive}, \cite{lippi2023enabling}, \cite{liu2020web-based}, \cite{lopez2023modeling}, \cite{malayjerdi2022combined}, \cite{monsalve2021novel}, \cite{novak2022digitalized}, \cite{oquendo2019dealing}, \cite{park2020digital}, \cite{pillai2023digital}, \cite{potteiger2023live}, \cite{redelinghuys2020six-layer}, \cite{reiche2021digital}, \cite{samak2023autodrive}, \cite{saraeian2022digital}, \cite{somma2023digital}, \cite{vermesan2021internet}, \cite{villalonga2021decision-making}, \cite{vogel-heuser2021approach}, \cite{wagner2023using}, \cite{wang2024construction}, \cite{wullink2024foundational}, \cite{zhang2022multi-scale}, \cite{zhang2021bi-level} \\
Cyber Physical Systems & \maindatabar{9} & \cite{alam2017c2ps}, \cite{clark2021chapter}, \cite{hatakeyama2018systems}, \cite{mahoro2023articulating}, \cite{marah2023architecture}, \cite{mavromatis2024umbrella}, \cite{priyanta2024is}, \cite{schluse2017experimentable}, \cite{stary2022privacy} \\
Cyber-Physical-Human Systems & \maindatabar{7} & \cite{dickopf2019holistic}, \cite{folds2019digital}, \cite{gil2024integrating}, \cite{parri2021framework}, \cite{parri2019jarvis}, \cite{pickering2023towards}, \cite{savur2019hrc-sos} \\
Enterprise Systems & \maindatabar{2} & \cite{kulkarni2019towards}, \cite{maheshwari2022digital} \\
\bottomrule
            \end{tabular}
            \end{table*}

%% file: tables/rq3/autonomyTable.tex
\begin{table*}[]
            \centering
            \caption{Levels of autonomy}
            \label{tab:autonomy-table}
            \begin{tabular}{@{}p{4cm}l p{11.5cm}@{}}
            \toprule
            \multicolumn{1}{c}{\textbf{Autonomy}} & 
            \multicolumn{1}{c}{\textbf{Frequency}} & 
            \multicolumn{1}{c}{\textbf{Studies}} \\ 
            \midrule
            Digital Twin & \maindatabar{66} & \cite{acharya2023twins}, \cite{alam2017c2ps}, \cite{altamiranda2019system}, \cite{ashtaritalkhestani2019architecture}, \cite{aziz2022empowering}, \cite{bao2024digital}, \cite{barden2022academic}, \cite{becue2018cyberfactory}, \cite{bellavista2023requirements}, \cite{binder2021utilizing}, \cite{chen2018digital}, \cite{clark2021chapter}, \cite{coupaye2023graph-based}, \cite{dahmen2022modeling}, \cite{demir2023vertically-integrated}, \cite{doubell2023digital}, \cite{duan2023digital}, \cite{ehemann2023digital}, \cite{esterle2021digital}, \cite{gil2023modeling}, \cite{gill2022method}, \cite{gollner2022collaborative}, \cite{hatakeyama2018systems}, \cite{hatledal2020co-simulation}, \cite{heininger2021capturing}, \cite{heithoff2023challenges}, \cite{howard2021greenhouse}, \cite{human2023design}, \cite{jiang2022novel}, \cite{jirsa2024use}, \cite{joseph2021aggregated}, \cite{kruger2022towards}, \cite{kutzke2021subsystem}, \cite{larsen2024towards}, \cite{lee2022simulation}, \cite{li2022cognitive}, \cite{li2024comprehensive}, \cite{lippi2023enabling}, \cite{liu2020web-based}, \cite{lopez2023modeling}, \cite{maheshwari2022digital}, \cite{mahoro2023articulating}, \cite{malayjerdi2022combined}, \cite{marah2023architecture}, \cite{mavromatis2024umbrella}, \cite{monsalve2021novel}, \cite{novak2022digitalized}, \cite{oquendo2019dealing}, \cite{park2020digital}, \cite{pillai2023digital}, \cite{potteiger2023live}, \cite{priyanta2024is}, \cite{redelinghuys2020six-layer}, \cite{reiche2021digital}, \cite{samak2023autodrive}, \cite{schluse2017experimentable}, \cite{somma2023digital}, \cite{stary2022privacy}, \cite{vermesan2021internet}, \cite{villalonga2021decision-making}, \cite{vogel-heuser2021approach}, \cite{wagner2023using}, \cite{wang2024construction}, \cite{wullink2024foundational}, \cite{zhang2022multi-scale}, \cite{zhang2021bi-level} \\
Digital Shadow & \maindatabar{6} & \cite{bertoni2022digital}, \cite{chavezbaliguat2023digital}, \cite{dobie2024network}, \cite{hofmeister2024cross-domain}, \cite{hofmeister2024semantic}, \cite{saraeian2022digital} \\
Human-Supervised Digital Twin & \maindatabar{4} & \cite{folds2019digital}, \cite{gil2024integrating}, \cite{pickering2023towards}, \cite{savur2019hrc-sos} \\
Human-Actuated Digital Twin & \maindatabar{3} & \cite{dickopf2019holistic}, \cite{parri2021framework}, \cite{parri2019jarvis} \\
Digital Model & \maindatabar{1} & \cite{kulkarni2019towards} \\
\bottomrule
            \end{tabular}
            \end{table*}

%% file: tables/rq3/dtServicesTable.tex
\begin{table*}[]
            \centering
            \caption{DT services supported}
            \label{tab:dt-services-table}
            \begin{tabular}{@{}p{4cm}l p{11.5cm}@{}}
            \toprule
            \multicolumn{1}{c}{\textbf{Service}} & 
            \multicolumn{1}{c}{\textbf{Frequency}} & 
            \multicolumn{1}{c}{\textbf{Studies}} \\ 
            \midrule
            Real-time monitoring & \maindatabar{79} & \cite{acharya2023twins}, \cite{alam2017c2ps}, \cite{altamiranda2019system}, \cite{ashtaritalkhestani2019architecture}, \cite{aziz2022empowering}, \cite{bao2024digital}, \cite{barden2022academic}, \cite{becue2018cyberfactory}, \cite{bellavista2023requirements}, \cite{bertoni2022digital}, \cite{binder2021utilizing}, \cite{chavezbaliguat2023digital}, \cite{chen2018digital}, \cite{clark2021chapter}, \cite{coupaye2023graph-based}, \cite{dahmen2022modeling}, \cite{demir2023vertically-integrated}, \cite{dickopf2019holistic}, \cite{dobie2024network}, \cite{doubell2023digital}, \cite{duan2023digital}, \cite{ehemann2023digital}, \cite{esterle2021digital}, \cite{folds2019digital}, \cite{gil2023modeling}, \cite{gil2024integrating}, \cite{gill2022method}, \cite{gollner2022collaborative}, \cite{hatakeyama2018systems}, \cite{hatledal2020co-simulation}, \cite{heininger2021capturing}, \cite{heithoff2023challenges}, \cite{hofmeister2024cross-domain}, \cite{hofmeister2024semantic}, \cite{howard2021greenhouse}, \cite{human2023design}, \cite{jiang2022novel}, \cite{jirsa2024use}, \cite{joseph2021aggregated}, \cite{kruger2022towards}, \cite{kutzke2021subsystem}, \cite{larsen2024towards}, \cite{lee2022simulation}, \cite{li2022cognitive}, \cite{li2024comprehensive}, \cite{lippi2023enabling}, \cite{liu2020web-based}, \cite{lopez2023modeling}, \cite{maheshwari2022digital}, \cite{mahoro2023articulating}, \cite{malayjerdi2022combined}, \cite{marah2023architecture}, \cite{mavromatis2024umbrella}, \cite{monsalve2021novel}, \cite{novak2022digitalized}, \cite{oquendo2019dealing}, \cite{park2020digital}, \cite{parri2021framework}, \cite{parri2019jarvis}, \cite{pickering2023towards}, \cite{pillai2023digital}, \cite{potteiger2023live}, \cite{priyanta2024is}, \cite{redelinghuys2020six-layer}, \cite{reiche2021digital}, \cite{samak2023autodrive}, \cite{saraeian2022digital}, \cite{savur2019hrc-sos}, \cite{schluse2017experimentable}, \cite{somma2023digital}, \cite{stary2022privacy}, \cite{vermesan2021internet}, \cite{villalonga2021decision-making}, \cite{vogel-heuser2021approach}, \cite{wagner2023using}, \cite{wang2024construction}, \cite{wullink2024foundational}, \cite{zhang2022multi-scale}, \cite{zhang2021bi-level} \\
Simulation & \maindatabar{77} & \cite{acharya2023twins}, \cite{alam2017c2ps}, \cite{altamiranda2019system}, \cite{ashtaritalkhestani2019architecture}, \cite{bao2024digital}, \cite{barden2022academic}, \cite{becue2018cyberfactory}, \cite{bellavista2023requirements}, \cite{bertoni2022digital}, \cite{binder2021utilizing}, \cite{chen2018digital}, \cite{clark2021chapter}, \cite{coupaye2023graph-based}, \cite{dahmen2022modeling}, \cite{demir2023vertically-integrated}, \cite{dickopf2019holistic}, \cite{dobie2024network}, \cite{doubell2023digital}, \cite{duan2023digital}, \cite{ehemann2023digital}, \cite{esterle2021digital}, \cite{folds2019digital}, \cite{gil2023modeling}, \cite{gil2024integrating}, \cite{gill2022method}, \cite{gollner2022collaborative}, \cite{hatakeyama2018systems}, \cite{hatledal2020co-simulation}, \cite{heininger2021capturing}, \cite{heithoff2023challenges}, \cite{hofmeister2024cross-domain}, \cite{hofmeister2024semantic}, \cite{howard2021greenhouse}, \cite{human2023design}, \cite{jiang2022novel}, \cite{jirsa2024use}, \cite{joseph2021aggregated}, \cite{kruger2022towards}, \cite{kulkarni2019towards}, \cite{kutzke2021subsystem}, \cite{larsen2024towards}, \cite{lee2022simulation}, \cite{li2022cognitive}, \cite{li2024comprehensive}, \cite{lippi2023enabling}, \cite{liu2020web-based}, \cite{lopez2023modeling}, \cite{maheshwari2022digital}, \cite{mahoro2023articulating}, \cite{malayjerdi2022combined}, \cite{marah2023architecture}, \cite{mavromatis2024umbrella}, \cite{monsalve2021novel}, \cite{novak2022digitalized}, \cite{oquendo2019dealing}, \cite{park2020digital}, \cite{parri2021framework}, \cite{parri2019jarvis}, \cite{pickering2023towards}, \cite{potteiger2023live}, \cite{priyanta2024is}, \cite{redelinghuys2020six-layer}, \cite{reiche2021digital}, \cite{samak2023autodrive}, \cite{saraeian2022digital}, \cite{savur2019hrc-sos}, \cite{schluse2017experimentable}, \cite{somma2023digital}, \cite{stary2022privacy}, \cite{vermesan2021internet}, \cite{villalonga2021decision-making}, \cite{vogel-heuser2021approach}, \cite{wagner2023using}, \cite{wang2024construction}, \cite{wullink2024foundational}, \cite{zhang2022multi-scale}, \cite{zhang2021bi-level} \\
Optimization & \maindatabar{68} & \cite{acharya2023twins}, \cite{alam2017c2ps}, \cite{altamiranda2019system}, \cite{ashtaritalkhestani2019architecture}, \cite{bao2024digital}, \cite{barden2022academic}, \cite{becue2018cyberfactory}, \cite{bellavista2023requirements}, \cite{bertoni2022digital}, \cite{binder2021utilizing}, \cite{chavezbaliguat2023digital}, \cite{chen2018digital}, \cite{clark2021chapter}, \cite{coupaye2023graph-based}, \cite{demir2023vertically-integrated}, \cite{dickopf2019holistic}, \cite{dobie2024network}, \cite{doubell2023digital}, \cite{duan2023digital}, \cite{ehemann2023digital}, \cite{esterle2021digital}, \cite{folds2019digital}, \cite{gil2023modeling}, \cite{gil2024integrating}, \cite{gill2022method}, \cite{hatakeyama2018systems}, \cite{hatledal2020co-simulation}, \cite{heininger2021capturing}, \cite{heithoff2023challenges}, \cite{howard2021greenhouse}, \cite{jiang2022novel}, \cite{jirsa2024use}, \cite{joseph2021aggregated}, \cite{kruger2022towards}, \cite{kulkarni2019towards}, \cite{kutzke2021subsystem}, \cite{larsen2024towards}, \cite{lee2022simulation}, \cite{li2022cognitive}, \cite{li2024comprehensive}, \cite{lippi2023enabling}, \cite{liu2020web-based}, \cite{maheshwari2022digital}, \cite{mahoro2023articulating}, \cite{malayjerdi2022combined}, \cite{marah2023architecture}, \cite{mavromatis2024umbrella}, \cite{monsalve2021novel}, \cite{novak2022digitalized}, \cite{oquendo2019dealing}, \cite{park2020digital}, \cite{pickering2023towards}, \cite{pillai2023digital}, \cite{potteiger2023live}, \cite{priyanta2024is}, \cite{redelinghuys2020six-layer}, \cite{samak2023autodrive}, \cite{saraeian2022digital}, \cite{schluse2017experimentable}, \cite{somma2023digital}, \cite{vermesan2021internet}, \cite{villalonga2021decision-making}, \cite{vogel-heuser2021approach}, \cite{wagner2023using}, \cite{wang2024construction}, \cite{wullink2024foundational}, \cite{zhang2022multi-scale}, \cite{zhang2021bi-level} \\
Prediction & \maindatabar{56} & \cite{acharya2023twins}, \cite{alam2017c2ps}, \cite{altamiranda2019system}, \cite{ashtaritalkhestani2019architecture}, \cite{bao2024digital}, \cite{barden2022academic}, \cite{becue2018cyberfactory}, \cite{bellavista2023requirements}, \cite{bertoni2022digital}, \cite{chavezbaliguat2023digital}, \cite{chen2018digital}, \cite{clark2021chapter}, \cite{coupaye2023graph-based}, \cite{dahmen2022modeling}, \cite{demir2023vertically-integrated}, \cite{dickopf2019holistic}, \cite{dobie2024network}, \cite{doubell2023digital}, \cite{duan2023digital}, \cite{ehemann2023digital}, \cite{esterle2021digital}, \cite{folds2019digital}, \cite{gollner2022collaborative}, \cite{hatakeyama2018systems}, \cite{hatledal2020co-simulation}, \cite{heininger2021capturing}, \cite{hofmeister2024cross-domain}, \cite{hofmeister2024semantic}, \cite{howard2021greenhouse}, \cite{jiang2022novel}, \cite{joseph2021aggregated}, \cite{kulkarni2019towards}, \cite{kutzke2021subsystem}, \cite{larsen2024towards}, \cite{li2022cognitive}, \cite{li2024comprehensive}, \cite{lippi2023enabling}, \cite{maheshwari2022digital}, \cite{mahoro2023articulating}, \cite{marah2023architecture}, \cite{novak2022digitalized}, \cite{oquendo2019dealing}, \cite{park2020digital}, \cite{parri2021framework}, \cite{parri2019jarvis}, \cite{pickering2023towards}, \cite{pillai2023digital}, \cite{priyanta2024is}, \cite{samak2023autodrive}, \cite{saraeian2022digital}, \cite{somma2023digital}, \cite{villalonga2021decision-making}, \cite{wagner2023using}, \cite{wang2024construction}, \cite{wullink2024foundational}, \cite{zhang2022multi-scale} \\
Visualization & \maindatabar{49} & \cite{acharya2023twins}, \cite{alam2017c2ps}, \cite{altamiranda2019system}, \cite{ashtaritalkhestani2019architecture}, \cite{aziz2022empowering}, \cite{bao2024digital}, \cite{barden2022academic}, \cite{bertoni2022digital}, \cite{chavezbaliguat2023digital}, \cite{chen2018digital}, \cite{coupaye2023graph-based}, \cite{dahmen2022modeling}, \cite{demir2023vertically-integrated}, \cite{dickopf2019holistic}, \cite{dobie2024network}, \cite{doubell2023digital}, \cite{duan2023digital}, \cite{ehemann2023digital}, \cite{esterle2021digital}, \cite{folds2019digital}, \cite{hatledal2020co-simulation}, \cite{hofmeister2024cross-domain}, \cite{hofmeister2024semantic}, \cite{howard2021greenhouse}, \cite{jirsa2024use}, \cite{joseph2021aggregated}, \cite{kruger2022towards}, \cite{larsen2024towards}, \cite{lee2022simulation}, \cite{li2024comprehensive}, \cite{liu2020web-based}, \cite{lopez2023modeling}, \cite{maheshwari2022digital}, \cite{mahoro2023articulating}, \cite{malayjerdi2022combined}, \cite{marah2023architecture}, \cite{mavromatis2024umbrella}, \cite{pickering2023towards}, \cite{potteiger2023live}, \cite{priyanta2024is}, \cite{redelinghuys2020six-layer}, \cite{samak2023autodrive}, \cite{savur2019hrc-sos}, \cite{schluse2017experimentable}, \cite{somma2023digital}, \cite{stary2022privacy}, \cite{villalonga2021decision-making}, \cite{wang2024construction}, \cite{wullink2024foundational} \\
Information Retrieval & \maindatabar{48} & \cite{acharya2023twins}, \cite{alam2017c2ps}, \cite{altamiranda2019system}, \cite{ashtaritalkhestani2019architecture}, \cite{aziz2022empowering}, \cite{becue2018cyberfactory}, \cite{bellavista2023requirements}, \cite{binder2021utilizing}, \cite{chavezbaliguat2023digital}, \cite{clark2021chapter}, \cite{coupaye2023graph-based}, \cite{dahmen2022modeling}, \cite{demir2023vertically-integrated}, \cite{dickopf2019holistic}, \cite{dobie2024network}, \cite{doubell2023digital}, \cite{ehemann2023digital}, \cite{gil2024integrating}, \cite{gollner2022collaborative}, \cite{hatakeyama2018systems}, \cite{heininger2021capturing}, \cite{heithoff2023challenges}, \cite{hofmeister2024cross-domain}, \cite{hofmeister2024semantic}, \cite{howard2021greenhouse}, \cite{human2023design}, \cite{jiang2022novel}, \cite{jirsa2024use}, \cite{kruger2022towards}, \cite{kulkarni2019towards}, \cite{kutzke2021subsystem}, \cite{larsen2024towards}, \cite{lee2022simulation}, \cite{li2022cognitive}, \cite{li2024comprehensive}, \cite{lippi2023enabling}, \cite{liu2020web-based}, \cite{maheshwari2022digital}, \cite{mavromatis2024umbrella}, \cite{novak2022digitalized}, \cite{oquendo2019dealing}, \cite{pillai2023digital}, \cite{redelinghuys2020six-layer}, \cite{reiche2021digital}, \cite{somma2023digital}, \cite{stary2022privacy}, \cite{vogel-heuser2021approach}, \cite{zhang2021bi-level} \\
Diagnosis & \maindatabar{39} & \cite{acharya2023twins}, \cite{alam2017c2ps}, \cite{altamiranda2019system}, \cite{ashtaritalkhestani2019architecture}, \cite{bao2024digital}, \cite{becue2018cyberfactory}, \cite{bellavista2023requirements}, \cite{clark2021chapter}, \cite{coupaye2023graph-based}, \cite{dahmen2022modeling}, \cite{dickopf2019holistic}, \cite{dobie2024network}, \cite{doubell2023digital}, \cite{duan2023digital}, \cite{esterle2021digital}, \cite{folds2019digital}, \cite{gil2023modeling}, \cite{gill2022method}, \cite{heithoff2023challenges}, \cite{howard2021greenhouse}, \cite{human2023design}, \cite{jiang2022novel}, \cite{kruger2022towards}, \cite{li2024comprehensive}, \cite{lippi2023enabling}, \cite{liu2020web-based}, \cite{lopez2023modeling}, \cite{marah2023architecture}, \cite{monsalve2021novel}, \cite{park2020digital}, \cite{parri2021framework}, \cite{parri2019jarvis}, \cite{reiche2021digital}, \cite{saraeian2022digital}, \cite{stary2022privacy}, \cite{villalonga2021decision-making}, \cite{wullink2024foundational}, \cite{zhang2022multi-scale}, \cite{zhang2021bi-level} \\
Event Detection & \maindatabar{28} & \cite{alam2017c2ps}, \cite{aziz2022empowering}, \cite{bao2024digital}, \cite{becue2018cyberfactory}, \cite{chen2018digital}, \cite{clark2021chapter}, \cite{coupaye2023graph-based}, \cite{dobie2024network}, \cite{doubell2023digital}, \cite{gollner2022collaborative}, \cite{hofmeister2024cross-domain}, \cite{hofmeister2024semantic}, \cite{human2023design}, \cite{joseph2021aggregated}, \cite{li2024comprehensive}, \cite{lippi2023enabling}, \cite{liu2020web-based}, \cite{mahoro2023articulating}, \cite{mavromatis2024umbrella}, \cite{park2020digital}, \cite{parri2021framework}, \cite{parri2019jarvis}, \cite{pickering2023towards}, \cite{pillai2023digital}, \cite{stary2022privacy}, \cite{villalonga2021decision-making}, \cite{wang2024construction}, \cite{zhang2021bi-level} \\
\bottomrule
            \end{tabular}
            \end{table*}

%% file: tables/rq3/hierarchicalModelingMethodsTable.tex
\begin{table*}[]
\centering
\setlength{\tabcolsep}{1em}
\caption{Modeling and simulation formalisms}
\label{tab:modeling-methods-structured-table}
\footnotesize
\begin{tabular}{@{}p{5cm} l p{10cm}@{}}
\toprule
\textbf{Category} & \textbf{Frequency} & \textbf{Studies} \\
\midrule
\textbf{Architectural and Structural} & \textbf{\maindatabar{31}} & \\
\;\;\corner{} Systems Modeling Language (SysML) & \subdatabar{13} & \cite{ashtaritalkhestani2019architecture}, \cite{dahmen2022modeling}, \cite{dickopf2019holistic}, \cite{gollner2022collaborative}, \cite{jiang2022novel}, \cite{kutzke2021subsystem}, \cite{lopez2023modeling}, \cite{parri2019jarvis}, \cite{parri2021framework}, \cite{pickering2023towards}, \cite{schluse2017experimentable}, \cite{wagner2023using}, \cite{zhang2022multi-scale} \\
\;\;\corner{} Unified Modeling Language (UML) & \subdatabar{12} & \cite{dahmen2022modeling}, \cite{duan2023digital}, \cite{gil2024integrating}, \cite{gill2022method}, \cite{gollner2022collaborative}, \cite{heithoff2023challenges}, \cite{hofmeister2024semantic}, \cite{jiang2022novel}, \cite{lee2022simulation}, \cite{parri2019jarvis}, \cite{parri2021framework}, \cite{vogel-heuser2021approach} \\
\;\;\corner{} Business Process Modeling (BPM) & \subdatabar{3} & \cite{binder2021utilizing}, \cite{kulkarni2019towards}, \cite{vogel-heuser2021approach} \\
\;\;\corner{} Building Information Modeling (BIM) & \subdatabar{3} & \cite{coupaye2023graph-based}, \cite{doubell2023digital}, \cite{larsen2024towards} \\
\;\;\corner{} Subject-Oriented Modeling (S-BPM) & \subdatabar{2} & \cite{heininger2021capturing}, \cite{stary2022privacy} \\
\;\;\corner{} State Models & \subdatabar{2} & \cite{kruger2022towards}, \cite{reiche2021digital} \\
\;\;\corner{} \textit{Other} & \subdatabar{8} & \cite{binder2021utilizing}, \cite{dahmen2022modeling}, \cite{dobie2024network}, \cite{gil2024integrating}, \cite{gollner2022collaborative}, \cite{kulkarni2019towards}, \cite{villalonga2021decision-making}, \cite{wagner2023using} \\
\textbf{Spatial and Visual Modeling} & \textbf{\maindatabar{24}} & \\
\;\;\corner{} Computer-Aided Design (CAD) & \subdatabar{12} & \cite{ashtaritalkhestani2019architecture}, \cite{becue2018cyberfactory}, \cite{coupaye2023graph-based}, \cite{duan2023digital}, \cite{ehemann2023digital}, \cite{jiang2022novel}, \cite{joseph2021aggregated}, \cite{liu2020web-based}, \cite{novak2022digitalized}, \cite{park2020digital}, \cite{reiche2021digital}, \cite{zhang2021bi-level} \\
\;\;\corner{} 3D Modeling & \subdatabar{10} & \cite{bao2024digital}, \cite{chavezbaliguat2023digital}, \cite{ehemann2023digital}, \cite{hatledal2020co-simulation}, \cite{malayjerdi2022combined}, \cite{mavromatis2024umbrella}, \cite{priyanta2024is}, \cite{samak2023autodrive}, \cite{somma2023digital}, \cite{vermesan2021internet} \\
\;\;\corner{} Geometric Models & \subdatabar{2} & \cite{duan2023digital}, \cite{ehemann2023digital} \\
\;\;\corner{} Parametric Models & \subdatabar{2} & \cite{li2024comprehensive}, \cite{wagner2023using} \\
\;\;\corner{} \textit{Other} & \subdatabar{6} & \cite{becue2018cyberfactory}, \cite{chavezbaliguat2023digital}, \cite{coupaye2023graph-based}, \cite{demir2023vertically-integrated}, \cite{ehemann2023digital}, \cite{priyanta2024is} \\
\textbf{Mathematical and Statistical} & \textbf{\maindatabar{23}} & \\
\;\;\corner{} Bayesian Networks (BN) & \subdatabar{5} & \cite{alam2017c2ps}, \cite{kutzke2021subsystem}, \cite{lippi2023enabling}, \cite{maheshwari2022digital}, \cite{vogel-heuser2021approach} \\
\;\;\corner{} General Mathematical Models & \subdatabar{5} & \cite{hatledal2020co-simulation}, \cite{howard2021greenhouse}, \cite{jiang2022novel}, \cite{kruger2022towards}, \cite{maheshwari2022digital} \\
\;\;\corner{} Fuzzy Logic & \subdatabar{2} & \cite{alam2017c2ps}, \cite{altamiranda2019system} \\
\;\;\corner{} Model Reference Adaptive Control (MRAC) & \subdatabar{2} & \cite{clark2021chapter}, \cite{kulkarni2019towards} \\
\;\;\corner{} \textit{Other} & \subdatabar{18} & \cite{altamiranda2019system}, \cite{barden2022academic}, \cite{bertoni2022digital}, \cite{chavezbaliguat2023digital}, \cite{dobie2024network}, \cite{esterle2021digital}, \cite{folds2019digital}, \cite{gil2023modeling}, \cite{gill2022method}, \cite{heininger2021capturing}, \cite{howard2021greenhouse}, \cite{jiang2022novel}, \cite{kulkarni2019towards}, \cite{lippi2023enabling}, \cite{maheshwari2022digital}, \cite{pillai2023digital}, \cite{saraeian2022digital}, \cite{vogel-heuser2021approach} \\
\textbf{Ontological and Knowledge Representation} & \textbf{\maindatabar{19}} & \\
\;\;\corner{} Web Ontology Language (OWL) & \subdatabar{7} & \cite{ashtaritalkhestani2019architecture}, \cite{bao2024digital}, \cite{gil2023modeling}, \cite{hofmeister2024semantic}, \cite{jiang2022novel}, \cite{li2024comprehensive}, \cite{liu2020web-based} \\
\;\;\corner{} AutomationML & \subdatabar{5} & \cite{ashtaritalkhestani2019architecture}, \cite{gil2023modeling}, \cite{gollner2022collaborative}, \cite{liu2020web-based}, \cite{novak2022digitalized} \\
\;\;\corner{} Resource Description Framework (RDF) & \subdatabar{3} & \cite{coupaye2023graph-based}, \cite{hofmeister2024semantic}, \cite{li2024comprehensive} \\
\;\;\corner{} Property Graphs (PGs) & \subdatabar{2} & \cite{coupaye2023graph-based}, \cite{mahoro2023articulating} \\
\;\;\corner{} Information Model & \subdatabar{2} & \cite{hatledal2020co-simulation}, \cite{reiche2021digital} \\
\;\;\corner{} \textit{Other} & \subdatabar{10} & \cite{coupaye2023graph-based}, \cite{demir2023vertically-integrated}, \cite{gil2023modeling}, \cite{hofmeister2024cross-domain}, \cite{hofmeister2024semantic}, \cite{li2022cognitive}, \cite{li2024comprehensive}, \cite{monsalve2021novel}, \cite{park2020digital}, \cite{pickering2023towards} \\
\textbf{Formal and State Based Methods} & \textbf{\maindatabar{14}} & \\
\;\;\corner{} Finite State Machines (FSM) & \subdatabar{5} & \cite{alam2017c2ps}, \cite{dahmen2022modeling}, \cite{liu2020web-based}, \cite{savur2019hrc-sos}, \cite{vogel-heuser2021approach} \\
\;\;\corner{} Fault Tree Analysis (FTA) & \subdatabar{3} & \cite{parri2019jarvis}, \cite{parri2021framework}, \cite{saraeian2022digital} \\
\;\;\corner{} \textit{Other} & \subdatabar{7} & \cite{chen2018digital}, \cite{hatledal2020co-simulation}, \cite{heininger2021capturing}, \cite{heithoff2023challenges}, \cite{larsen2024towards}, \cite{oquendo2019dealing}, \cite{parri2019jarvis} \\
\textbf{AI and Machine Learning} & \textbf{\maindatabar{13}} & \\
\;\;\corner{} Machine Learning & \subdatabar{4} & \cite{dobie2024network}, \cite{esterle2021digital}, \cite{folds2019digital}, \cite{jiang2022novel} \\
\;\;\corner{} Reinforcement Learning (RL) & \subdatabar{2} & \cite{clark2021chapter}, \cite{kulkarni2019towards} \\
\;\;\corner{} Genetic Algorithms (GA) & \subdatabar{2} & \cite{kutzke2021subsystem}, \cite{park2020digital} \\
\;\;\corner{} \textit{Other} & \subdatabar{5} & \cite{altamiranda2019system}, \cite{bao2024digital}, \cite{chen2018digital}, \cite{saraeian2022digital}, \cite{villalonga2021decision-making} \\
\textbf{Continuous Simulation} & \textbf{\maindatabar{12}} & \\
\;\;\corner{} System Dynamics Models (SDM) & \subdatabar{4} & \cite{folds2019digital}, \cite{gill2022method}, \cite{kulkarni2019towards}, \cite{pickering2023towards} \\
\;\;\corner{} Kinematic Models & \subdatabar{3} & \cite{duan2023digital}, \cite{gil2023modeling}, \cite{schluse2017experimentable} \\
\;\;\corner{} General Physics Models & \subdatabar{2} & \cite{demir2023vertically-integrated}, \cite{hatakeyama2018systems} \\
\;\;\corner{} Finite Element Method (FEM) & \subdatabar{2} & \cite{demir2023vertically-integrated}, \cite{li2024comprehensive} \\
\;\;\corner{} \textit{Other} & \subdatabar{4} & \cite{altamiranda2019system}, \cite{demir2023vertically-integrated}, \cite{gil2023modeling}, \cite{monsalve2021novel} \\
\textbf{Agent-Based Simulation} & \textbf{\maindatabar{10}} & \\
\;\;\corner{} Multi Agent System (MAS) & \subdatabar{9} & \cite{clark2021chapter}, \cite{heininger2021capturing}, \cite{howard2021greenhouse}, \cite{jirsa2024use}, \cite{liu2020web-based}, \cite{marah2023architecture}, \cite{samak2023autodrive}, \cite{vogel-heuser2021approach}, \cite{zhang2021bi-level} \\
\;\;\corner{} Agent Based Modeling (ABM) & \subdatabar{2} & \cite{barden2022academic}, \cite{clark2021chapter} \\
\;\;\corner{} \textit{Other} & \subdatabar{1} & \cite{marah2023architecture} \\
\textbf{Discrete-Event Simulation} & \textbf{\maindatabar{8}} & \\
\;\;\corner{} Discrete Event Simulation (DES) & \subdatabar{4} & \cite{bertoni2022digital}, \cite{clark2021chapter}, \cite{demir2023vertically-integrated}, \cite{villalonga2021decision-making} \\
\;\;\corner{} Discrete Event System Specification (DEVS) & \subdatabar{2} & \cite{lee2022simulation}, \cite{oquendo2019dealing} \\
\;\;\corner{} \textit{Other} & \subdatabar{3} & \cite{lee2022simulation}, \cite{wang2024construction}, \cite{zhang2022multi-scale} \\
\bottomrule
\end{tabular}
\end{table*}

%% file: tables/rq4/emergenceTable.tex
\begin{table*}[]
            \centering
            \caption{Emergence type}
            \label{tab:emergence-type-table}
            \begin{tabular}{@{}p{4cm}l p{11.5cm}@{}}
            \toprule
            \multicolumn{1}{c}{\textbf{Emergence}} & 
            \multicolumn{1}{c}{\textbf{Frequency}} & 
            \multicolumn{1}{c}{\textbf{Studies}} \\ 
            \midrule
            Not Addressed & \maindatabar{28} & \cite{acharya2023twins}, \cite{ashtaritalkhestani2019architecture}, \cite{aziz2022empowering}, \cite{becue2018cyberfactory}, \cite{binder2021utilizing}, \cite{chavezbaliguat2023digital}, \cite{doubell2023digital}, \cite{duan2023digital}, \cite{gil2023modeling}, \cite{gollner2022collaborative}, \cite{hatakeyama2018systems}, \cite{hatledal2020co-simulation}, \cite{heithoff2023challenges}, \cite{hofmeister2024semantic}, \cite{howard2021greenhouse}, \cite{human2023design}, \cite{kruger2022towards}, \cite{kutzke2021subsystem}, \cite{li2024comprehensive}, \cite{liu2020web-based}, \cite{mahoro2023articulating}, \cite{marah2023architecture}, \cite{monsalve2021novel}, \cite{redelinghuys2020six-layer}, \cite{reiche2021digital}, \cite{villalonga2021decision-making}, \cite{vogel-heuser2021approach}, \cite{zhang2021bi-level} \\
Simple & \maindatabar{16} & \cite{gil2024integrating}, \cite{lee2022simulation}, \cite{li2022cognitive}, \cite{lopez2023modeling}, \cite{maheshwari2022digital}, \cite{novak2022digitalized}, \cite{park2020digital}, \cite{parri2021framework}, \cite{pillai2023digital}, \cite{potteiger2023live}, \cite{priyanta2024is}, \cite{samak2023autodrive}, \cite{somma2023digital}, \cite{vermesan2021internet}, \cite{wagner2023using}, \cite{zhang2022multi-scale} \\
Weak & \maindatabar{30} & \cite{alam2017c2ps}, \cite{altamiranda2019system}, \cite{bao2024digital}, \cite{barden2022academic}, \cite{bellavista2023requirements}, \cite{bertoni2022digital}, \cite{chen2018digital}, \cite{clark2021chapter}, \cite{coupaye2023graph-based}, \cite{demir2023vertically-integrated}, \cite{dickopf2019holistic}, \cite{dobie2024network}, \cite{ehemann2023digital}, \cite{esterle2021digital}, \cite{gill2022method}, \cite{heininger2021capturing}, \cite{hofmeister2024cross-domain}, \cite{larsen2024towards}, \cite{lippi2023enabling}, \cite{malayjerdi2022combined}, \cite{mavromatis2024umbrella}, \cite{oquendo2019dealing}, \cite{parri2019jarvis}, \cite{pickering2023towards}, \cite{saraeian2022digital}, \cite{savur2019hrc-sos}, \cite{schluse2017experimentable}, \cite{stary2022privacy}, \cite{wang2024construction}, \cite{wullink2024foundational} \\
Strong & \maindatabar{6} & \cite{dahmen2022modeling}, \cite{folds2019digital}, \cite{jiang2022novel}, \cite{jirsa2024use}, \cite{joseph2021aggregated}, \cite{kulkarni2019towards} \\
\bottomrule
            \end{tabular}
            \end{table*}

%% file: tables/rq5/reliabilityTable.tex
\begin{table*}[]
            \centering
            \caption{Reliability}
            \label{tab:reliability-table}
            \begin{tabular}{@{}p{4cm}l p{11.5cm}@{}}
            \toprule
            \multicolumn{1}{c}{\textbf{Context}} & 
            \multicolumn{1}{c}{\textbf{Frequency}} & 
            \multicolumn{1}{c}{\textbf{Studies}} \\ 
            \midrule
            Not Addressed & \maindatabar{26} & \cite{bao2024digital}, \cite{bertoni2022digital}, \cite{binder2021utilizing}, \cite{clark2021chapter}, \cite{coupaye2023graph-based}, \cite{dahmen2022modeling}, \cite{demir2023vertically-integrated}, \cite{dickopf2019holistic}, \cite{ehemann2023digital}, \cite{gil2023modeling}, \cite{gil2024integrating}, \cite{gollner2022collaborative}, \cite{hatakeyama2018systems}, \cite{jiang2022novel}, \cite{li2022cognitive}, \cite{li2024comprehensive}, \cite{maheshwari2022digital}, \cite{mahoro2023articulating}, \cite{reiche2021digital}, \cite{samak2023autodrive}, \cite{schluse2017experimentable}, \cite{somma2023digital}, \cite{stary2022privacy}, \cite{wagner2023using}, \cite{wang2024construction}, \cite{zhang2022multi-scale} \\
Mentioned & \maindatabar{8} & \cite{barden2022academic}, \cite{becue2018cyberfactory}, \cite{heininger2021capturing}, \cite{jirsa2024use}, \cite{lee2022simulation}, \cite{marah2023architecture}, \cite{pickering2023towards}, \cite{wullink2024foundational} \\
Architecturally Addressed & \maindatabar{41} & \cite{acharya2023twins}, \cite{alam2017c2ps}, \cite{altamiranda2019system}, \cite{ashtaritalkhestani2019architecture}, \cite{aziz2022empowering}, \cite{bellavista2023requirements}, \cite{chavezbaliguat2023digital}, \cite{chen2018digital}, \cite{dobie2024network}, \cite{doubell2023digital}, \cite{duan2023digital}, \cite{esterle2021digital}, \cite{folds2019digital}, \cite{gill2022method}, \cite{hatledal2020co-simulation}, \cite{heithoff2023challenges}, \cite{hofmeister2024cross-domain}, \cite{hofmeister2024semantic}, \cite{howard2021greenhouse}, \cite{human2023design}, \cite{joseph2021aggregated}, \cite{kruger2022towards}, \cite{kulkarni2019towards}, \cite{larsen2024towards}, \cite{lippi2023enabling}, \cite{liu2020web-based}, \cite{lopez2023modeling}, \cite{mavromatis2024umbrella}, \cite{monsalve2021novel}, \cite{novak2022digitalized}, \cite{parri2021framework}, \cite{parri2019jarvis}, \cite{pillai2023digital}, \cite{potteiger2023live}, \cite{priyanta2024is}, \cite{redelinghuys2020six-layer}, \cite{savur2019hrc-sos}, \cite{vermesan2021internet}, \cite{villalonga2021decision-making}, \cite{vogel-heuser2021approach}, \cite{zhang2021bi-level} \\
Explicitly Modeled & \maindatabar{2} & \cite{kutzke2021subsystem}, \cite{oquendo2019dealing} \\
Evaluated or Validated & \maindatabar{3} & \cite{malayjerdi2022combined}, \cite{park2020digital}, \cite{saraeian2022digital} \\
\bottomrule
            \end{tabular}
            \end{table*}

%% file: tables/rq5/securityTable.tex
\begin{table*}[]
            \centering
            \caption{Security}
            \label{tab:security-table}
            \begin{tabular}{@{}p{4cm}l p{11.5cm}@{}}
            \toprule
            \multicolumn{1}{c}{\textbf{Context}} & 
            \multicolumn{1}{c}{\textbf{Frequency}} & 
            \multicolumn{1}{c}{\textbf{Studies}} \\ 
            \midrule
            Not Addressed & \maindatabar{32} & \cite{bertoni2022digital}, \cite{chavezbaliguat2023digital}, \cite{chen2018digital}, \cite{clark2021chapter}, \cite{dahmen2022modeling}, \cite{dickopf2019holistic}, \cite{ehemann2023digital}, \cite{folds2019digital}, \cite{gil2023modeling}, \cite{gil2024integrating}, \cite{gollner2022collaborative}, \cite{hofmeister2024cross-domain}, \cite{hofmeister2024semantic}, \cite{howard2021greenhouse}, \cite{kulkarni2019towards}, \cite{kutzke2021subsystem}, \cite{lee2022simulation}, \cite{li2022cognitive}, \cite{lippi2023enabling}, \cite{lopez2023modeling}, \cite{maheshwari2022digital}, \cite{novak2022digitalized}, \cite{oquendo2019dealing}, \cite{park2020digital}, \cite{pillai2023digital}, \cite{priyanta2024is}, \cite{samak2023autodrive}, \cite{savur2019hrc-sos}, \cite{schluse2017experimentable}, \cite{wagner2023using}, \cite{wang2024construction}, \cite{zhang2022multi-scale} \\
Mentioned & \maindatabar{24} & \cite{alam2017c2ps}, \cite{altamiranda2019system}, \cite{ashtaritalkhestani2019architecture}, \cite{bao2024digital}, \cite{barden2022academic}, \cite{bellavista2023requirements}, \cite{binder2021utilizing}, \cite{demir2023vertically-integrated}, \cite{doubell2023digital}, \cite{duan2023digital}, \cite{esterle2021digital}, \cite{gill2022method}, \cite{hatakeyama2018systems}, \cite{jirsa2024use}, \cite{joseph2021aggregated}, \cite{li2024comprehensive}, \cite{mahoro2023articulating}, \cite{marah2023architecture}, \cite{monsalve2021novel}, \cite{reiche2021digital}, \cite{saraeian2022digital}, \cite{vogel-heuser2021approach}, \cite{wullink2024foundational}, \cite{zhang2021bi-level} \\
Architecturally Addressed & \maindatabar{19} & \cite{acharya2023twins}, \cite{coupaye2023graph-based}, \cite{dobie2024network}, \cite{hatledal2020co-simulation}, \cite{heininger2021capturing}, \cite{heithoff2023challenges}, \cite{human2023design}, \cite{jiang2022novel}, \cite{kruger2022towards}, \cite{larsen2024towards}, \cite{liu2020web-based}, \cite{parri2021framework}, \cite{parri2019jarvis}, \cite{pickering2023towards}, \cite{potteiger2023live}, \cite{redelinghuys2020six-layer}, \cite{somma2023digital}, \cite{vermesan2021internet}, \cite{villalonga2021decision-making} \\
Explicitly Modeled & \maindatabar{2} & \cite{becue2018cyberfactory}, \cite{stary2022privacy} \\
Evaluated or Validated & \maindatabar{3} & \cite{aziz2022empowering}, \cite{malayjerdi2022combined}, \cite{mavromatis2024umbrella} \\
\bottomrule
            \end{tabular}
            \end{table*}

%% file: tables/rq6/trlTable.tex
\begin{table*}[]
            \centering
            \caption{TRL}
            \label{tab:trl-table}
            \begin{tabular}{@{}p{4cm}l p{11.5cm}@{}}
            \toprule
            \multicolumn{1}{c}{\textbf{TRL}} & 
            \multicolumn{1}{c}{\textbf{Frequency}} & 
            \multicolumn{1}{c}{\textbf{Studies}} \\ 
            \midrule
            Initial & \maindatabar{20} & \cite{becue2018cyberfactory}, \cite{folds2019digital}, \cite{kruger2022towards}, \cite{li2022cognitive}, \cite{lippi2023enabling}, \cite{lopez2023modeling}, \cite{maheshwari2022digital}, \cite{mahoro2023articulating}, \cite{oquendo2019dealing}, \cite{parri2021framework}, \cite{parri2019jarvis}, \cite{pillai2023digital}, \cite{samak2023autodrive}, \cite{saraeian2022digital}, \cite{somma2023digital}, \cite{stary2022privacy}, \cite{vermesan2021internet}, \cite{wagner2023using}, \cite{wang2024construction}, \cite{wullink2024foundational} \\
Proof-of-Concept & \maindatabar{16} & \cite{acharya2023twins}, \cite{alam2017c2ps}, \cite{altamiranda2019system}, \cite{barden2022academic}, \cite{demir2023vertically-integrated}, \cite{dobie2024network}, \cite{esterle2021digital}, \cite{gollner2022collaborative}, \cite{hatakeyama2018systems}, \cite{hatledal2020co-simulation}, \cite{heininger2021capturing}, \cite{hofmeister2024cross-domain}, \cite{joseph2021aggregated}, \cite{redelinghuys2020six-layer}, \cite{reiche2021digital}, \cite{villalonga2021decision-making} \\
Demo Prototype & \maindatabar{35} & \cite{aziz2022empowering}, \cite{bao2024digital}, \cite{bellavista2023requirements}, \cite{bertoni2022digital}, \cite{chavezbaliguat2023digital}, \cite{chen2018digital}, \cite{clark2021chapter}, \cite{dahmen2022modeling}, \cite{dickopf2019holistic}, \cite{doubell2023digital}, \cite{duan2023digital}, \cite{gil2023modeling}, \cite{gil2024integrating}, \cite{gill2022method}, \cite{heithoff2023challenges}, \cite{howard2021greenhouse}, \cite{human2023design}, \cite{jiang2022novel}, \cite{jirsa2024use}, \cite{kulkarni2019towards}, \cite{kutzke2021subsystem}, \cite{larsen2024towards}, \cite{lee2022simulation}, \cite{li2024comprehensive}, \cite{liu2020web-based}, \cite{marah2023architecture}, \cite{monsalve2021novel}, \cite{pickering2023towards}, \cite{potteiger2023live}, \cite{priyanta2024is}, \cite{savur2019hrc-sos}, \cite{schluse2017experimentable}, \cite{vogel-heuser2021approach}, \cite{zhang2022multi-scale}, \cite{zhang2021bi-level} \\
Deployed Prototype & \maindatabar{8} & \cite{ashtaritalkhestani2019architecture}, \cite{binder2021utilizing}, \cite{coupaye2023graph-based}, \cite{ehemann2023digital}, \cite{hofmeister2024semantic}, \cite{malayjerdi2022combined}, \cite{novak2022digitalized}, \cite{park2020digital} \\
Operational & \maindatabar{1} & \cite{mavromatis2024umbrella} \\
\bottomrule
            \end{tabular}
            \end{table*}

%% file: tables/rq6/contributionTypeTable.tex
\begin{table*}[]
            \centering
            \caption{Contribution type}
            \label{tab:contribution-type-table}
            \begin{tabular}{@{}p{4cm}l p{11.5cm}@{}}
            \toprule
            \multicolumn{1}{c}{\textbf{Contribution}} & 
            \multicolumn{1}{c}{\textbf{Frequency}} & 
            \multicolumn{1}{c}{\textbf{Studies}} \\ 
            \midrule
            Technical & \maindatabar{60} & \cite{acharya2023twins}, \cite{alam2017c2ps}, \cite{aziz2022empowering}, \cite{bao2024digital}, \cite{barden2022academic}, \cite{bellavista2023requirements}, \cite{bertoni2022digital}, \cite{chavezbaliguat2023digital}, \cite{chen2018digital}, \cite{clark2021chapter}, \cite{dahmen2022modeling}, \cite{demir2023vertically-integrated}, \cite{dickopf2019holistic}, \cite{doubell2023digital}, \cite{duan2023digital}, \cite{ehemann2023digital}, \cite{gil2023modeling}, \cite{gil2024integrating}, \cite{gollner2022collaborative}, \cite{hatledal2020co-simulation}, \cite{heininger2021capturing}, \cite{heithoff2023challenges}, \cite{hofmeister2024cross-domain}, \cite{hofmeister2024semantic}, \cite{howard2021greenhouse}, \cite{jiang2022novel}, \cite{jirsa2024use}, \cite{kulkarni2019towards}, \cite{kutzke2021subsystem}, \cite{larsen2024towards}, \cite{lee2022simulation}, \cite{li2022cognitive}, \cite{li2024comprehensive}, \cite{lippi2023enabling}, \cite{liu2020web-based}, \cite{lopez2023modeling}, \cite{maheshwari2022digital}, \cite{mahoro2023articulating}, \cite{marah2023architecture}, \cite{monsalve2021novel}, \cite{novak2022digitalized}, \cite{oquendo2019dealing}, \cite{park2020digital}, \cite{parri2021framework}, \cite{parri2019jarvis}, \cite{pickering2023towards}, \cite{pillai2023digital}, \cite{potteiger2023live}, \cite{priyanta2024is}, \cite{reiche2021digital}, \cite{samak2023autodrive}, \cite{saraeian2022digital}, \cite{savur2019hrc-sos}, \cite{schluse2017experimentable}, \cite{somma2023digital}, \cite{stary2022privacy}, \cite{villalonga2021decision-making}, \cite{vogel-heuser2021approach}, \cite{wagner2023using}, \cite{zhang2021bi-level} \\
Conceptual & \maindatabar{13} & \cite{altamiranda2019system}, \cite{becue2018cyberfactory}, \cite{dobie2024network}, \cite{esterle2021digital}, \cite{folds2019digital}, \cite{hatakeyama2018systems}, \cite{human2023design}, \cite{joseph2021aggregated}, \cite{kruger2022towards}, \cite{redelinghuys2020six-layer}, \cite{vermesan2021internet}, \cite{wang2024construction}, \cite{wullink2024foundational} \\
Case Study & \maindatabar{7} & \cite{ashtaritalkhestani2019architecture}, \cite{binder2021utilizing}, \cite{coupaye2023graph-based}, \cite{gill2022method}, \cite{malayjerdi2022combined}, \cite{mavromatis2024umbrella}, \cite{zhang2022multi-scale} \\
\bottomrule
            \end{tabular}
            \end{table*}

%% file: tables/rq6/hierarchicalEvaluationTable.tex
\begin{table*}[]
\centering
\setlength{\tabcolsep}{1em}
\caption{Validation and evaluation approaches}
\label{tab:evaluation-structured-table}
\footnotesize
\begin{tabular}{@{}p{5cm} l p{10cm}@{}}
\toprule
\textbf{Evaluation Category} & \textbf{Frequency} & \textbf{Studies} \\
\midrule
\textbf{Validation} & \textbf{\maindatabar{72}} & \\
\;\;\corner{} Prototyping & \subdatabar{36} & \cite{aziz2022empowering}, \cite{bao2024digital}, \cite{bellavista2023requirements}, \cite{chavezbaliguat2023digital}, \cite{dahmen2022modeling}, \cite{doubell2023digital}, \cite{duan2023digital}, \cite{ehemann2023digital}, \cite{gil2023modeling}, \cite{gollner2022collaborative}, \cite{heininger2021capturing}, \cite{heithoff2023challenges}, \cite{hofmeister2024semantic}, \cite{howard2021greenhouse}, \cite{jiang2022novel}, \cite{jirsa2024use}, \cite{larsen2024towards}, \cite{li2022cognitive}, \cite{li2024comprehensive}, \cite{liu2020web-based}, \cite{lopez2023modeling}, \cite{marah2023architecture}, \cite{monsalve2021novel}, \cite{novak2022digitalized}, \cite{oquendo2019dealing}, \cite{park2020digital}, \cite{parri2019jarvis}, \cite{parri2021framework}, \cite{pickering2023towards}, \cite{reiche2021digital}, \cite{samak2023autodrive}, \cite{saraeian2022digital}, \cite{somma2023digital}, \cite{stary2022privacy}, \cite{villalonga2021decision-making}, \cite{wagner2023using} \\
\;\;\corner{} Empirical Simulation & \subdatabar{16} & \cite{barden2022academic}, \cite{chen2018digital}, \cite{clark2021chapter}, \cite{demir2023vertically-integrated}, \cite{dickopf2019holistic}, \cite{hatledal2020co-simulation}, \cite{hofmeister2024cross-domain}, \cite{kulkarni2019towards}, \cite{lee2022simulation}, \cite{lippi2023enabling}, \cite{maheshwari2022digital}, \cite{pillai2023digital}, \cite{potteiger2023live}, \cite{schluse2017experimentable}, \cite{vogel-heuser2021approach}, \cite{zhang2021bi-level} \\
\;\;\corner{} Architectural/Conceptual Design & \subdatabar{13} & \cite{altamiranda2019system}, \cite{becue2018cyberfactory}, \cite{dobie2024network}, \cite{esterle2021digital}, \cite{folds2019digital}, \cite{hatakeyama2018systems}, \cite{human2023design}, \cite{joseph2021aggregated}, \cite{kruger2022towards}, \cite{redelinghuys2020six-layer}, \cite{vermesan2021internet}, \cite{wang2024construction}, \cite{wullink2024foundational} \\
\;\;\corner{} Laboratory Experiments & \subdatabar{4} & \cite{acharya2023twins}, \cite{gil2024integrating}, \cite{priyanta2024is}, \cite{savur2019hrc-sos} \\
\;\;\corner{} Mathematical Analysis & \subdatabar{3} & \cite{alam2017c2ps}, \cite{kutzke2021subsystem}, \cite{mahoro2023articulating} \\
\textbf{Evaluation} & \textbf{\maindatabar{8}} & \\
\;\;\corner{} Industrial Case Study & \subdatabar{7} & \cite{ashtaritalkhestani2019architecture}, \cite{binder2021utilizing}, \cite{coupaye2023graph-based}, \cite{gill2022method}, \cite{malayjerdi2022combined}, \cite{mavromatis2024umbrella}, \cite{zhang2022multi-scale} \\
\;\;\corner{} Action Research & \subdatabar{1} & \cite{bertoni2022digital} \\
\bottomrule
\end{tabular}
\end{table*}

%% file: tables/rq6/standards.tex
\begin{table*}[]
            \centering
            \caption{Standards}
            \label{tab:standards-table}
            \begin{tabular}{@{}p{4cm}l p{11.5cm}@{}}
            \toprule
            \multicolumn{1}{c}{\textbf{Standard}} & 
            \multicolumn{1}{c}{\textbf{Frequency}} & 
            \multicolumn{1}{c}{\textbf{Studies}} \\ 
            \midrule
            Open Platform Communications Unified Architecture (OPC UA) & \maindatabar{13} & \cite{acharya2023twins}, \cite{ashtaritalkhestani2019architecture}, \cite{binder2021utilizing}, \cite{dobie2024network}, \cite{gollner2022collaborative}, \cite{howard2021greenhouse}, \cite{jirsa2024use}, \cite{joseph2021aggregated}, \cite{liu2020web-based}, \cite{novak2022digitalized}, \cite{redelinghuys2020six-layer}, \cite{reiche2021digital}, \cite{villalonga2021decision-making} \\
IEC 63278: Asset Administration Shell & \maindatabar{8} & \cite{acharya2023twins}, \cite{ashtaritalkhestani2019architecture}, \cite{gil2023modeling}, \cite{gill2022method}, \cite{gollner2022collaborative}, \cite{jirsa2024use}, \cite{reiche2021digital}, \cite{vogel-heuser2021approach} \\
Reference Architectural Model Industrie 4.0 (RAMI 4.0) & \maindatabar{4} & \cite{binder2021utilizing}, \cite{gill2022method}, \cite{human2023design}, \cite{park2020digital} \\
Other & \maindatabar{26} & \cite{alam2017c2ps}, \cite{altamiranda2019system}, \cite{ashtaritalkhestani2019architecture}, \cite{barden2022academic}, \cite{bellavista2023requirements}, \cite{binder2021utilizing}, \cite{dickopf2019holistic}, \cite{dobie2024network}, \cite{gollner2022collaborative}, \cite{hatledal2020co-simulation}, \cite{heininger2021capturing}, \cite{heithoff2023challenges}, \cite{howard2021greenhouse}, \cite{human2023design}, \cite{jiang2022novel}, \cite{li2022cognitive}, \cite{malayjerdi2022combined}, \cite{monsalve2021novel}, \cite{novak2022digitalized}, \cite{parri2021framework}, \cite{parri2019jarvis}, \cite{pickering2023towards}, \cite{savur2019hrc-sos}, \cite{stary2022privacy}, \cite{vermesan2021internet}, \cite{vogel-heuser2021approach} \\
\bottomrule
            \end{tabular}
            \end{table*}

%% file: tables/rq6/dtOrSoSRelated.tex
\begin{table*}[]
            \centering
            \caption{Standards usage context (DT vs. SoS)}
            \label{tab:dt-or-sos-related-table}
            \begin{tabular}{@{}p{4cm}l p{11.5cm}@{}}
            \toprule
            \multicolumn{1}{c}{\textbf{Context}} & 
            \multicolumn{1}{c}{\textbf{Frequency}} & 
            \multicolumn{1}{c}{\textbf{Studies}} \\ 
            \midrule
            DT & \maindatabar{18} & \cite{acharya2023twins}, \cite{alam2017c2ps}, \cite{bellavista2023requirements}, \cite{binder2021utilizing}, \cite{dickopf2019holistic}, \cite{gil2023modeling}, \cite{gill2022method}, \cite{gollner2022collaborative}, \cite{hatledal2020co-simulation}, \cite{heininger2021capturing}, \cite{heithoff2023challenges}, \cite{jirsa2024use}, \cite{joseph2021aggregated}, \cite{liu2020web-based}, \cite{malayjerdi2022combined}, \cite{reiche2021digital}, \cite{stary2022privacy}, \cite{villalonga2021decision-making} \\
SoS & \maindatabar{10} & \cite{altamiranda2019system}, \cite{barden2022academic}, \cite{howard2021greenhouse}, \cite{li2022cognitive}, \cite{monsalve2021novel}, \cite{park2020digital}, \cite{pickering2023towards}, \cite{redelinghuys2020six-layer}, \cite{savur2019hrc-sos}, \cite{vermesan2021internet} \\
Both & \maindatabar{6} & \cite{ashtaritalkhestani2019architecture}, \cite{dobie2024network}, \cite{human2023design}, \cite{jiang2022novel}, \cite{novak2022digitalized}, \cite{vogel-heuser2021approach} \\
\bottomrule
            \end{tabular}
            \end{table*}

%% file: tables/rq7/hierarchicalProgrammingLanguagesTable.tex
\begin{table*}[]
\centering
\setlength{\tabcolsep}{1em}
\caption{Programming languages and data formats}
\label{tab:programming-languages-structured-table}
\footnotesize
\begin{tabular}{@{}p{5cm} l p{10cm}@{}}
\toprule
\textbf{Category} & \textbf{Frequency} & \textbf{Studies} \\
\midrule
\textbf{General Purpose} & \textbf{\maindatabar{36}} & \\
\;\;\corner{} Python & \subdatabar{22} & \cite{bao2024digital}, \cite{barden2022academic}, \cite{bellavista2023requirements}, \cite{chavezbaliguat2023digital}, \cite{doubell2023digital}, \cite{duan2023digital}, \cite{gil2023modeling}, \cite{jirsa2024use}, \cite{lippi2023enabling}, \cite{liu2020web-based}, \cite{maheshwari2022digital}, \cite{malayjerdi2022combined}, \cite{marah2023architecture}, \cite{mavromatis2024umbrella}, \cite{monsalve2021novel}, \cite{park2020digital}, \cite{potteiger2023live}, \cite{samak2023autodrive}, \cite{saraeian2022digital}, \cite{savur2019hrc-sos}, \cite{vogel-heuser2021approach}, \cite{wagner2023using} \\
\;\;\corner{} Java & \subdatabar{14} & \cite{alam2017c2ps}, \cite{ashtaritalkhestani2019architecture}, \cite{aziz2022empowering}, \cite{bellavista2023requirements}, \cite{clark2021chapter}, \cite{gil2023modeling}, \cite{gil2024integrating}, \cite{hatledal2020co-simulation}, \cite{li2024comprehensive}, \cite{marah2023architecture}, \cite{parri2019jarvis}, \cite{parri2021framework}, \cite{vogel-heuser2021approach}, \cite{wagner2023using} \\
\;\;\corner{} JavaScript & \subdatabar{8} & \cite{bao2024digital}, \cite{barden2022academic}, \cite{doubell2023digital}, \cite{duan2023digital}, \cite{hofmeister2024semantic}, \cite{liu2020web-based}, \cite{priyanta2024is}, \cite{samak2023autodrive} \\
\;\;\corner{} C++ & \subdatabar{4} & \cite{hatledal2020co-simulation}, \cite{mavromatis2024umbrella}, \cite{park2020digital}, \cite{samak2023autodrive} \\
\;\;\corner{} C\# & \subdatabar{3} & \cite{lee2022simulation}, \cite{park2020digital}, \cite{redelinghuys2020six-layer} \\
\;\;\corner{} C & \subdatabar{1} & \cite{hatledal2020co-simulation} \\
\;\;\corner{} Xtend & \subdatabar{1} & \cite{oquendo2019dealing} \\
\;\;\corner{} Jython & \subdatabar{1} & \cite{wagner2023using} \\
\textbf{Data Representation} & \textbf{\maindatabar{12}} & \\
\;\;\corner{} XML & \subdatabar{9} & \cite{ashtaritalkhestani2019architecture}, \cite{binder2021utilizing}, \cite{dahmen2022modeling}, \cite{jiang2022novel}, \cite{jirsa2024use}, \cite{kutzke2021subsystem}, \cite{monsalve2021novel}, \cite{oquendo2019dealing}, \cite{redelinghuys2020six-layer} \\
\;\;\corner{} JSON & \subdatabar{5} & \cite{acharya2023twins}, \cite{aziz2022empowering}, \cite{dahmen2022modeling}, \cite{jirsa2024use}, \cite{vogel-heuser2021approach} \\
\textbf{Markup and Styling} & \textbf{\maindatabar{4}} & \\
\;\;\corner{} HTML & \subdatabar{4} & \cite{bao2024digital}, \cite{doubell2023digital}, \cite{hofmeister2024semantic}, \cite{samak2023autodrive} \\
\;\;\corner{} CSS & \subdatabar{4} & \cite{bao2024digital}, \cite{doubell2023digital}, \cite{hofmeister2024semantic}, \cite{samak2023autodrive} \\
\bottomrule
\end{tabular}
\end{table*}

%% file: tables/rq7/hierarchicalFrameworksTable.tex
\begin{table*}[]
\centering
\setlength{\tabcolsep}{1em}
\caption{Tools and frameworks}
\label{tab:frameworks-structured-table}
\footnotesize
\begin{tabular}{@{}p{5cm} l p{10cm}@{}}
\toprule
\textbf{Category} & \textbf{Frequency} & \textbf{Studies} \\
\midrule
\textbf{Modeling \& Simulation} & \textbf{\maindatabar{35}} & \\
\;\;\corner{} MATLAB & \subdatabar{10} & \cite{ashtaritalkhestani2019architecture}, \cite{bertoni2022digital}, \cite{chen2018digital}, \cite{kutzke2021subsystem}, \cite{larsen2024towards}, \cite{lopez2023modeling}, \cite{novak2022digitalized}, \cite{reiche2021digital}, \cite{schluse2017experimentable}, \cite{zhang2022multi-scale} \\
\;\;\corner{} Simulink & \subdatabar{4} & \cite{ashtaritalkhestani2019architecture}, \cite{lopez2023modeling}, \cite{novak2022digitalized}, \cite{zhang2022multi-scale} \\
\;\;\corner{} Modelica & \subdatabar{4} & \cite{ashtaritalkhestani2019architecture}, \cite{howard2021greenhouse}, \cite{larsen2024towards}, \cite{zhang2022multi-scale} \\
\;\;\corner{} Gazebo & \subdatabar{4} & \cite{esterle2021digital}, \cite{mavromatis2024umbrella}, \cite{savur2019hrc-sos}, \cite{schluse2017experimentable} \\
\;\;\corner{} Tecnomatix & \subdatabar{3} & \cite{gill2022method}, \cite{redelinghuys2020six-layer}, \cite{schluse2017experimentable} \\
\;\;\corner{} AnyLogic & \subdatabar{3} & \cite{howard2021greenhouse}, \cite{joseph2021aggregated}, \cite{marah2023architecture} \\
\;\;\corner{} CARLA Simulator & \subdatabar{2} & \cite{malayjerdi2022combined}, \cite{potteiger2023live} \\
\;\;\corner{} Java Agent Development Framework (JADE) & \subdatabar{2} & \cite{marah2023architecture}, \cite{vogel-heuser2021approach} \\
\;\;\corner{} Virtual Robotics Experimentation Platform (V-REP) & \subdatabar{2} & \cite{savur2019hrc-sos}, \cite{schluse2017experimentable} \\
\;\;\corner{} \textit{Other} & \subdatabar{22} & \cite{acharya2023twins}, \cite{alam2017c2ps}, \cite{dahmen2022modeling}, \cite{gil2023modeling}, \cite{gollner2022collaborative}, \cite{hatledal2020co-simulation}, \cite{heithoff2023challenges}, \cite{howard2021greenhouse}, \cite{larsen2024towards}, \cite{li2022cognitive}, \cite{lopez2023modeling}, \cite{marah2023architecture}, \cite{monsalve2021novel}, \cite{novak2022digitalized}, \cite{oquendo2019dealing}, \cite{park2020digital}, \cite{parri2019jarvis}, \cite{potteiger2023live}, \cite{priyanta2024is}, \cite{saraeian2022digital}, \cite{savur2019hrc-sos}, \cite{vogel-heuser2021approach} \\
\textbf{Data Management} & \textbf{\maindatabar{19}} & \\
\;\;\corner{} MongoDB & \subdatabar{6} & \cite{aziz2022empowering}, \cite{dobie2024network}, \cite{larsen2024towards}, \cite{somma2023digital}, \cite{villalonga2021decision-making}, \cite{zhang2021bi-level} \\
\;\;\corner{} PostgreSQL & \subdatabar{3} & \cite{doubell2023digital}, \cite{human2023design}, \cite{mavromatis2024umbrella} \\
\;\;\corner{} InfluxDB & \subdatabar{3} & \cite{larsen2024towards}, \cite{li2024comprehensive}, \cite{mavromatis2024umbrella} \\
\;\;\corner{} Redis & \subdatabar{3} & \cite{li2024comprehensive}, \cite{liu2020web-based}, \cite{zhang2021bi-level} \\
\;\;\corner{} Prometheus & \subdatabar{2} & \cite{bellavista2023requirements}, \cite{mavromatis2024umbrella} \\
\;\;\corner{} Protégé & \subdatabar{2} & \cite{gil2024integrating}, \cite{liu2020web-based} \\
\;\;\corner{} MySQL & \subdatabar{2} & \cite{li2024comprehensive}, \cite{liu2020web-based} \\
\;\;\corner{} \textit{Other} & \subdatabar{9} & \cite{chavezbaliguat2023digital}, \cite{clark2021chapter}, \cite{dahmen2022modeling}, \cite{dobie2024network}, \cite{hofmeister2024semantic}, \cite{jirsa2024use}, \cite{li2024comprehensive}, \cite{pickering2023towards}, \cite{zhang2021bi-level} \\
\textbf{Geospatial \& Visualization} & \textbf{\maindatabar{19}} & \\
\;\;\corner{} Unity & \subdatabar{5} & \cite{chen2018digital}, \cite{esterle2021digital}, \cite{gil2023modeling}, \cite{samak2023autodrive}, \cite{schluse2017experimentable} \\
\;\;\corner{} WebGL & \subdatabar{2} & \cite{duan2023digital}, \cite{li2024comprehensive} \\
\;\;\corner{} Microsoft Kinect & \subdatabar{2} & \cite{joseph2021aggregated}, \cite{savur2019hrc-sos} \\
\;\;\corner{} \textit{Other} & \subdatabar{14} & \cite{barden2022academic}, \cite{bertoni2022digital}, \cite{chavezbaliguat2023digital}, \cite{coupaye2023graph-based}, \cite{duan2023digital}, \cite{hofmeister2024semantic}, \cite{human2023design}, \cite{joseph2021aggregated}, \cite{li2024comprehensive}, \cite{malayjerdi2022combined}, \cite{mavromatis2024umbrella}, \cite{pickering2023towards}, \cite{savur2019hrc-sos}, \cite{somma2023digital} \\
\textbf{Digital Twin \& IoT} & \textbf{\maindatabar{15}} & \\
\;\;\corner{} Eclipse Ditto & \subdatabar{4} & \cite{acharya2023twins}, \cite{aziz2022empowering}, \cite{larsen2024towards}, \cite{marah2023architecture} \\
\;\;\corner{} Robot Operating System (ROS) & \subdatabar{4} & \cite{mavromatis2024umbrella}, \cite{pickering2023towards}, \cite{samak2023autodrive}, \cite{savur2019hrc-sos} \\
\;\;\corner{} Eclipse Arrowhead & \subdatabar{2} & \cite{acharya2023twins}, \cite{aziz2022empowering} \\
\;\;\corner{} FIWARE & \subdatabar{2} & \cite{coupaye2023graph-based}, \cite{somma2023digital} \\
\;\;\corner{} Thing’in & \subdatabar{2} & \cite{coupaye2023graph-based}, \cite{mahoro2023articulating} \\
\;\;\corner{} \textit{Other} & \subdatabar{6} & \cite{acharya2023twins}, \cite{dickopf2019holistic}, \cite{gil2023modeling}, \cite{jirsa2024use}, \cite{joseph2021aggregated}, \cite{marah2023architecture} \\
\textbf{Systems Eng. \& Architecture} & \textbf{\maindatabar{11}} & \\
\;\;\corner{} Enterprise Architect & \subdatabar{2} & \cite{binder2021utilizing}, \cite{kutzke2021subsystem} \\
\;\;\corner{} Cameo Systems Modeler & \subdatabar{2} & \cite{dickopf2019holistic}, \cite{wagner2023using} \\
\;\;\corner{} Metasonic Suite & \subdatabar{2} & \cite{heininger2021capturing}, \cite{stary2022privacy} \\
\;\;\corner{} \textit{Other} & \subdatabar{7} & \cite{dobie2024network}, \cite{larsen2024towards}, \cite{lopez2023modeling}, \cite{mavromatis2024umbrella}, \cite{pickering2023towards}, \cite{stary2022privacy}, \cite{wagner2023using} \\
\textbf{App/Web Technologies} & \textbf{\maindatabar{10}} & \\
\;\;\corner{} .NET Framework & \subdatabar{2} & \cite{lee2022simulation}, \cite{park2020digital} \\
\;\;\corner{} \textit{Other} & \subdatabar{10} & \cite{aziz2022empowering}, \cite{chavezbaliguat2023digital}, \cite{doubell2023digital}, \cite{duan2023digital}, \cite{esterle2021digital}, \cite{larsen2024towards}, \cite{lee2022simulation}, \cite{li2022cognitive}, \cite{liu2020web-based}, \cite{park2020digital} \\
\textbf{Cloud, Edge, and DevOps} & \textbf{\maindatabar{8}} & \\
\;\;\corner{} Docker & \subdatabar{5} & \cite{bellavista2023requirements}, \cite{hofmeister2024semantic}, \cite{mavromatis2024umbrella}, \cite{monsalve2021novel}, \cite{pickering2023towards} \\
\;\;\corner{} Kubernetes & \subdatabar{2} & \cite{bellavista2023requirements}, \cite{mavromatis2024umbrella} \\
\;\;\corner{} Azure & \subdatabar{2} & \cite{larsen2024towards}, \cite{pickering2023towards} \\
\;\;\corner{} \textit{Other} & \subdatabar{4} & \cite{bellavista2023requirements}, \cite{demir2023vertically-integrated}, \cite{mavromatis2024umbrella}, \cite{redelinghuys2020six-layer} \\
\textbf{AI, Data Analytics \& ML} & \textbf{\maindatabar{7}} & \\
\;\;\corner{} Grafana & \subdatabar{3} & \cite{bellavista2023requirements}, \cite{esterle2021digital}, \cite{mavromatis2024umbrella} \\
\;\;\corner{} Jupyter Lab & \subdatabar{2} & \cite{chavezbaliguat2023digital}, \cite{larsen2024towards} \\
\;\;\corner{} \textit{Other} & \subdatabar{3} & \cite{joseph2021aggregated}, \cite{malayjerdi2022combined}, \cite{mavromatis2024umbrella} \\
\bottomrule
\end{tabular}
\end{table*}

%% file: sections/discussion.tex
\section{Discussion}\label{sec:discussion}

We now discuss the key takeaways of our study and recommend research directions to prospective researchers.

\subsection{Orthogonal analysis}

We analyzed the extracted data for orthogonal findings, i.e., interesting patterns that emerge from combinations of individual factors reported in \secref{sec:results}. We generated contingency tables for each pair of categories from \secref{sec:results} and looked for statistically significant patterns under a Chi-square test at $\alpha=0.95$. Of the roughly 300 contingency tables, 15 demonstrated statistical significance. In this subsection, we discuss the most interesting ones. The complete collection of contingency tables, as well as the generation script are available in the replication package.

\subsubsection{Constituent units vs DT class}\label{sec:orthogonal-constituents-vs-dt}

We observed a statistically significant relationship when cross-tabulating the types of constituent units and class of DTs in our sample ($p=4.764e{-20}$).
First, we found higher than expected number of DTs of physical systems. This tendency hints at traditional DT engineering methods and systems being prevalent in current SoTS or, alternatively, SoTS as a field primarily originating from traditional DT systems.
Second, we found higher than expected number of human-supervised and human-actuated DTs of cyber-physical-human systems, i.e., CPS with direct impact on the human, such as human-robot collaborative experimentation systems~\cite{savur2019hrc-sos} and horticulture systems~\cite{pickering2023towards}. In fact, human supervision and human actuation appears only in CPS with human involvement. This tendency hints at the recognized differences in the human's role in operating the supposed central DT of a SoTS, especially when the underlying infrastructure impacts humans.

\subsubsection{DT class vs evolution}\label{sec:orthogonal-dt-evolution}
We observed a statistically significant relationship between the class of DTs in SoTS and the supported evolution of the underlying SoS ($p=1.032e{-02}$). Specifically, we found higher than expected number of traditional DTs that ignore evolution altogether or even implement it partially. Evolution is a key aspect of SoS and the lacking support limits the long-term applicability of SoTS. Evolution of DTs has gained recognition in mainstream DT research, mostly focusing on developing taxonomies~\cite{david2023towards} and languages~\cite{mertens2024continuous}, with applications, for example, in smart ecosystems~\cite{michael2024digital}. Such research directions are in high demand in order to render the long-term usage and utility of SoTS viable.

\subsubsection{On the relationships of SoS properties}

We found numerous statistically significant relationships between pairs of SoS properties. These relationships typically reveal SoS traits that tend to be implemented or be missing together. It is plausible to assume that SoS properties are not refined enough in SoTS and are lumped together when designing SoTS. This is corroborated by the observation that the traditional DT engineering paradigm seems to be the prevalent in the design and development of SoTS (\secref{sec:orthogonal-constituents-vs-dt}).

Some of the characteristic examples include evolution and interoperability.
Considerations for evolution have a strong relationship with the support for dynamic reconfiguration ($p=2.644e{-07}$), emergent behavior ($p=3.046e{-05}$), and interdependence ($p=4.286e{-02}$). These strong relationships imply potential issues in the evolution potential of SoTS, leading to effects already discussed in \secref{sec:orthogonal-dt-evolution}.
SoS interoperability shows a strong relationship with the distribution ($p=3.223e{-07}$), interdependence ($p=1.873e{-05}$), and evolution ($p=1.724e{-02}$) properties of SoS in that these properties appear together with interoperability.

In contrast, it is the lack of considerations for emergent behavior that has a statistically significant relationship with the support for interoperability ($p=1.724e{-02}$). We conjecture that establishing interoperability is a sizable challenge even without unpredictable emergent behavior and the presence of emergence may limit the degree of interoperability in SoTS. This is corroborated, e.g., by observation from the DT literature that DT interoperability is currently achieved at lower technical levels~\cite{david2024interoperability}.

\subsubsection{Intent vs emergence}

We observed a statistically significant relationship when cross-tabulating the intents of SoTS and the type of emergence they support ($p=1.368e{-04}$). We find higher than expected numbers of support for simple emergence when the intent is digitally twinning an SoS; and higher than expected numbers of no support for emergence when the intent is combining DTs into an SoS. 

Strong emergence is challenging to plan, engineer, and even recognize in an SoS. Not surprisingly, strong emergence is seldom encountered in current SoTS. However, we observe slightly higher than expected numbers when the intent of the SoTS is combining DTs into SoS. We conjecture that the strong engineering foundations of traditional DT engineering allows for addressing challenges of strong emergence.

\subsubsection{TRL vs intent and contribution type}

We observed a statistically significant relationship between the TRL and the intent of the SoTS ($p=4.878e{-02}$). We see more than the expected number of proof-of-concept (low TRL) contributions when the intent is combining DTs into an SoS. At the same time, we see more than the expected number of initial (lowest TRL) level contributions when the intent is twinning an existing SoS. This demonstrates slightly higher technical maturity in cases when SoS are twinned.

Considering the relationship of TRL with contribution type ($p=6.297e{-07}$), the highest TRL that deviates from expected numbers is the deployed prototype level, where we find more case study type publications than expected. This makes sense as case studies are inquiries into contemporary phenomena and require a realistic study environment. The number of technical contributions exceeds the expectations at the demonstrative prototype level, indicating a recognized need for tangible demonstrators when researching SoTS.

\subsection{Key takeaways}

We now summarize the key takeaways of our study.

\subsubsection{On the need for SoTS architectures}

One of the key challenges in DT engineering is the relative lack of established architectures~\cite{ferko2022architecting}. Our empirical inquiry suggests that this issue inherited in SoTS, as evidenced by \tabref{tab:challenges-table} identifying the lack of architectures and lack of standards as recurring design challenges. As shown in \tabref{tab:intents-table}, the intent of SoTS is typically the organization of DTs into SoS, which hints at the need for specialized architectures that are flexible enough to accommodate SoS dynamics. This hypothesis is corroborated by \tabref{tab:challenges-table} identifying key SoS-related operational challenges of SoTS, such as interoperability---in two instances, in fact: operative interoperability and data interoperability, the two discussed in nearly 40\% the sampled studies.

The prevalence of acknowledged and directed SoS types in \tabref{tab:sots-type-table} (found in over 70\% of SoTS) highlight that current SoTS indeed struggle to support dynamical architectures. Collaborative and virtual SoS, i.e., more dynamical flavors of SoS are encountered in less than 30\% of the cases. Indeed, this might be the artifact of the lack of architectural specifications and standards.

The good news for prospective researchers is that among the most typical modeling formalisms, we often find structural and architectural ones. As shown in \tabref{tab:modeling-methods-structured-table}, SysML and UML Class Diagrams are frequently encountered, which may hint at attempts at structural definitions of SoTS.

Developing SoTS architectures, therefore, should be a priority for prospective researchers. Such architectural specifications will indirectly contribute to the maturity of research and the maturity of systems as well---two areas current SoTS struggle with (see \tabref{tab:evaluation-structured-table} and \tabref{tab:trl-table}). We suggest research into microservice architectures~\cite{bellavista2023requirements}, possibly bundled with the FMI/FMU standard for co-simulation\cite{bottjer2023review}, as well as interoperability of DTs which has shown to be an important enabler of SoTS~\cite{david2024interoperability}. For these efforts, our classification framework in \secref{sec:classification} should provide valuable input.

We recommend researchers and practitioners to develop architectural specifications and reference implementations for SoTS to ease their engineering and to allow higher levels of maturity in their research and development.

\subsubsection{Managing emergent behavior in SoS by DTs}

The essential trait of SoS is the emergent behavior they exhibit. Yet, as witnessed by \tabref{tab:sots-type-table}, state-of-the-art SoTS techniques are mostly limited to acknowledged and directed flavors of SoS. Our hypothesis is that augmented with DTs, SoTS can achieve more. The uniquely tight coupling of cyber and physical components in DTs allow for leveraging them to understand and manage emergent behavior. The idea of active experimentation with the physical system to infer simulation models dates back in the '70s~\cite{zeigler2018theory}, and it is living its renaissance thanks to DTs~\cite{mittal2023towards,barat2022digital}. Active experimentation is the purposeful modification of the twinned system in a way that it exhibits interesting configurations from which valuable information can be extracted. Such ideas have been explored, e.g., in the control of uncrewed aerial systems~\cite{guo2023intelligent}, computer vision for autonomous vehicles~\cite{pun2023neural}, and AI simulation~\cite{liu2024ai}. Purposeful experimentation will help SoTS engineers to characterize emergent behavior better and learn about the environment of the SoTS.

Even after purposeful experimentation, some uncertainty about the behavior of the SoTS remains. To manage these unknown unknowns~\cite{ramasesh2014conceptual}, we recommend researching computing techniques that have the potential to react to unknown unknowns better, e.g., faster-than-real-time simulations to react to emergence faster or to anticipate it on a short time horizon; and using sound modeling techniques, such as goal modeling (e.g., via I*~\cite{goncalves2018systematic} and KAOS~\cite{goncalves2025systematic}) to codify the expected behavior of SoTS.

We recommend researchers and adopters to leverage DT capabilities to understand and manage emergent behavior in SoS, e.g., by purposeful experimentation with physical systems, or by improving run-time modeling\&simulation capabilities.

\subsubsection{The lack of standards is a key challenge in SoTS}

\input{tables/rq1/challenges_table}

Standardization is an overlooked aspect of engineering SoTS. We found that less than half of the sampled studies rely on any sort of standard (\xofyp{36}{80}{}, see \tabref{tab:standards-table}), and these standards are not primarily DT or SoS related. In most cases, we find (business) data management and exchange standards, e.g., OPC UA, the Asset Administration Shell (IEC 63278), and RAMI 4.0. These standards are among the recognized ones to support the engineering DTs in the lack of more suitable standards~\cite{shao2024manufacturing}. Among the challenges of designing SoTS (\tabref{tab:challenges-table}), standards are explicitly mentioned in a number of studies. The previous point on architecting SoTS also raises the need for technical standards~\cite{ferko2022architecting}.
Another, strong evidence of the need for standards are the application domains in which SoTS are used. As shown in \tabref{tab:domains-table}, some of the typical application domains include automotive systems and smart cities---both of which enforce rigorous standards and will likely do so for SoTS. The lack of standards hinders the adoption of SoTS in these domains, and likely in others too.

Unfortunately, the limitations of the only ISO-grade DT standard (ISO 23247) to support dynamical systems are well-known~\cite{liu2024ai}; and standardization of SoS is an afterthought. According to \textcite{shao2024manufacturing}, two new extensions to the ISO 23247 standard are expected to appear in the coming years: digital thread for DTs (Part 5) and DT composition (Part 6). These extensions are well-positioned to address the key challenges of SoTS, including interoperability and synchronization among DTs. 

\subsubsection{Empirical evaluations are lagging behind}

We observe a relatively high ratio of technical contributions compared to conceptual works in our sample (see \tabref{tab:contribution-type-table}). This is, of course, partly the result of our study design which excluded works with shallow and superficial contributions. Thus, the ratio of technical and conceptual contributions may not be representative to the overall field of SoTS. \tabref{tab:trl-table} reports that more than half of the sampled studies are beyond a demo prototype TRL. \figref{fig:trl-v-cont} shows a more detailed view of the TRL of the various contribution types. As expected, conceptual contributions are situated at lower levels of TRL (initial and proof-of-concept, i.e., TRLs 1--4), while the distribution of technical contributions peaks at a demonstrated prototype level (i.e., TRL 5--6), with occasional instances at the deployed prototype level (i.e., TRL 7--8). The few case studies we found are predominantly situated at the deployed prototype level, with one instance at the operational level of maturity (i.e., TRL 9).

The apparent existence of mature SoTS provides excellent opportunities for empirical inquiries. We encourage such investigations and suggest prospective researchers to consider reporting in case report and exemplar formats~\cite{mezard2007where}, e.g., in the industry and practice tracks of conferences, which are as reputable as foundations tracks. In terms of methods, we recommend case studies~\cite{wohlin2021case}, engineering research (also known as design science)~\cite{dresch2015design}, action research~\cite{cohen2017research}, and ethnography~\cite{sharp2010using} for human-focused studies (e.g., when researching the role of the human in a SoTS).

Such empirical inquiries will indirectly contribute to improved research maturity, e.g., by naturally improving the ratio of evaluative assessments over validation types. The latter is currently the prevalent assessment method, by far (90\% vs 10\%), as evidenced by \tabref{tab:evaluation-structured-table}, but ranked lower on the methodological list of \textcite{petersen2015guidelines}.

%% file: tables/rq1/challenges_table.tex
\begin{table*}[]
\centering
\setlength{\tabcolsep}{1em}
\caption{Challenges}
\label{tab:challenges-table}
\footnotesize
\begin{tabular}{@{}p{4cm} l p{11cm}@{}}
\toprule
\textbf{Challenge} & \textbf{Frequency} & \textbf{Studies} \\
\midrule
\textbf{Operational Challenges} & \textbf{\maindatabar{60}} & \\
\;\;\corner{} Interoperability & \subdatabar{26} & \cite{acharya2023twins}, \cite{alam2017c2ps}, \cite{chen2018digital}, \cite{dahmen2022modeling}, \cite{dobie2024network}, \cite{esterle2021digital}, \cite{gollner2022collaborative}, \cite{heithoff2023challenges}, \cite{hofmeister2024cross-domain}, \cite{jiang2022novel}, \cite{jirsa2024use}, \cite{kulkarni2019towards}, \cite{larsen2024towards}, \cite{li2022cognitive}, \cite{lippi2023enabling}, \cite{marah2023architecture}, \cite{park2020digital}, \cite{parri2019jarvis}, \cite{pickering2023towards}, \cite{pillai2023digital}, \cite{samak2023autodrive}, \cite{schluse2017experimentable}, \cite{somma2023digital}, \cite{vermesan2021internet}, \cite{villalonga2021decision-making}, \cite{vogel-heuser2021approach} \\
\;\;\corner{} Synchronization & \subdatabar{11} & \cite{acharya2023twins}, \cite{altamiranda2019system}, \cite{ashtaritalkhestani2019architecture}, \cite{bertoni2022digital}, \cite{coupaye2023graph-based}, \cite{duan2023digital}, \cite{esterle2021digital}, \cite{li2022cognitive}, \cite{monsalve2021novel}, \cite{novak2022digitalized}, \cite{pillai2023digital} \\
\;\;\corner{} Real-Time Constraints & \subdatabar{9} & \cite{becue2018cyberfactory}, \cite{gill2022method}, \cite{hofmeister2024cross-domain}, \cite{hofmeister2024semantic}, \cite{joseph2021aggregated}, \cite{malayjerdi2022combined}, \cite{park2020digital}, \cite{priyanta2024is}, \cite{zhang2021bi-level} \\
\;\;\corner{} Uncertainty & \subdatabar{8} & \cite{bellavista2023requirements}, \cite{bertoni2022digital}, \cite{clark2021chapter}, \cite{coupaye2023graph-based}, \cite{demir2023vertically-integrated}, \cite{oquendo2019dealing}, \cite{parri2021framework}, \cite{wang2024construction} \\
\;\;\corner{} Emergent Behaviors & \subdatabar{7} & \cite{barden2022academic}, \cite{chen2018digital}, \cite{dahmen2022modeling}, \cite{gil2024integrating}, \cite{kruger2022towards}, \cite{li2022cognitive}, \cite{liu2020web-based} \\
\;\;\corner{} Cost & \subdatabar{6} & \cite{ehemann2023digital}, \cite{gill2022method}, \cite{hatakeyama2018systems}, \cite{hatledal2020co-simulation}, \cite{mavromatis2024umbrella}, \cite{pickering2023towards} \\
\;\;\corner{} Data Interoperability & \subdatabar{6} & \cite{doubell2023digital}, \cite{kruger2022towards}, \cite{li2024comprehensive}, \cite{mahoro2023articulating}, \cite{park2020digital}, \cite{somma2023digital} \\
\;\;\corner{} Lifecycle Management & \subdatabar{4} & \cite{altamiranda2019system}, \cite{aziz2022empowering}, \cite{esterle2021digital}, \cite{heithoff2023challenges} \\
\;\;\corner{} Adoption & \subdatabar{4} & \cite{becue2018cyberfactory}, \cite{demir2023vertically-integrated}, \cite{gill2022method}, \cite{pickering2023towards} \\
\;\;\corner{} Decision Making & \subdatabar{4} & \cite{alam2017c2ps}, \cite{barden2022academic}, \cite{clark2021chapter}, \cite{zhang2021bi-level} \\
\;\;\corner{} Reconfiguration & \subdatabar{4} & \cite{clark2021chapter}, \cite{kruger2022towards}, \cite{oquendo2019dealing}, \cite{redelinghuys2020six-layer} \\
\;\;\corner{} Processing Efficiency & \subdatabar{3} & \cite{ehemann2023digital}, \cite{marah2023architecture}, \cite{saraeian2022digital} \\
\textbf{Design Challenges} & \textbf{\maindatabar{33}} & \\
\;\;\corner{} Complexity & \subdatabar{12} & \cite{bao2024digital}, \cite{dickopf2019holistic}, \cite{duan2023digital}, \cite{ehemann2023digital}, \cite{gill2022method}, \cite{lee2022simulation}, \cite{malayjerdi2022combined}, \cite{marah2023architecture}, \cite{pillai2023digital}, \cite{saraeian2022digital}, \cite{schluse2017experimentable}, \cite{zhang2022multi-scale} \\
\;\;\corner{} Lack Of Standards & \subdatabar{11} & \cite{acharya2023twins}, \cite{binder2021utilizing}, \cite{coupaye2023graph-based}, \cite{dickopf2019holistic}, \cite{gill2022method}, \cite{hatledal2020co-simulation}, \cite{hofmeister2024cross-domain}, \cite{howard2021greenhouse}, \cite{jirsa2024use}, \cite{larsen2024towards}, \cite{vogel-heuser2021approach} \\
\;\;\corner{} Compatibility With Legacy Systems & \subdatabar{7} & \cite{dobie2024network}, \cite{ehemann2023digital}, \cite{gill2022method}, \cite{howard2021greenhouse}, \cite{lippi2023enabling}, \cite{liu2020web-based}, \cite{lopez2023modeling} \\
\;\;\corner{} Regulatory Constraints & \subdatabar{3} & \cite{malayjerdi2022combined}, \cite{mavromatis2024umbrella}, \cite{wullink2024foundational} \\
\;\;\corner{} Lack Of Frameworks/Architectures & \subdatabar{3} & \cite{ashtaritalkhestani2019architecture}, \cite{howard2021greenhouse}, \cite{villalonga2021decision-making} \\
\;\;\corner{} Collaboration & \subdatabar{3} & \cite{barden2022academic}, \cite{demir2023vertically-integrated}, \cite{mavromatis2024umbrella} \\
\;\;\corner{} Sociotechnical Integration & \subdatabar{2} & \cite{folds2019digital}, \cite{mavromatis2024umbrella} \\
\;\;\corner{} Knowledge Representation & \subdatabar{2} & \cite{gil2023modeling}, \cite{vogel-heuser2021approach} \\
\textbf{Non-Functional Properties} & \textbf{\maindatabar{22}} & \\
\;\;\corner{} Scalability & \subdatabar{6} & \cite{bertoni2022digital}, \cite{chavezbaliguat2023digital}, \cite{clark2021chapter}, \cite{howard2021greenhouse}, \cite{pillai2023digital}, \cite{vermesan2021internet} \\
\;\;\corner{} Reliability & \subdatabar{4} & \cite{altamiranda2019system}, \cite{aziz2022empowering}, \cite{hofmeister2024semantic}, \cite{kutzke2021subsystem} \\
\;\;\corner{} Privacy & \subdatabar{4} & \cite{heininger2021capturing}, \cite{heithoff2023challenges}, \cite{stary2022privacy}, \cite{vermesan2021internet} \\
\;\;\corner{} Usability & \subdatabar{3} & \cite{chavezbaliguat2023digital}, \cite{mavromatis2024umbrella}, \cite{wagner2023using} \\
\;\;\corner{} Fidelity & \subdatabar{3} & \cite{folds2019digital}, \cite{potteiger2023live}, \cite{saraeian2022digital} \\
\;\;\corner{} Safety & \subdatabar{2} & \cite{joseph2021aggregated}, \cite{savur2019hrc-sos} \\
\;\;\corner{} Security & \subdatabar{2} & \cite{becue2018cyberfactory}, \cite{dobie2024network} \\
\bottomrule
\end{tabular}
\end{table*}

%% file: sections/conclusion.tex
\section{Conclusion}\label{sec:conclusion}

In this paper, we reported the results of our systematic literature review on systems of twinned systems, i.e., systems that combine the principles of digital twins and system of systems. Screening over 2\,500 potential studies, we selected and analyzed 80 of them.

Our findings indicate that systems of twinned systems are in an early stage of maturity. Some key contemporary challenges in systems of twinned systems include the lack of architectural specifications, standards, and the ignorance of human factors.
We invite researchers to contribute to these core challenges. Such efforts will enable better management of unexpected emergent behavior---a typical problem in system of systems that digital twins can aid. To aid the most critical challenge---the development of flexible systems of twinned systems architectures---we devise a conceptual reference framework to situate digital twins and system of systems in systems of twinned systems.

In future work, we will focus on developing reference architectures, supporting methods and technology, and, finally, the proper evaluation of our reference framework in a real cyber-physical swarm demonstrator.